\documentclass[
	superscriptaddress,
	twocolumn,
	amsmath,
	amssymb,
	nobibnotes,
	aps,
	pra,
	10pt
]{revtex4}

\usepackage[T1]{fontenc}
\usepackage[utf8]{inputenc}
\usepackage[english]{babel}
\usepackage{microtype}
\usepackage{lmodern}

\usepackage{mathtools}
\usepackage{braket}
\newtheorem{thm}{Theorem}

\usepackage{array}
\usepackage{booktabs}
\usepackage{multirow}
\usepackage{dcolumn}
\usepackage{placeins}
\usepackage{xcolor}

\begin{document}

\title{Optimized auxiliary oscillators for the simulation of general open quantum systems}

\author{F. Mascherpa}
\email[]{fabio.mascherpa@uni-ulm.de}
\affiliation{Institut f\"ur Theoretische Physik, Universit\"at Ulm, Albert-Einstein-Allee 11, D-89069 Ulm, Germany}
\author{A. Smirne}
\email[]{andrea.smirne@unimi.it}
\affiliation{Institut f\"ur Theoretische Physik, Universit\"at Ulm, Albert-Einstein-Allee 11, D-89069 Ulm, Germany}
\affiliation{Università degli Studi di Milano, Dipartimento di Fisica, Via Celoria 16, I-20133 Milano, Italy}
\author{A. D. Somoza}
\affiliation{Institut f\"ur Theoretische Physik, Universit\"at Ulm, Albert-Einstein-Allee 11, D-89069 Ulm, Germany}
\author{P. Fern\'andez-Acebal}
\affiliation{Institut f\"ur Theoretische Physik, Universit\"at Ulm, Albert-Einstein-Allee 11, D-89069 Ulm, Germany}
\author{S. Donadi}
\affiliation{Institut f\"ur Theoretische Physik, Universit\"at Ulm, Albert-Einstein-Allee 11, D-89069 Ulm, Germany}
\affiliation{Frankfurt Institute for Advanced Studies (FIAS), Ruth-Moufang-Stra{\ss}e 1, D-60438 Frankfurt am Main, Germany}
\author{D. Tamascelli}
\affiliation{Institut f\"ur Theoretische Physik, Universit\"at Ulm, Albert-Einstein-Allee 11, D-89069 Ulm, Germany}
\affiliation{Università degli Studi di Milano, Dipartimento di Fisica, Via Celoria 16, I-20133 Milano, Italy}
\author{S. F. Huelga}
\email[]{susana.huelga@uni-ulm.de}
\author{M. B. Plenio}
\email[]{martin.plenio@uni-ulm.de}
\affiliation{Institut f\"ur Theoretische Physik, Universit\"at Ulm, Albert-Einstein-Allee 11, D-89069 Ulm, Germany}

%\date{\today}

\begin{abstract}
A method for the systematic construction of few-body damped harmonic oscillator networks accurately reproducing the effect of general bosonic environments in open quantum systems is presented. Under the sole assumptions of a Gaussian environment and regardless of the system coupled to it, an algorithm to determine the parameters of an equivalent set of interacting damped oscillators obeying a Markovian quantum master equation is introduced. By choosing a suitable coupling to the system and minimizing an appropriate distance between the two-time correlation function of this effective bath and that of the target environment, the error induced in the reduced dynamics of the system is brought under rigorous control. The interactions among the effective modes provide remarkable flexibility in replicating non-Markovian effects on the system even with a small number of oscillators, and the resulting Lindblad equation for the system and the modes may therefore be integrated at a very reasonable computational cost using standard methods for Markovian problems, even in strongly non-perturbative coupling regimes and at arbitrary temperatures including zero. We apply the method to an exactly solvable problem in order to demonstrate its accuracy, and present two studies based on current research in the context of coherent transport in biological aggregates and organic photovoltaics as more realistic examples of its use and potential; performance and versatility are highlighted, and theoretical and numerical advantages over existing methods, as well as possible future improvements, are discussed.
\end{abstract}

\maketitle

\section{Introduction}

Any physical system in nature may be studied theoretically in complete isolation from its surroundings. However, since interactions with uncontrolled environmental degrees of freedom are unavoidable in practice, this condition is never actually realized. The effects of said degrees of freedom on the dynamics and general properties of a system are especially important in quantum mechanics, where the time and energy scales involved are likely to make interactions between the system and the surrounding environment a key actor in their own right in the physics at play. The goal of the theory of open quantum systems is to determine the behavior and investigate the physical properties of systems both in and out of equilibrium by properly accounting for environmental effects and other external influences (e.g.\ driving forces) using appropriate analytical or numerical methods~\citep{BreuerPetruccione, GardinerZoller, Weiss, RivasHuelga}.

The starting point of such methods may be either a microscopic model for the system and the environment, such as the spin-boson~\citep{Leggett_SpinBoson, BreuerPetruccione}, Caldeira--Leggett~\citep{CaldeiraLeggettModel} or more complex models, or an effective description of the system alone with the effects of the environment implicitly taken into account via a quantum master equation~\citep{GoriniKossakowskiSudarshan, GoriniFrigerioVerriKossakowskiSudarshan, Lindblad, Nakajima, Zwanzig, Prigogine, Shibata_TCL}. The former setup leads to a wide variety of potentially more complete and general treatments, but this greater range of attainable results and predictions comes at moderate to high computational costs~\citep{NEGF_Review, Tanimura_HEOM, Makri_QUAPIletter, ChinPlenio_TEDOPA, PriorPlenio_TEDOPA, DiosiStrunz_QuantumStateDiffusion, Piilo_NonMarkovianQuantumJumpsPRL}; the latter construction is typically less expensive but applies to a constrained class of physical settings, since it either delivers accurate results only in a few well-defined limiting cases~\citep{GoriniFrigerioVerriKossakowskiSudarshan, Davies_WeakCoupling, DumckeSpohn_WeakCoupling} or relies on equations which are difficult to derive for general systems~\citep{Nakajima, Zwanzig, Prigogine, Shibata_TCL, SmirneVacchini_NakajimaVsTCL, BreuerKapplerPetruccione_TCL}. Provided that the necessary assumptions on the system-environment interaction are satisfied, however, efficient methods for the solution of the master equation are widely available~\citep{Minchev_MatExp, PlenioKnight_MCWF, GisinPercival_QuantumStateDiffusion}.

Much theoretical research in recent decades has focused on the study of complex non-Markovian environments~\citep{RivasHuelgaPlenio_NonMarkovianity, BreuerVacchini_NonMarkovianity, DeVegaAlonso_NonMarkovianity, LiHallWiseman_NonMarkovianity}, for which analytical results are hard to obtain except for specific models, and numerical simulation may become very challenging depending on the physical regime of interest. For thermal bosonic environments, the most commonly studied category by far, numerical methods developed for a general treatment of non-Markovian problems include e.g.\ Hierarchical Equations of Motion (HEOM)~\citep{Tanimura_HEOM, TanimuraKubo_HEOM}, Quasi-Adiabatic Path Integrals (QUAPI)~\citep{Makri_QUAPIletter, TopalerMakri_QUAPI, MakriMakarov_QUAPI1, MakriMakarov_QUAPI2}, Nonequilibrium Green's Function (NEGF) techniques~\citep{Danielewicz_NEGF, NEGF_Review}, Non-Markovian Quantum State Diffusion (NMQSD) and similar stochastic methods~\citep{Diosi_QuantumStateDiffusion, Strunz_QuantumStateDiffusion, DiosiStrunz_QuantumStateDiffusion, Piilo_NonMarkovianQuantumJumpsPRA, Piilo_NonMarkovianQuantumJumpsPRL}, Time-Evolving Matrix Product Operators (TEMPO)~\citep{Lovett_TEMPO} or simulated evolution of the state using the time-adaptive Density Matrix Renormalization Group (t-DMRG)~\citep{Daley_tDMRG, Guifre_tDMRG, Schollwock_tDMRG} in combination with convenient exact mappings of the environment e.g.\ into one-dimensional oscillator chains well suited for such numerical methods, as in the Time-Evolving Density with Orthogonal Polynomials Algorithm (TEDOPA)~\citep{ChinPlenio_TEDOPA, PriorPlenio_TEDOPA, TamascelliSmirne_ThermalizedTEDOPA}, to name a few. These methods are often referred to as numerically exact, in the sense that they are designed to address problems from the bottom up, requiring only numerical approximations (e.g.\ Hilbert space truncation, discretized integrals or finite expansions of relevant functions) in order to keep the costs manageable, but otherwise posing no physical restrictions on the models themselves; these numerical errors can sometimes be bounded rigorously, e.g.\ for TEDOPA~\citep{WoodsCramerPleino_TEDOPAerrorbars, WoodsPlenio_LiebRobinsonBounds} or HEOM~\citep{SpinBosonBounds}. Finite bosonic environments~\citep{PiiloManiscalco_HarmonicNetworks} can also be used as an approximate treatment for simulation times short enough to prevent recurrence in the dynamics.

An alternative route for the numerical study of such nontrivial open-system problems is to model environmental effects on a system by splitting them into coherent, information-preserving contributions and purely dissipative Markovian damping. Then one can devise effective models in which the system of interest is coupled explicitly to a finite auxiliary system acting as the non-Markovian core of the environment, and dissipation is accounted for through Markovian damping of these auxiliary degrees of freedom. This is the idea underlying approaches such as the pseudomode method~\citep{Imamoglu_pseudomodes, Garraway_pseudomodes, Dalton_pseudomodes, Lemmer_SpinBoson}, the reaction-coordinate mapping~\citep{IlesSmith_ReactionCoordinate, IlesSmith_StructuredEnvironments, Lambert_SpinBosonStudy} or other techniques based on the same concept but differing in the ansatz used to create the effective environment and the techniques to solve for the dynamics~\citep{Fruchtman_Perturbative, Falci_1overf, Schwarz_LindbladDrivenDiscretizedLeads, Luchnikov_TensorNetworkLindblad, Faccioli_QTFT}. Such remappings of open-system problems can be very convenient numerically, but are not always grounded in a mathematically rigorous and physically sound relation between the original and effective environments, making assessment of their accuracy somewhat challenging.

In this paper, we present a new approach to general open quantum systems interacting with Gaussian bosonic environments. Our method combines the simplicity and efficiency of simulating a small set of effective degrees of freedom with analytical equivalence relations between the structure and parameters of this auxiliary system and the exact properties of the microscopic environment. Even in cases in which no exact equivalence holds, the physical error from replacing a unitary environment by a dissipative one is kept to a bare minimum and under rigorous control.

Our scheme is based on a quantitatively certified recipe to construct networks of interacting, damped harmonic oscillators specifically designed to mimic any given target environment as specified by its spectral density and temperature. The reduced dynamics is then computed by solving a time-homogeneous quantum master equation of the Gorini--Kossakowski--Sudarshan--Lindblad (GKSL) type~\citep{GoriniKossakowskiSudarshan, GoriniFrigerioVerriKossakowskiSudarshan, Lindblad} for the system coupled to these effective harmonic modes and tracing them out at the end. The theoretical foundation underlying this construction lies in a recently proved equivalence theorem between unitary and non-unitary Gaussian environments in open quantum systems~\citep{TSO_Theorem}, which states that the reduced dynamics of a system coupled to an environment of either type is identical if the single-time averages and two-time correlation functions of the environment operators relevant to the interaction are the same. Exploiting this notion, we introduce a systematic procedure by which the effective environment is tailored to reproduce the correlation function of the target environment with an accuracy quantitatively controlled through known error bounds for Gaussian environments~\citep{SpinBosonBounds}. The advantages of the method proposed are the simple yet versatile structure of the effective environments, which can emulate a broad range of nontrivial unitary environments using small numbers of auxiliary modes, the small, controlled error in the resulting effective dynamics, a high flexibility in the physical regimes which can be studied at comparably low costs, such as high and low temperature and strong as well as weak coupling, and numerical simplicity, since the simulations only require solving a Lindblad equation.

We have organized the presentation of our results as follows: in Section~\ref{sec:Theorem} we will outline the theoretical background and state the equivalence theorem from Ref.~\citep{TSO_Theorem} lying at the core of our method; Section~\ref{sec:TSO} details the procedure by which an effective environment corresponding to a nontrivial microscopic one may be constructed, and includes an analysis of the theoretical implications and approximations involved; a demonstration of our scheme on the spin-boson model as an exactly solvable test system, with accuracy and performance reports as well as a profile of the numerical advantages and disadvantages of the method in different physical regimes, is given in Section~\ref{sec:SpinBoson}; Section~\ref{sec:Applications} contains two applications of the method to systems in structured environments relevant to current research, namely optical signatures of coherent effects in biomolecular aggregates and the propagation of excitations in organic polymers with photovoltaic properties; in Section~\ref{sec:Discussion} we discuss the current state of the method, focusing on its scope and applicability, numerical and conceptual strengths and limitations as well as some possible improvements; finally, Section~\ref{sec:Conclusions} summarizes our conclusions and future prospects.

\section{Theoretical foundations and scope of the method}
\label{sec:Theorem}

The non-perturbative method we are going to introduce relies on the equivalence theorem between unitary and dissipative environments stated and proved in Ref.~\citep{TSO_Theorem}; in order to set the stage for discussing our work, we will now introduce the relevant notation, outline the physical context in which the theorem applies and state it explicitly for reference within the paper.

\subsection{Gaussian unitary environments}

A wide array of open quantum system (OQS) problems, ranging e.g.\ from quantum Brownian motion~\citep{BreuerPetruccione, CaldeiraLeggettModel} to dissipative cavity and circuit electrodynamics~\citep{YurkeDenker_QuantumCircuits, RibeiroVieira_Transport} or the study of charge and energy transfer in noisy natural or artificial aggregates~\citep{Scholes_QBioNature, HuelgaPlenio_QBio}, can be modeled microscopically by coupling the system of interest to an infinite collection of harmonic oscillators: the full system-environment Hamiltonian takes the form
\begin{equation}\label{eq:H}
	H\coloneqq H_S\otimes\mathbb{I}_E+\mathbb{I}_S\otimes H_E+H_I
\end{equation}
where $H_S$ is the free Hamiltonian of the system,
\[
	H_E\coloneqq\int^\infty_0\!\!\!\!\!\mathrm{d}\omega\,
	\hbar\omega a^\dagger_\omega a_\omega
\]
is the free Hamiltonian of the environment, expressed in terms of creation and annihilation operators obeying the continuum canonical commutation relations $[a_\omega,a_{\omega'}]=0$, $[a_\omega,a^\dagger_{\omega'}]=\delta(\omega-\omega')$, and the two are coupled through a general interaction term of the form~\citep{BreuerPetruccione}
\[
	H_I\coloneqq\sum_kA_{Sk}\otimes G_{Ek}
\]
with $A_{Sk}$ and $G_{Ek}$ operators acting on the system and the environment, respectively. In the following, we will consider these to be Hermitian, without loss of generality~\citep{RivasHuelga}.

The global state $\rho$ of the system and the environment at time $t$ is determined by the Liouville--Von Neumann equation
\begin{equation}\label{eq:U_full}
	\frac{\mathrm{d}}{\mathrm{d}t}\rho(t)=-\frac{i}{\hbar}[H,\rho(t)]
\end{equation}
and the initial state $\rho_0\coloneqq\rho(0)$; the reduced state $\rho_S$ of the system at time $t$ is obtained by taking the partial trace over the environmental degrees of freedom:
\begin{equation}
	\rho_S(t)=\mathrm{Tr}_E[\rho(t)].
\end{equation}

We are interested in the reduced dynamics of systems interacting with Gaussian environments, i.e.\ with $H_I$ linear in $a_\omega$ and $a^\dagger_\omega$ and factorizing initial conditions $\rho_0=\rho_{0S}\otimes\rho_{0E}$ with $\rho_{0E}$ a Gaussian state. Then the oscillators can be traced out exactly, and the reduced dynamics of the system only depends on the single- and two-time environmental averages $\langle G_{Ek}(t)\rangle_E$ and $C^E_{kk'}(t+\tau,\tau)\coloneqq\langle G_{Ek}(t+\tau)G_{Ek'}(\tau)\rangle_E$ as given by the evolution of the oscillators with no coupling to the system:
\begin{align}\label{eq:1tAverage_U}
	\langle G_{Ek}(t)\rangle_E&=\mathrm{Tr}_E[U^\dagger_E(t)G_{Ek}U_E(t)
	\rho_{0E}]
	\\ \label{eq:2tAverage_U}
	C^E_{kk'}(t+\tau,\tau)&=
	\mathrm{Tr}_E[U_E^\dagger(t+\tau)G_{Ek}U_E(t)G_{Ek'}U_E(\tau)\rho_{0E}]
\end{align}
with $U_E(t)\coloneqq e^{-iH_Et/\hbar}$.

\subsection{Gaussian dissipative environments}

Considering infinite environments evolving unitarily in the absence of a coupled system is one way to bring about dissipation and decoherence in the evolution of the latter when the coupling is nonzero. Alternatively, one may consider finite environments which evolve non-unitarily according to a quantum master equation (QME). In this case, one may start from a different combined Hamiltonian
\begin{equation}\label{eq:H'}
	H'\coloneqq H_S\otimes\mathbb{I}_R+\mathbb{I}_S\otimes H_R+H'_I
\end{equation}
and a QME describing the evolution of the dissipative environment when decoupled from the system:
\begin{equation}\label{eq:L_free}
	\frac{\mathrm{d}}{\mathrm{d}t}\rho_R(t)=\mathcal{L}_R[\rho_R(t)],
\end{equation}
where the new quantum Liouville superoperator
\[
	\mathcal{L}_R[\rho_R]\coloneqq
	-\frac{i}{\hbar}[H_R,\rho_R]+\mathcal{D}_R[\rho_R]
\]
on the right-hand side includes a dissipator
\[
	\mathcal{D}_R[\rho_R]\coloneqq\sum^m_{i,j=1}\Gamma_{ij}\left(
	L_{Ri}\rho_RL^\dagger_{Rj}-\frac{1}{2}\left\{L^\dagger_{Rj}L_{Ri},
	\rho_R\right\}\right)
\]
with a positive semidefinite rate matrix $\Gamma_{ij}$. This makes the dynamics non-unitary but ensures a completely positive and trace-preserving evolution at all positive times; the rate matrix $\Gamma_{ij}$, which we take to be constant, can always be brought into diagonal form by changing the basis of operators $L_{Ri}$~\citep{BreuerPetruccione}, giving the quantum dynamical semigroup master equation for Markovian open systems first derived by Gorini, Kossakowski, Sudarshan and Lindblad~\citep{GoriniKossakowskiSudarshan, GoriniFrigerioVerriKossakowskiSudarshan, Lindblad}. We will refer to this master equation simply as the Lindblad equation throughout this paper.

The full state of the system and a non-unitary environment evolves according to the QME
\begin{equation}\label{eq:L_full}
	\frac{\mathrm{d}}{\mathrm{d}t}\rho(t)=\mathcal{L}[\rho(t)]
\end{equation}
where
\[
	\mathcal{L}[\rho]\coloneqq-\frac{i}{\hbar}[H',\rho]+\mathcal{D}[\rho]
\]
is the complete quantum Liouvillian for the system and the environment and $\mathcal{D}\coloneqq\mathbb{I}\otimes\mathcal{D}_R$ embeds the dissipator $\mathcal{D}_R$ into the full Hilbert space of the problem.

For harmonic environments coupled linearly to the system, i.e.\ for
\[
	H_R\coloneqq\sum_n\hbar\omega_nb^\dagger_nb_n
\]
with $[b_m,b_n]=0$, $[b_m,b^\dagger_n]=\delta_{mn}$ and
\[
	H'_I\coloneqq\sum_lA_{Sl}\otimes F_{Rl}
\]
with $F_{Rl}$ linear in the creation and annihilation operators, if one also takes the Lindblad operators $L_{Ri}$ linear in $b_n$ and $b^\dagger_n$ and initial conditions $\rho_0=\rho_{0S}\otimes\rho_{0R}$ with a Gaussian $\rho_{0R}$, then the reduced dynamics of the system will only depend on the environment through $\langle F_{Rl}(t)\rangle_R$ and $C^R_{ll'}(t+\tau,\tau)\coloneqq\langle F_{Rl}(t+\tau)F_{Rl'}(\tau)\rangle_R$, again considering the free dynamics of the environment with no system attached, like in the unitary case:
\begin{align}\label{eq:1tAverage_L}
	\langle F_{Rl}(t)\rangle_R&=\mathrm{Tr}_R[F_{Rl}e^{\mathcal{L}_Rt}
	[\rho_{0R}]]
	\\ \label{eq:2tAverage_L}
	C^R_{ll'}(t+\tau,\tau)&=
	\mathrm{Tr}_R[F_{Rl}e^{\mathcal{L}_Rt}[F_{Rl'}e^{\mathcal{L}_R\tau}
	[\rho_{0R}]]].
\end{align}
Note that the two-time correlation function~\eqref{eq:2tAverage_L} has the form one would obtain by applying the quantum regression hypothesis~\citep{Lax_QRT}, which must be handled with some care in general but is true by construction for the Lindblad-damped environments relevant to our work. No approximation is required or implied at this stage~\citep{TSO_Theorem}.

\subsection{Equivalence between unitary and non-unitary environments}

While it is clear that if two unitary Gaussian environments share the same averages $\langle G_{Ek}(t)\rangle_E$ and correlation functions $C^E_{kk'}(t+\tau,\tau)$ at all times they will give rise to the same reduced dynamics if coupled to a system, this is not obvious if one or both environments are not unitary. In Ref.~\citep{TSO_Theorem} it was shown, using the unitary dilation formalism for Lindblad equations~\citep{GardinerZoller}, that this still holds for non-unitary environments under the same conditions. We restate this result here for reference.

Define the reduced dynamics
\begin{equation}
	\rho^U_S(t)\coloneqq\mathrm{Tr}_E[\rho(t)]
\end{equation}
for some system $S$ coupled to a unitary environment and evolving according to Eq.~\eqref{eq:U_full} from factorizing initial conditions with the environment starting in a Gaussian state, and the reduced dynamics
\begin{equation}
	\rho^L_S(t)\coloneqq\mathrm{Tr}_R[\rho(t)]
\end{equation}
for the same system coupled to a non-unitary environment and evolving according to Eq.~\eqref{eq:L_full} from factorizing initial conditions with the environment starting in a Gaussian state.

Both environments are taken to be harmonic and coupled to the system through the same set of $A_{Sk}$ operators in $H_I$ and $H'_I$, with the corresponding $G_{Ek}$ and $F_{Rk}$ as well as the Lindblad operators $L_{Ri}$ of the non-unitary environment linear in the relevant creation and annihilation operators. The initial state of the system is taken to be the same.

\begin{thm}\label{TSO_Theorem}
\citep{TSO_Theorem} Under the above assumptions, if
\[
	\langle F_{Rk}(t)\rangle_R=\langle G_{Ek}(t)\rangle_E\quad\forall k,t
\]
and
\[
	C^R_{kk'}(t+\tau,\tau)=C^E_{kk'}(t+\tau,\tau)\quad\forall k,k',t,\tau,
\]
then
\[
	\rho^L_S(t)=\rho^U_S(t)\quad\forall t.
\]
\end{thm}

This theorem is the cornerstone of our method; for the sake of clarity and an easier understanding of the rest of this paper, some remarks are in order.

First of all, it is important to stress that Gaussianity is a key ingredient of Theorem~\ref{TSO_Theorem}, because in principle all correlation functions up to infinite order would have to be equal for two environments to have the same effect on a system, but for Gaussian environments the single- and two-time functions generate all the others. This restricts the initial state of the environment to the Gaussian family; in this work, we will only consider system-environment product states which are Gaussian in the environmental variables as initial states, leaving the free Hamiltonian, interaction operators and initial density matrix of the system arbitrary.

Furthermore, we restrict our study to systems coupled to bosonic baths in this paper but a result analogous to Theorem~\ref{TSO_Theorem} was recently proved for fermionic environments as well~\citep{Chen_FermionicTSO}, making an extension of our work to fermionic open-system problems possible.

Finally, for physical reasons discussed in Section~\ref{sec:Discussion} and thoroughly analyzed in Ref.~\citep{Talkner_NoQRT}, in general a finite network of damped harmonic oscillators does not yield a two-time correlation function exactly equal to that of an infinite bath in a Gaussian equilibrium state, so we will apply the theorem in approximate form by looking for effective parameters such that $C^R_{kk'}(t+\tau,\tau)\approx C^E_{kk'}(t+\tau,\tau)$ and hence $\rho^L_S(t)\approx\rho^U_S(t)$ (single-time expectation values of coupling operators are typically zero or can be set to zero and will no longer be dealt with in this work), relying on the fact that the error in the former approximate relation rigorously bounds that in the latter, as established in previous work~\citep{SpinBosonBounds}.

Other than these caveats, no further problems arise in terms of applicability; in particular, temperature and coupling strength between system and environment pose no theoretical or computational limits in principle.

In the next sections we will show how one may exploit the theorem to systematically construct simple networks of damped harmonic oscillators, which can stand in for complex, highly non-Markovian thermal baths at any temperature, by comparing the associated correlation functions~\eqref{eq:2tAverage_L} and~\eqref{eq:2tAverage_U}. This procedure is independent of the system and the effective environments obtained through it can then be coupled arbitrarily strongly to any system of interest. Standard simulation methods for Lindblad equations may then be used to obtain the reduced dynamics at potentially very low computational costs.

\section{Systematic construction of effective environments}
\label{sec:TSO}

From now on, we will consider unitary environments with Gaussian stationary states, such as thermal baths, and assume them to be initialized in such states, so that
\[
	C^E_{kk'}(t+\tau,\tau)=C^E_{kk'}(t,0),
\]
which will be denoted by $C^E_{kk'}(t)$ in the following.

Any harmonic oscillator network obeying a Lindblad equation of the form~\eqref{eq:L_free}, with $\mathcal{L}_R$ quadratic in $b_n$ and $b^\dagger_n$, must also start from a stationary $\rho_{0R}$ in order to give a time-homogeneous correlation function matrix $C^R_{kk'}(t)\coloneqq C^R_{kk'}(t+\tau,\tau)=C^R_{kk'}(t,0)$ for operators of the form
\begin{equation}\label{eq:InteractionOperators}
	F_{Rk}=\sum_n(c_{nk}b_n+c^*_{nk}b^\dagger_n).
\end{equation}
The condition for $\rho_{0R}$ to be a stationary state of Eq.~\eqref{eq:L_free} is
\begin{equation}\label{eq:StationaryLindblad}
	\mathcal{L}_R[\rho_{0R}]=0.
\end{equation}
For the initial state of our effective environments, we will therefore need a Gaussian $\rho_{0R}$ satisfying this property.

\subsection{Ansatz and correlation function structure}

The correlation functions $C^R_{kk'}(t)$ of the auxiliary environment depend on all parameters appearing in $\mathcal{L}_R$, $\rho_{0R}$ and the operators $F_{Rk}$: unrestricted geometries and initial states allow for more generality at the expense of keeping potentially redundant parameters in the model and restricting the range of properties that can be easily calculated; to strike a balance between simplicity and versatility, we will now take an ansatz for the configuration and initial density matrix of the surrogate oscillator network such that the quantities of interest have a simple expression with little loss of generality; for a more extensive discussion of the technical details, we refer the reader to Appendix~\ref{app:TSO_details}.

We choose a free Hamiltonian $H_R$ corresponding to a chain of $N$ oscillators with a hopping interaction between nearest neighbors:
\begin{equation}\label{eq:H_R}
	H_R\coloneqq\sum^N_{n=1}\hbar\Omega_n b^\dagger_n b_n
	+\sum^{N-1}_{n=1}\hbar g_n\left(b_nb^\dagger_{n+1}+b^\dagger_nb_{n+1}\right),
\end{equation}
where the couplings $g_n$, as well as one of the coefficients $c_{nk}$ in the interaction operators $F_{Rk}$ appearing in $C^R_{kk'}(t)$, can be assumed real without loss of generality if the $F_{Rk}$ are nonlocal, i.e.\ acting on all effective modes (see Appendix~\ref{app:TSO_details}). We complete the QME by adding local thermal dissipators at zero temperature acting on each oscillator:
\begin{multline}\label{eq:L_R_0K}
	\frac{\mathrm{d}}{\mathrm{d}t}\rho_R(t)=-\frac{i}{\hbar}[H_R,\rho_R(t)]
	\\
	+\sum^N_{n=1}\Gamma_n\left(b_n\rho_R(t)b^\dagger_n
	-\frac{1}{2}\left\{b^\dagger_nb_n,\rho_R(t)\right\}\right)
\end{multline}
so that the stationary initial state satisfying Eq.~\eqref{eq:StationaryLindblad} is just the overall vacuum state
\begin{equation}\label{eq:Rho_0R}
	\rho_{0R}=\bigotimes^N_{n=1}\ket{0}\!\bra{0}_n.
\end{equation}

Note that a zero-temperature master equation for the effective environment does not restrict the temperature of the target environments it can simulate; the effect of a nonzero temperature in the target bath will simply be encoded in the parameters of the oscillator network, as is done in other approaches~\citep{MayKuehn, DeVegaBanuls_Thermofield, DiosiGisinStrunz_NonMarkovianity, Ritschel_AbsorptionSpectra, TamascelliSmirne_ThermalizedTEDOPA}.

The QME~\eqref{eq:L_R_0K} and initial condition~\eqref{eq:Rho_0R} lead to two decoupled sets of linear equations for $\langle b_n(t)\rangle_R$ and $\langle b^\dagger_n(t)\rangle_R$ related by Hermitian conjugation. For $\langle b_n(t)\rangle_R$ one has
\begin{equation}\label{eq:<a(t)>}
	\frac{\mathrm{d}}{\mathrm{d}t}\langle b_n(t)\rangle_R
	=\sum^N_{m=1}M_{nm}\langle b_m(t)\rangle_R,
\end{equation}
where
\begin{equation}\label{eq:MatrixM}
	M_{nm}:=\left(-\frac{\Gamma_n}{2}-i\Omega_n\right)\delta_{nm}
	-i(g_m\delta_{n\,m+1}+g_{m-1}\delta_{n\,m-1}).
\end{equation}
The two-time correlation function $\langle b_n(t)b^\dagger_m(0)\rangle_R$ also evolves according to Eq.~\eqref{eq:<a(t)>} as a direct consequence of the quantum regression hypothesis, which holds by construction in this context and states that correlation functions $\langle A(t+\tau)B(t)\rangle$ obey the same equations of motion as the single-time expectation values $\langle A(t+\tau)\rangle$~\citep{GardinerZoller, Carmichael}. This is equivalent to the statement that they can be written in the explicit form given in Eq.~\eqref{eq:2tAverage_L}. Integrating Eq.~\eqref{eq:<a(t)>} for $\langle b_n(t)b^\dagger_m(0)\rangle_R$ and plugging the result (as well as its conjugate $\langle b^\dagger_n(t)b_m(0)\rangle_R$, which is identically zero for our initial state~\eqref{eq:Rho_0R}) into the expression of $C^R_{kk'}(t)$ in terms of the operators $F_{Rk}$ as given in Eq.~\eqref{eq:InteractionOperators}, one finds
\begin{equation}\label{eq:CR}
	C^R_{kk'}(t)=\sum^N_{n=1}(w_n)_{kk'}e^{\lambda_nt}
\end{equation}
where $\lambda_n$ are the eigenvalues of the matrix $M$ defined in Eq.~\eqref{eq:MatrixM}, which we assume to be non-degenerate for simplicity (see Appendix~\ref{app:TSO_details} for further discussions), and
\begin{equation}\label{eq:ws}
	(w_n)_{kk'}=\sum^N_{l,m=1}c_{lk}c^*_{mk'}u^n_lv^n_m
\end{equation}
are complex coefficients obtained from the definition of the operators $F_{Rk}$ and the left and right eigenvectors $\mathbf{u}^n$ and $\mathbf{v}^n$ corresponding to each $\lambda_n$, normalized in such a way that $\sum^N_{l=1}v^m_lu^n_l=\delta_{mn}$. This exponential structure is a consequence of the Lindblad dynamics of the effective environment, which is a requirement of Theorem~\ref{TSO_Theorem}.

\subsection{Transformation to Surrogate Oscillators}

Consider an OQS problem described by a microscopic model of the form~\eqref{eq:H}; for simplicity we will now assume a single interaction term, to which there corresponds a correlation function $C^E(t)$. Our goal is to find the matrix elements $M_{mn}$ and operator coefficients $c_n$ of some operator $F_R$ as given in Eq.~\eqref{eq:InteractionOperators} such that the resulting effective correlation function $C^R(t)$ is as close as possible to $C^E(t)$.

The form of $C^R(t)$ in terms of $M_{mn}$ and $c_n$ is given by Eq.~\eqref{eq:CR}, where the eigenvalues $\lambda_n$ and weights $w_n$ can be thought of as functions of the free parameters $\Omega_k$, $g_k$, $\Gamma_k$ and---only the $w_n$---$c_k$ with $n,k=1,\dots,N$, where $N$ is the number of oscillators making up the effective bath.

In order to determine the values of these free parameters such that $C^R(t)\approx C^E(t)$, we proceed in two steps. First, we perform a nonlinear fit on $C^E(t)$ using $N$ damped exponentials with complex coefficients
\begin{equation}\label{eq:Fit}
	C^E(t)\longrightarrow\tilde{C}^E(t)=\sum^N_{n=1}\tilde{w}_ne^{\tilde{\lambda}_nt},
\end{equation}
for instance using Prony analysis~\citep{Marple_Prony}. Note that the sum of the coefficients $\tilde{w}_n$ is real and positive, since $\sum^N_{n=1}\tilde{w}_n=\tilde{C}^E(0)$ must be equal to $C^E(0)=\left\langle G^2_E(0)\right\rangle_E$, which is positive because $G_E$ is Hermitian.

\begin{figure*}
	\centering
	\input{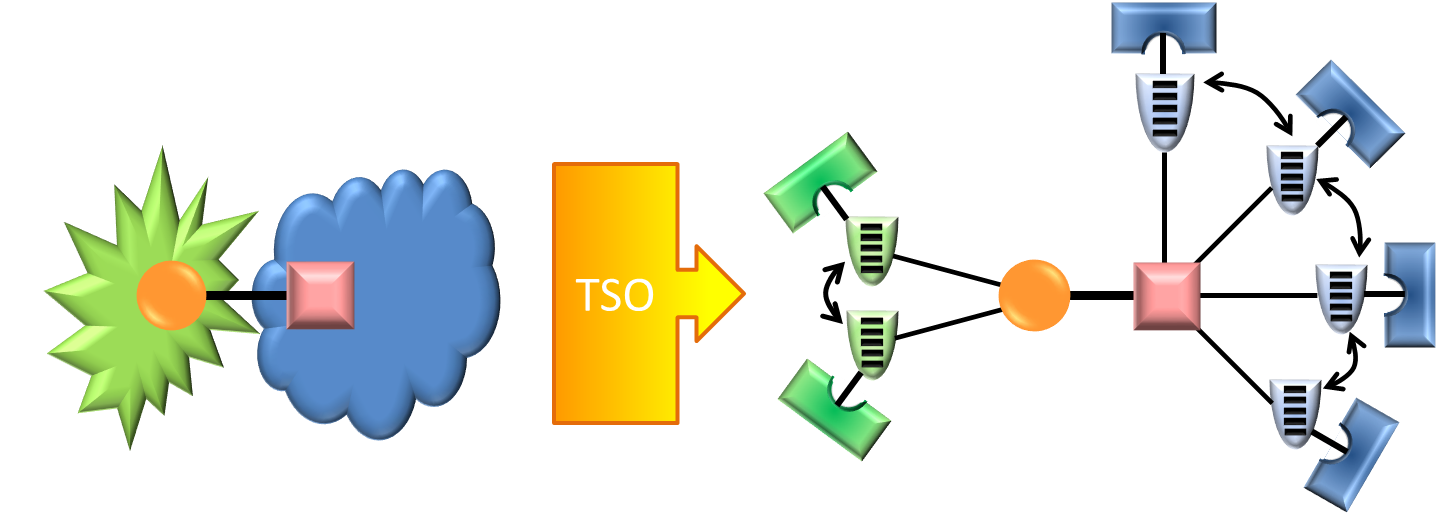}
	\caption{\label{Fig:TwoSystems}\textbf{Simulation of composite systems.} When applying the transformation to surrogate oscillators (TSO) to interacting systems coupled to local environments, each environment is replaced by the corresponding effective one, regardless of the properties of the system attached to it. Our procedure leads to modular structures which do not require a rederivation of the effective parameters when couplings among separate open systems are introduced. This makes the method suitable for the treatment of polymers with local environments surrounding each fundamental unit, as will be shown in a later section.}

	\begin{minipage}{.48\textwidth}
		\centering
		\input{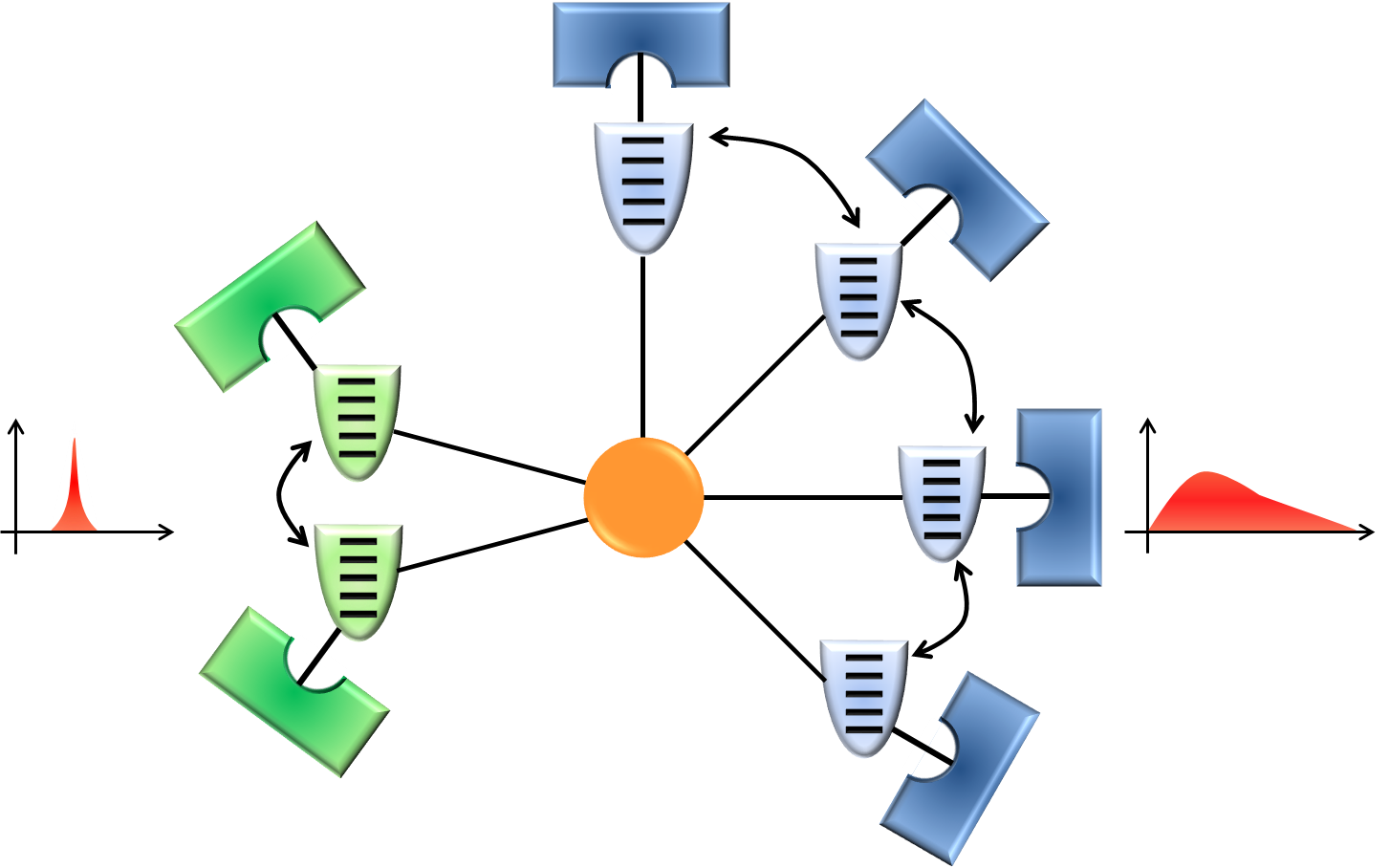}
		\caption{\label{Fig:SystemBath}\textbf{Simulation of structured environments.} A system coupled to an environment with spectral density $J(\omega)=J'(\omega)+J''(\omega)$, with $J'(\omega)$ a broad background and $J''(\omega)$ a sharp resonance as shown in the small plots, mapped to two distinct effective environments, one with $N=4$ and another with $N=2$ oscillators. We will encounter a similar structure in our example applications in Section~\ref{sec:Applications}.}
	\end{minipage}
	\hfill
	\begin{minipage}{.48\textwidth}
		\centering
		\input{Jeff_T1}
		\caption{\label{Fig:OhmicT1CF}Fourier transform of the Ohmic correlation function $C^E_\beta(t)$ with $\beta\Omega_c=1$ (solid orange line) and of the corresponding $C^R_\beta(t)$ from the TSO with parameters given in Table~\ref{tb:OhmicT1parameters} (dashed blue line). The inset shows the difference as a function of frequency.}
	\end{minipage}
\end{figure*}

Then we solve the problem of matching or getting as close as possible to the $\tilde{\lambda}_n$ and $\tilde{w}_n$ with the $\lambda_n(\Omega_k,g_k,\Gamma_k)$ and $w_n(\Omega_k,g_k,\Gamma_k,c_k)$ from the effective environment. This is in general a highly nontrivial inversion problem involving an underdetermined, non-convex system of nonlinear equations of mixed degrees, and can be hard to solve: there is a trade-off between this complexity and the accuracy of the initial fit, with an optimum at small numbers ($N\leqslant 5$ in all our applications) of interacting oscillators. Neither existence nor uniqueness of solutions are guaranteed for this inversion problem and physical requirements such as positivity of the rates $\Gamma_n$ need to be taken into account as well, so it is typically necessary to minimize some distance between $C^R(t)$ and $C^E(t)$ instead of exactly matching the best fit $\tilde{C}^E(t)$; this change in the correlation function is the only error introduced into the problem by the use of an effective environment.

In some cases with $N\leqslant3$, it is possible to invert the equations exactly and obtain valid effective bath parameters; we have listed a few explicit solutions in Appendix~\ref{app:ExactSols} and will put some of them to use in our example applications. For general $N$, we devised a variational recipe to carry out our Transformation to Surrogate Oscillators (TSO) systematically. This is described in detail in Appendix~\ref{app:TSO_details}. The whole procedure can be summarized as follows:

\begin{itemize}
	\item Fit $C^E(t)$ with $N$ complex exponentials $e^{\tilde{\lambda}_nt}$ with complex coefficients $\tilde{w}_n$ such that $\sum^N_{n=1}\tilde{w}_n>0$.
	\item Sample random points in a suitably sized open set $(0,g_\mathrm{max})^{N-1}$, to be used as coupling constants.
	\item Substitute each $(N-1)$-tuple $(g_1,\dots,g_{N-1})$ into the equations relating the complex eigenvalues $\lambda_n$ to the $\tilde{\lambda}_n$ and solve: this will give rates $\Gamma_n$ and frequencies $\Omega_n$ such that the eigenvalues match; accept only solutions with all $\Gamma_n>0$.
	\item Compute the left and right eigenvectors of the matrices $M$ corresponding to each solution found, plug them into Eq.~\eqref{eq:ws} and minimize a distance (e.g.\ the Manhattan distance $d_\mathrm{Man}(\mathbf{w},\mathbf{\tilde{w}})\coloneqq\sum^N_{n=1}|w_n-\tilde{w}_n|$) between $\mathbf{w}$ and $\mathbf{\tilde{w}}$ by varying the $c_n$.
	\item Assess overall accuracy of the solutions found and rank the corresponding $C^R(t)$ according to a meaningful figure of merit, such as the integral $\int^t_0\mathrm{d}t'\int^{t'}_0\mathrm{d}t''|C^R(t'-t'')-C^E(t'-t'')|$ from Ref.~\citep{SpinBosonBounds}.
	\item If the accuracy of all effective correlation functions obtained is deemed insufficient, repeat with one more mode.
\end{itemize}

Effective environments obtained through this procedure can then be used to simulate the reduced dynamics of any model in which the interaction with the bath is mediated by the same bath operator $F_R$: the TSO is carried out once and for all irrespective of the system coupled to the environment given, and the effective environment can be used in any problem involving the same correlation function $C^E(t)$. We also wish to remark that for composite systems with multiple local environments, the procedure applies to each independent correlation function individually and yields local effective environments to be coupled to the corresponding parts of the system in the same way as the original ones, with no further complications arising: this feature is sketched in Fig.~\ref{Fig:TwoSystems} and will be demonstrated in Section~\ref{sec:Applications}.

By the same token, complex correlation functions requiring many exponentials for an accurate fit can be treated by breaking down the effective environment into smaller clusters of interacting modes, with each cluster accounting for a different component of $C^E(t)$---or equivalently, the underlying spectral density $J(\omega)$ of the unitary environment---as shown in Fig.~\ref{Fig:SystemBath}. Note that decoupling all oscillators from each other, i.e.\ taking all $g_n=0$ (which corresponds to requiring all $\tilde{w}_n$ to be real and positive in Eq.~\eqref{eq:Fit}), one recovers the noninteracting pseudomodes of Ref.~\citep{Garraway_pseudomodes} as a limiting case.

\subsection{Working example: Ohmic spectral density}

To better illustrate the technique explained in the previous subsection, let us now demonstrate how our transformation works with an explicit example.

Consider an arbitrary quantum system coupled to an infinite environment in thermal equilibrium through the position operator of each oscillator (for a more succinct notation, we will leave the tensor products implicit and use natural units $\hbar=1$, $k_B=1$ from now on):
\begin{equation}
	H=H_S+\int^\infty_0\!\!\!\!\!\mathrm{d}\omega\,
	\omega a^\dagger_\omega a_\omega+A_S\int^\infty_0\!\!\!\!\!\mathrm{d}\omega\,
	g(\omega)(a_\omega+a^\dagger_\omega).
\end{equation}
This type of coupling for microscopic models is one of the most common in the OQS literature~\citep{BreuerPetruccione, Weiss, CaldeiraLeggettModel, Leggett_SpinBoson, DeVegaAlonso_NonMarkovianity}. For a thermal initial state at inverse temperature $\beta=1/T$, the correlation function of the interaction operator $G_E=\int^\infty_0\mathrm{d}\omega\,g(\omega)(a_\omega+a^\dagger_\omega)$ is
\begin{equation}\label{eq:ThermalCF}
	\begin{split}
		C^E_\beta(t)&=\langle G_E(t)G_E(0)\rangle_{\beta E}
		\\
		&=\int^\infty_0\!\!\frac{\mathrm{d}\omega}{\pi}J(\omega)
		\left(\coth\left(\frac{\beta\omega}{2}\right)\cos(\omega t)-i\sin(\omega t)\right)
	\end{split}
\end{equation}
where the spectral density $J(\omega)$ is related to the frequency-dependent coupling strength $g(\omega)$ through
\[
	J(\omega)=\pi g^2(\omega)
\]
and typically given as a starting point for studying the problem. Spectral densities are real and positive by definition, and are often categorized according to the power of $\omega$ best approximating their behavior near the origin, where they are always zero; a $J(\omega)\propto\omega^s$ is called Ohmic if $s=1$, and super-(sub-)Ohmic if $s>1$ ($s<1$). The spectral density and temperature uniquely determine $C^E_\beta(t)$ and, consequently, the effect of the environment on the system.

\begin{table*}
	\centering
	\begin{tabular*}{.65\textwidth}{@{\extracolsep{\fill}}cccccccc}
	\toprule
	& Mode 1	& &	Mode 2	& &	Mode 3	& &	Mode 4 \\
	\midrule
	$\Omega_n$	&	$0.512683$	& &	$2.53779$	& &	$4.53293$	& &	$0.151433$ \\[0.5ex]
	$g_n$ & &	$1.82454$	& &	$3.20774$	& &	$1.60194$ & \\[0.5ex]
	$\Gamma_n$ & $0.056336$	& &	$4.42709$	& &	$15.7371$	& &	$0.110104$ \\[0.5ex]
	\multirow{2}{*}{$c_n$} & $-0.962917$	& &	$-0.227707$	& &	$0.231179$	& &
	\multirow{2}{*}{$0.818093$} \\
	& $+0.819128i$	& &	$+0.0701249i$	& &	$-0.137866i$	& & \\
	\bottomrule
	\end{tabular*}
	\caption{\label{tb:OhmicT1parameters}Effective parameters for $N=4$ surrogate modes corresponding to the correlation function of an Ohmic bath at temperature $T=\Omega_c$ (Eq.~\eqref{eq:OhmicC}). All parameters have dimensions of frequency and are given in units of $\Omega_c$; the last $c_n$ is real.}
\end{table*}

Note that for unitary environments the correlation function is Hermitian in time, i.e.\ its real part is even and its imaginary part is odd, as can be seen clearly from Eq.~\eqref{eq:ThermalCF}. This implies that its Fourier transform
\begin{equation}\label{eq:C_Fourier}
	\begin{split}
		C^E_\beta(\omega)&=\int^\infty_{-\infty}\!\!\!\!\!\mathrm{d}t\,
		C^E_\beta(t)e^{i\omega t}
		\\
		&=\left(1+\coth\left(\frac{\beta\omega}{2}\right)\right)
		\!(J(\omega)\theta(\omega)-J(-\omega)\theta(-\omega)),
	\end{split}
\end{equation}
where $\theta(\omega)$ is the Heaviside step function, is always real; at temperature $T=0$, it is just $2J(\omega)\theta(\omega)$. In fact, $C^E_\beta(\omega)/2$ may itself be regarded as a spectral density defined over a new environment, which comprises both positive- and negative-frequency modes and gives the correlation function $C^E_\beta(t)$ if initialized in the vacuum state~\citep{MayKuehn}: this allows one to effectively rephrase arbitrary-temperature OQS problems as zero-temperature ones if it is convenient to do so, a possibility exploited by thermofield-based and other numerical methods~\citep{DeVegaBanuls_Thermofield, DiosiGisinStrunz_NonMarkovianity, Ritschel_AbsorptionSpectra, TamascelliSmirne_ThermalizedTEDOPA}.

For non-unitary environments, in which time evolution is not an invertible map, correlation functions $C^R(t)$ are only defined at positive times; we extend the definition to negative times by imposing the same symmetry
\[
	C^R(-t)\coloneqq C^{R*}(t)\quad\forall t>0
\]
in order to be able to compare exact and effective correlation functions in the frequency domain instead of inspecting their real and imaginary parts separately.

Consider an Ohmic $J(\omega)$ with an exponential cutoff:
\begin{equation}\label{eq:OhmicJ}
	J(\omega)=\pi\omega e^{-\frac{\omega}{\Omega_c}}.
\end{equation}
Ohmic spectral densities define a very important class of environments entering the study of many systems, such as a particle undergoing quantum Brownian motion, or microscopic models leading to a Lindblad equation for a harmonic oscillator in a weakly coupled high-temperature environment~\citep{BreuerPetruccione, CaldeiraLeggettModel, FordLewisOConnell_DampedOscillator}. The thermal correlation function $C^E_\beta(t)$ corresponding to the spectral density defined in Eq.~\eqref{eq:OhmicJ} can be determined analytically as
\begin{equation}\label{eq:OhmicC}
	\begin{split}
		C^E_\beta(t)&=\frac{\Omega^2_c}{(1+i\Omega_ct)^2}
		\\
		&+\frac{1}{\beta^2}
		\left(\psi'\left(1+\frac{1+i\Omega_ct}{\beta\Omega_c}\right)
		+\psi'\left(1+\frac{1-i\Omega_ct}{\beta\Omega_c}\right)\right)
	\end{split}
\end{equation}
where $\psi'(z)\coloneqq\frac{1}{\Gamma(z)}\frac{\mathrm{d}\Gamma(z)}{\mathrm{d}z}$ is the polygamma function of order one.

Performing our TSO on this correlation function at temperature $T=\Omega_c$ according to the recipe described in the previous subsection, for $N=4$ we determined the parameters given in Table~\ref{tb:OhmicT1parameters}; Fig.~\ref{Fig:OhmicT1CF} shows the Fourier-transformed effective correlation function $C^R_\beta(\omega)$ obtained using these parameters and the target $C^E_\beta(\omega)$ for comparison. As can be seen from the plot, four interacting oscillators were enough to obtain a very accurate $C^R_\beta(t)$, with a peak in the error around $\omega=0$ reaching about 2\% of the function value (see the inset of Fig.~\ref{Fig:OhmicT1CF}). This error affects the correlation function at very long times compared to its decay time, so we expect it to have a minor impact on the transient reduced dynamics of the system and become potentially more important at very long times. In all our tests, a small region around the origin was consistently found to be the part of the frequency domain where a general $C^E_\beta(\omega)$ is hardest to match: this is because any $C^R_\beta(\omega)$ is analytical around zero by construction, whereas $C^E_\beta(\omega)$ has discontinuous derivatives, as can be checked from Eq.~\eqref{eq:ThermalCF}. We stress again that the nonzero temperature is encoded in the effective parameters and not in the initial state, allowing us to treat very different temperature regimes at comparable costs, as will be made clearer in the next sections.

\begin{figure*}
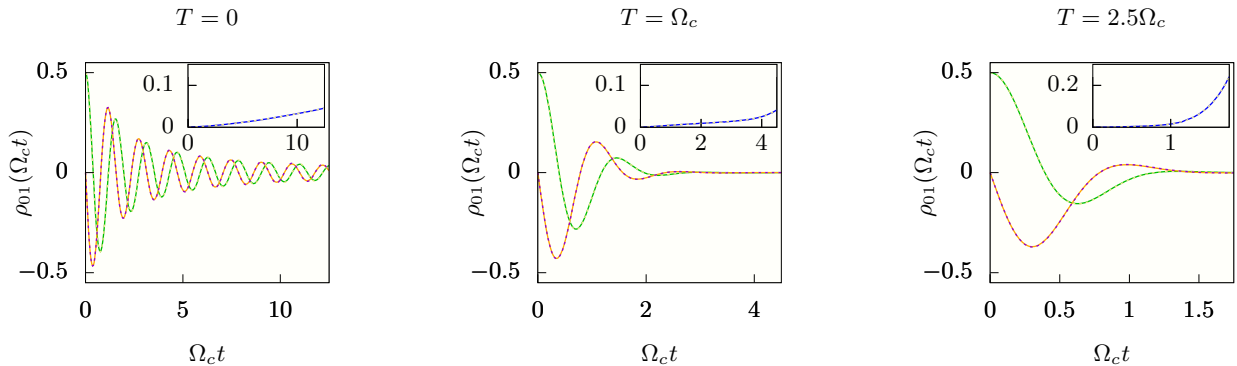

	\centering
	\begin{minipage}{.329\textwidth}
		\centering
		\input{SpinT0}
	\end{minipage}
	\hfill
	\begin{minipage}{.329\textwidth}
		\centering
		\input{SpinT1}
	\end{minipage}
	\hfill
	\begin{minipage}{.329\textwidth}
		\centering
		\input{SpinT2}
	\end{minipage}
	\caption{\label{Fig:SpinBoson}\textbf{Spin-boson results.} Time evolution of the density matrix of a qubit starting from the state $\rho_{0S}=\ket{+}\!\bra{+}$, in an Ohmic environment at three different temperatures. The solid lines show the real and imaginary parts of the coherence $\rho_{01}(t)$ obtained by our simulation of the equivalent Lindblad equations, the dashed lines show the analytical solution and the insets show the error as defined in Eq.~\eqref{eq:Error}. Both populations are identically $1/2$ throughout the evolution and are not shown.}
\end{figure*}

\section{A test case: the spin-boson model}
\label{sec:SpinBoson}

We now turn to the second part of our approach: computing the reduced dynamics of a system by coupling it to the effective environment and solving the relevant Lindblad master equation~\eqref{eq:L_full}.

In order to demonstrate and quantitatively validate the method, we will show here the results we obtained for a system for which an analytical solution is known: the purely dephasing spin-boson model~\citep{BreuerPetruccione, Leggett_SpinBoson, CaldeiraLeggett_PureDephasing}. The Hamiltonian for this system is
\begin{equation}\label{eq:SpinBosonH}
	H=\frac{\omega_0}{2}\sigma_z+\int^\infty_0\!\!\!\!\!\mathrm{d}\omega\,
	\omega a^\dagger_\omega a_\omega+\frac{k}{2}\sigma_z\int^\infty_0\!\!\!\!\!
	\mathrm{d}\omega\,g(\omega)(a_\omega+a^\dagger_\omega)
\end{equation}
and we consider again the Ohmic spectral density defined in Eq.~\eqref{eq:OhmicJ}. In this model, the system and interaction Hamiltonians commute and are both diagonal in the system basis, so the populations $p_0\coloneqq\rho_{00}$ and $p_1\coloneqq\rho_{11}$ are conserved by the evolution. Any coherence in this basis present in the initial state, on the other hand, is erased according to the law (see Ref.~\citep{BreuerPetruccione} for a derivation)
\begin{equation}\label{eq:SpinBosonDecoherence}
	\rho_{01}(t)=\rho^*_{10}(t)=e^{-i\omega_0t+k^2\Gamma(t)}\rho_{01}(0)
\end{equation}
with
\[
	\Gamma(t)\coloneqq\int^\infty_0\!\!\frac{\mathrm{d}\omega}{\pi}
	J(\omega)\coth\left(\frac{\beta\omega}{2}\right)\frac{\cos(\omega t)-1}{\omega^2}.
\]

Using the cutoff frequency of the environment $\Omega_c$ as our energy scale, we set the parameter values $\omega_0=4\Omega_c$ and $k=1$, corresponding to a strong-coupling regime. Comparable coupling strengths appear e.g.\ in the study of superconducting quantum transmission lines~\citep{Peropadre_UltrastrongCoupling}. The system is initialized in the pure state $\rho_{0S}=\ket{+}\!\bra{+}$, with $\ket{+}\coloneqq\frac{1}{\sqrt{2}}(\ket{0}+\ket{1})$ in terms of the eigenstates of $\sigma_z$, and we simulated the reduced dynamics at three different temperatures $T=0$, $T=\Omega_c$ and $T=\frac{5}{2}\Omega_c$. Recall that the effective bath is always at zero temperature; different temperatures of the original environment require different surrogate baths. We found accurate effective correlation functions with $N=4$ for the first two cases, and with $N=5$ for the high-temperature regime; the parameters are given in Appendix~\ref{app:Parameters}, and the errors of the two correlation functions at nonzero temperatures are similar to the zero-temperature case already discussed.

\subsection{Results, accuracy and performance}

We solved the effective Lindblad equations for all three cases using the QME integrator provided by the Python OQS package QuTiP~\citep{QuTiP1, QuTiP2}, which implements a twelfth-order Adams-Moulton discrete integration algorithm.

From the results shown in Fig.~\ref{Fig:SpinBoson}, we see that our simulations with effective correlation functions give quantitatively good results for the coherence $\rho_{01}(t)$ at all times and temperatures (the populations $p_0$ and $p_1$ are both equal to $1/2$ throughout the evolution), as the overlap between the numerical (solid lines) and exact (dashed lines) solutions shows. The pure quantum decoherence at $T=0$ induces an algebraic decay asymptotically proportional to $t^{-k^2}$, while at $T>0$ the damping becomes exponential; a stronger effective coupling regime, which is determined both by $k$ and the strength of thermal effects, induces faster relaxation in the system dynamics.

The plots in the insets show the error figure
\begin{equation}\label{eq:Error}
	E_f(t)\coloneqq\frac{|f(t)-f_\mathrm{Num}(t)|}{|f(t)|+|f_\mathrm{Num}(t)|},
\end{equation}
which is identical for $f=\Re[\rho_{01}]$ (solid line) and $f=\Im[\rho_{01}]$ (dashed line). This is a better estimator for the accuracy of the simulations than e.g.\ the absolute difference $|f(t)-f_\mathrm{Num}(t)|$ because it removes the bias coming from changes in the relaxation time due to temperature, allowing us to compare all regimes on an equal footing. The error, as measured by $E_f(t)$, remains of the order of a few percent until the system has almost reached equilibrium and is comparable for the three regimes probed, mirroring the similar relative errors we had in all three effective correlation functions. The latter observation can be understood as follows: as higher temperatures or larger coupling constants increase the effects of the bath on the system, the error being carried from the correlation function into the reduced dynamics is magnified accordingly; on the other hand, these stronger effective regimes shorten the relaxation time of the system, so the cumulative effect of the error over time is not as severe as when the coupling is weaker or the temperature is lower.

From these results, we conclude that the method is quite reliable and stable provided that the effective correlation functions used are reasonably accurate, and that this accuracy does not command significantly greater effort or complexity in the TSO at higher temperatures and is completely independent of the system and the coupling strength. Furthermore, it is worth noting that any method based on approximating the environment alters its correlation function and is therefore prone to the same kind of error as ours, but we use a rigorously motivated and physically meaningful quantifier to optimize our correlation functions and keep it under control.

The computational cost of the simulations depends on the local dimension at which each effective mode is truncated and on the spread between the total evolution time and any faster timescales in the problem at hand, though the memory requirements scale faster with the complexity of the problem than the computation times do; to obtain our converged results for this system, which required $\sim 4$ levels for most oscillators and a maximum of $7$ for one mode in each simulation, all running times were below 10 minutes on a laptop. This cost grows rapidly with the number of effective modes, the local dimensions needed for convergence and  the size of the system itself; on the other hand, temperature and coupling strength have a limited impact on these factors: a very strong coupling or high temperature will require higher local dimensions but also cause very rapid relaxation to equilibrium, making long simulation times unnecessary. Moreover, when the coupling is stronger and more levels are needed for convergence, this typically affects one particular mode much more than the others, leading to an effective polynomial rather than exponential scaling in the coupling strength and temperature.

\section{Physically relevant applications}
\label{sec:Applications}

\begin{figure*}
	\centering
	\input{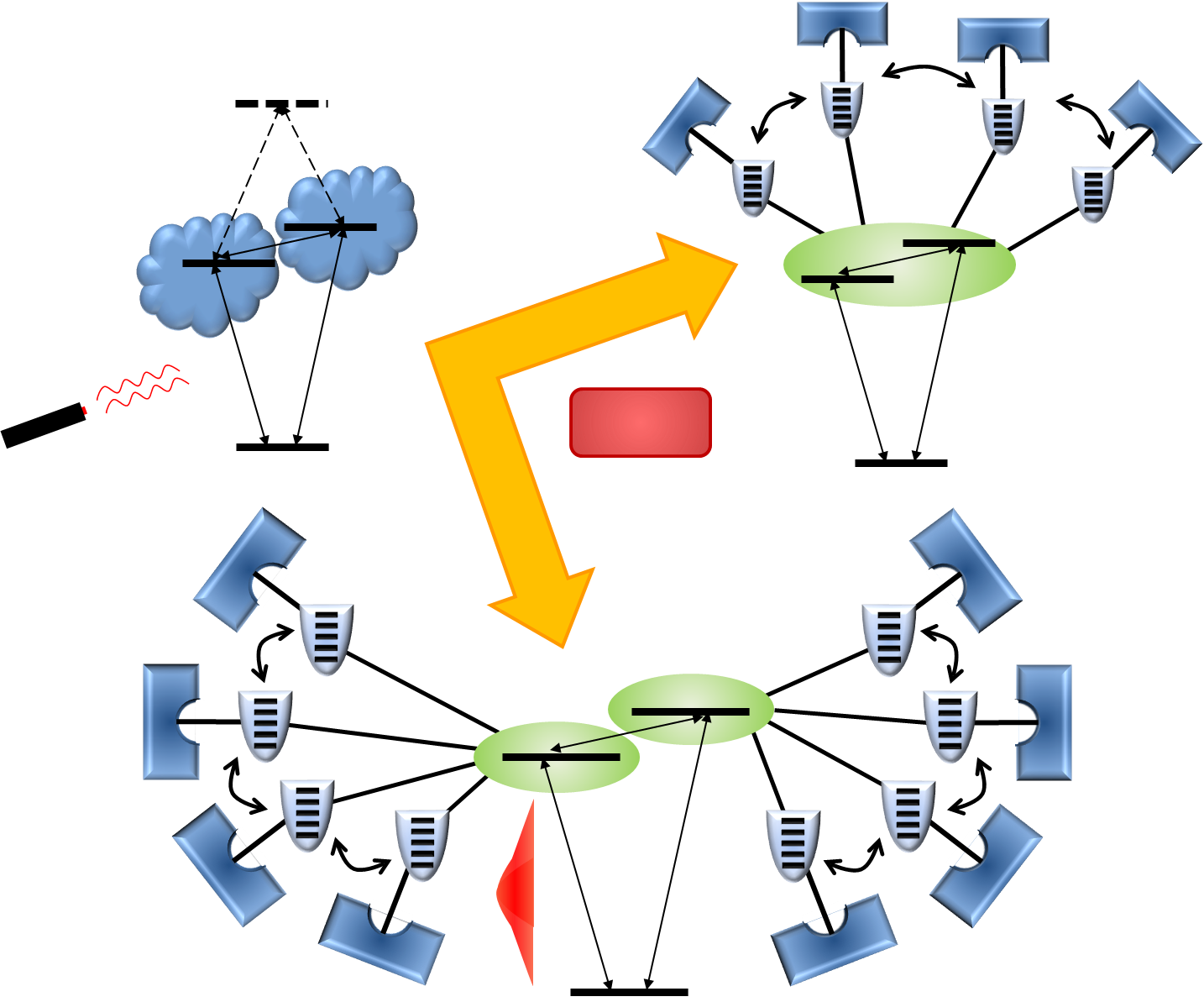}
	\caption{\label{Fig:Dimer}\textbf{The dimer model.} \textbf{a)} The physical picture with the 0-, 1- and 2-excitation subspaces (spanned by the ground state $\ket{g}$, the local excited states $\ket{E_1}$ and $\ket{E_2}$ of the two sites and the doubly excited state $\ket{E_{12}}$, respectively) and independent local environments interacting with each site. \textbf{b)} The equivalent model after the TSO and the transformation to common-mode and relative coordinates as in Eq.~\eqref{eq:Dimer_H_TNO}, used to compute the single-excitation dynamics (the common-mode effective bath is omitted because it does not contribute to the result). \textbf{c)} The model with the two identical surrogate environments used to obtain the absorption spectrum of the dimer. The surrogate mode parameters are the same in all setups.}
\end{figure*}

Few- and many-body systems non-perturbatively coupled to non-Markovian environments with structured spectral densities are ubiquitous in many fields ranging from biological physics and chemistry~\citep{HuelgaPlenio_QBio, Pelzer_Transport, DeSio_OPV} to condensed matter~\citep{RibeiroVieira_Transport, Haase_Metrology}, thermodynamics~\citep{UzdinLevyKosloff_QHE, MitchisonPlenio_NonEquilibrium}, nanomaterial science and sensing~\citep{JelezkoPlenio_NV} and quantum metrology~\citep{ChinHuelgaPlenio_Metrology, SmirneKolodynski_Metrology, HaaseSmirneKolodynski_Metrology}, and have prompted much research in theoretical modeling and numerical simulation methods for general OQS. In this section, we will demonstrate how our approach may be used to solve and make predictions on models at the forefront of current research, by presenting the results we obtained in two different applications.

In the first of the two examples, we will show some results for experimentally measurable optical properties in a model inspired by studies on coherent charge and energy transfer in biological molecular aggregates~\citep{Scholes_QBioNature}. After that, we will consider a polymer model of the type relevant in research on organic photovoltaic materials~\citep{Clark_OPV, Tamura_OPV}, and report simulation results for excitation transport dynamics in such a system under the assumptions of strongly coupled, non-Markovian local environments interacting with each monomer. The former example demonstrates the use of our simulation technique to gain physical insight by direct comparison of the results with observable data, while the latter gives an idea of its potential in terms of performance by addressing a problem beyond the reach of other current state-of-the-art methods.

\subsection{Optical spectra in molecular aggregates}

We considered a simple dimer model with system parameters in the range of those found in biomolecular aggregates participating in excitation energy transfer~\citep{Plenio_QBio, PlenioAlmeidaHuelga_Dimer}, coupled to an environment with a realistic spectral density derived from models common in the literature~\citep{AdolphsRenger}. We first simulated the reduced dynamics of this system at liquid nitrogen temperature ($T=77\,\mathrm{K}$), comparing the results with a simulation done using the numerically exact TEDOPA~\citep{PriorPlenio_TEDOPA, TamascelliSmirne_ThermalizedTEDOPA, Tamascelli_RSVD, Kohn_RSVD}, and then computed its absorption spectrum for the same temperature as well as for $T=0\,\mathrm{K}$ and $T=300\,\mathrm{K}$. Two different simulation techniques were used to integrate the effective Lindblad equation for the dynamics and the absorption spectra, and the spectra were calculated for two different environmental spectral densities and compared in order to identify the optical signatures setting them apart: in particular, we sought to determine the differences between spectra obtained in the presence or absence of a strongly coupled vibrational mode in addition to a broad background spectral density.

\subsubsection{Details of the model and reduced dynamics}

Following Ref.~\citep{PlenioAlmeidaHuelga_Dimer}, we considered a free dimer Hamiltonian
\begin{multline}
	H_S=E_g\ket{g}\!\bra{g}+\sum^2_{n=1}E_n\ket{E_n}\!\bra{E_n}
	+(E_1+E_2)\ket{E_{12}}\!\bra{E_{12}}
	\\
	+J(\ket{E_1}\!\bra{E_2}+\ket{E_2}\!\bra{E_1}),
\end{multline}
where the two monomers have on-site energies $E_1=E_g+12\,328\,\mathrm{cm}^{-1}$ and $E_2=E_g+12\,472\,\mathrm{cm}^{-1}$ and interact through a hopping coefficient $J=70.7\,\mathrm{cm}^{-1}$, and $\ket{E_{12}}$ is the state with both monomers excited. We only considered the ground state $\ket{g}$ and the single-excitation manifold spanned by the states $\ket{E_n}$, ignoring the doubly excited state $\ket{E_{12}}$ since its contribution is typically negligible both in excitation transport phenomenology and most absorption experiments~\citep{MayKuehn}. Then, setting $E_g=0$ as our reference energy, we are left with an effective two-level Hamiltonian for the single-excitation manifold
\begin{equation}\label{eq:Dimer_FreeH}
	H_\mathrm{1ex}=\sum^2_{n=1}E_n\ket{E_n}\!\bra{E_n}
	+J(\ket{E_1}\!\bra{E_2}+\ket{E_2}\!\bra{E_1}),
\end{equation}
whose eigenstates $\ket{\varepsilon_{1,2}}$ have an energy gap of $\Delta=201.8\,\mathrm{cm}^{-1}$, with the ground state dynamically decoupled and only contributing to expectation values or correlation functions of operators explicitly dependent on it.

The local excited states $\ket{E_n}$ interact with separate environments, which account for the molecular vibrations (both within the system and in the protein scaffold around it) and the presence of a solvent. We model these degrees of freedom by coupling the monomers to independent thermal baths with the same spectral density and temperature; the physical model is sketched in Fig.~\ref{Fig:Dimer} (a).

We first studied the problem for a spectral density consisting of two contributions: a broad background noise spectrum in the super-Ohmic form first introduced by Adolphs and Renger~\citep{AdolphsRenger}
\begin{equation}\label{eq:J_AdRe}
	J_\mathrm{AR}(\omega)\coloneqq\frac{\pi}{2\cdot 9!}\sum^2_{a=1}\rho_a
	\frac{\omega^5}{\Omega^4_{\mathrm{AR}a}}e^{-\sqrt{\omega/\Omega_{\mathrm{AR}a}}},
\end{equation}
where the two cutoff frequencies are $(\Omega_{\mathrm{AR}1},\Omega_{\mathrm{AR}2})=(0.557,1.936)\,\mathrm{cm}^{-1}$ and the weights of the two terms are $(\rho_1,\rho_2)=\frac{288}{5}(\frac{8}{13},\frac{5}{13})$, and a strongly coupled vibrational mode represented by adding an antisymmetrized Lorentzian peak
\begin{equation}\label{eq:J_AntiL}
	J_\mathrm{AL}(\Omega,\Gamma,S;\omega)
	\coloneqq S\frac{8\Gamma\Omega(4\Omega^2+\Gamma^2)\omega}{
	(4(\omega-\Omega)^2+\Gamma^2)(4(\omega+\Omega)^2+\Gamma^2)}.
\end{equation}
For this sharp spectral feature, we set $\Omega=227.5\,\mathrm{cm}^{-1}$, slightly above resonance with the system, a width $\Gamma=20\,\mathrm{cm}^{-1}$ corresponding to a decay time $(\Gamma/2)^{-1}\sim 0.5\,\mathrm{ps}$, and a Huang--Rhys factor $S=0.0379$ placing it in a moderate-coupling regime with the system. The reorganization energies corresponding to the background and the full environment are
\begin{align*}
	\lambda_\mathrm{AR}&=\int^\infty_0\!\!\frac{\mathrm{d}\omega}{\pi}
	\frac{J_\mathrm{AR}(\omega)}{\omega}=\sum^2_{a=1}
	\rho_a\Omega_{\mathrm{AR}a}=19.93\,\mathrm{cm}^{-1},
	\\
	\lambda&=\int^\infty_0\!\!\frac{\mathrm{d}\omega}{\pi}
	\frac{J_1(\omega)}{\omega}=\lambda_\mathrm{AR}+S\Omega
	=28.55\,\mathrm{cm}^{-1}.
\end{align*}

\begin{figure*}
	\centering
	\begin{minipage}{.48\textwidth}
		\input{C_77K}
		\caption{\label{Fig:C_77K}Exact (solid orange line) and effective (dashed blue line) correlation function for $J(\omega)$ at $T=77\,\mathrm{K}$; four modes were used for the background and two for the peak. The inset shows the TSO error. Note the shape related to the spectral density by Eq.~\eqref{eq:C_Fourier}, in particular the super-Ohmic dip at frequencies near zero and the local maxima at $\pm\Omega$.}
	\end{minipage}
	\hfill
	\begin{minipage}{.48\textwidth}
		\centering
		\input{TSOvsTEDOPA}
		\caption{\label{Fig:TSOvsTEDOPA}Short-time reduced dynamics in the single-excitation subspace of our dimer model with initial state $\rho_{0S}=\ket{\tilde{+}}\!\bra{\tilde{+}}$ and spectral density defined in the text at $T=77\,\mathrm{K}$, as simulated by our effective Lindblad equation (solid lines, colors as in legend) and TEDOPA (dashed lines). The inset shows the difference between the results.}
	\end{minipage}
\end{figure*}

In order to compute the reduced dynamics of the system in the single-excitation subspace, the total Hamiltonian of our problem
\begin{multline}\label{eq:DimerH}
	H_\mathrm{tot}=H_\mathrm{1ex}+\sum^2_{n=1}
	\int^\infty_0\!\!\!\!\!\mathrm{d}\omega_n\left(\omega_n
	a^\dagger_{\omega_n} a_{\omega_n}\right.
	\\
	+g(\omega_n)\left.\ket{E_n}\!\bra{E_n}
	(a_{\omega_n}+a^\dagger_{\omega_n})\right),
\end{multline}
can be rewritten in terms of the `common-mode' and `relative' creation and annihilation operators parametrized by a single frequency $A^{(\dagger)}_\omega=\frac{a^{(\dagger)}_{\omega_1}+a^{(\dagger)}_{\omega_2}}{\sqrt{2}}$ and $a^{(\dagger)}_\omega=\frac{a^{(\dagger)}_{\omega_1}-a^{(\dagger)}_{\omega_2}}{\sqrt{2}}$:
\begin{multline}\label{eq:Dimer_H}
	H=H_\mathrm{1ex}+\int^\infty_0\!\!\!\!\!\mathrm{d}\omega\,
	\omega\left(a^\dagger_\omega a_\omega+A^\dagger_\omega A_\omega\right)
	\\
	+\frac{1}{\sqrt{2}}(\ket{E_1}\!\bra{E_1}-\ket{E_2}\!\bra{E_2})
	\int^\infty_0\!\!\!\!\!\mathrm{d}\omega\,g(\omega)
	(a_\omega+a^\dagger_\omega)
	\\
	+\frac{1}{\sqrt{2}}(\mathbb{I}_S-\ket{g}\!\bra{g})
	\int^\infty_0\!\!\!\!\!\mathrm{d}\omega\,g(\omega)
	(A_\omega+A^\dagger_\omega).
\end{multline}
The common-mode environment only interacts with the single-excitation subspace through the last term, which is proportional to the identity in that subspace. Therefore, it can be ignored in any calculation not involving the ground state: for such applications, the Hamiltonian then reduces to
\begin{multline}\label{eq:Dimer_H_TNO}
	H=H_\mathrm{1ex}+\int^\infty_0\!\!\!\!\!\mathrm{d}\omega\,
	\omega a^\dagger_\omega a_\omega
	\\
	+\frac{1}{\sqrt{2}}(\ket{E_1}\!\bra{E_1}-\ket{E_2}\!\bra{E_2})
	\int^\infty_0\!\!\!\!\!\mathrm{d}\omega\,g(\omega)
	(a_\omega+a^\dagger_\omega)
\end{multline}
in terms of the relative modes only, and the dynamics factorizes between the two subspaces unless coherences between them are present in the initial state. A sketch of the model after this rearrangement of the environmental modes and the TSO is given in Fig.~\ref{Fig:Dimer} (b).

We computed the reduced dynamics in the single-excitation subspace for an initial coherent superposition of energy eigenstates $\rho_{0S}=\ket{\tilde{+}}\!\bra{\tilde{+}}$, where
\[
	\ket{\tilde{+}}\coloneqq\frac{\ket{\varepsilon_1}
	+\ket{\varepsilon_2}}{\sqrt{2}},
\]
for the spectral density $J(\omega)=J_\mathrm{AR}(\omega)+J_\mathrm{AL}(\Omega,\Gamma,S;\omega)$ considered. To this end, we determined effective parameters corresponding to the two terms of $J(\omega)$ and temperatures $T=0\,\mathrm{K}$, $T=77\,\mathrm{K}$ ($53.5\,\mathrm{cm}^{-1}$) and $T=300\,\mathrm{K}$ ($208.5\,\mathrm{cm}^{-1}$), performing the TSO separately on the Adolphs--Renger background, Eq.~\eqref{eq:J_AdRe}, and the antisymmetrized Lorentzian peak, Eq.~\eqref{eq:J_AntiL}. This corresponds to assigning a separate effective environment to each additive part of the spectral density $J(\omega)$ and can be a convenient strategy to break down structured spectra, as mentioned in an earlier section and shown in Fig.~\ref{Fig:SystemBath}. The Adolphs--Renger correlation function required $N=4$ oscillators at all three temperatures, and the Lorentzian mode was replaced by one effective oscillator at $T=0$ and two interacting ones at $T>0$ using the exact methods for $N=1,2$ described in Appendix~\ref{app:ExactSols}. All parameters of the effective environments are given in Appendix~\ref{app:Parameters}. The environmental correlation function at $T=77\,\mathrm{K}$, the temperature for which we computed the dynamics, is plotted along with the effective correlation function from the TSO in Fig.~\ref{Fig:C_77K}. The other temperatures will be considered in the calculation of absorption spectra for the model dimer.

Since the amount of memory required for a direct integration of the effective Lindblad equation would become too large for the system coupled to six effective modes with the local dimensions needed for convergence, we carried out the simulations using the quantum jump or Monte Carlo Wave Function (MCWF) method for pure states~\citep{DumZollerRitsch_MCWF, DalibardCastinMolmer_MCWF, PlenioKnight_MCWF} instead (the memory cost of MCWF scales linearly with the total Hilbert space dimension $\mathcal{N}$ for sparse Lindblad superoperators such as ours, while a master equation integrator requires at least $O(\mathcal{N}^2)$). The simulation was performed using another QuTiP code, since the package also provides MCWF routines. Our averages converged after as few as 1000 trajectories (this is due to the quantum jumps in the evolution directly affecting only the modes but not the system, since the latter has no Lindblad damping of its own, and thus partly canceling in the trace); we computed twice as many trajectories as a check but found no visible differences. The results of our simulation are shown in Fig.~\ref{Fig:TSOvsTEDOPA} along with those obtained by using TEDOPA: again, the accuracy of our effective correlation function---with errors of the order of 1\% as in the previous section---translates to a satisfactory result for the reduced dynamics throughout the time window considered, which is almost enough for the system to reach equilibrium (no comparison was possible for times longer that about $1.3\,\mathrm{ps}$ due to the rapidly increasing cost of the TEDOPA simulation at later times). The numerical cost is also remarkably low: for converged local dimensions, the simulation required under $200\,\mathrm{MB}$ of memory per thread and could therefore have been carried out on a desktop or laptop computer. To achieve higher parallelization of the work, however, we used the JUSTUS cluster at Ulm University: on a 16-core cluster node, the reduced dynamics up to $t=2\,\mathrm{ps}$ took 22 minutes to compute and the scaling is linear in the total simulation time. For comparison, TEDOPA took around 60 minutes to reach $t=1.3\,\mathrm{ps}$ on the same hardware and started scaling superlinearly in the simulation time at around that point.

\subsubsection{Absorption spectra}

Absorption experiments probe the linear response of the system; light from a laser source can be described as interacting with the local dipole moment operators $\vec{\mu}_n\coloneqq\vec{d}_n\ket{E_n}\!\bra{g}$, where $\vec{d}_n$ is the classical dipole moment of the $n$-th site, in a perturbative manner~\citep{Carmichael, MayKuehn}. Then the spectrum is obtained from the one-sided Fourier transform of the correlation function of the total dipole operator $\vec{\mu}\coloneqq\sum^2_{n=1}\vec{\mu}_n$ over the initial stationary state
\begin{equation}\label{eq:AbsorptionInitialState}
	\rho_{0\mathrm{Abs}}\coloneqq\ket{g}\!\bra{g}\rho_\beta,
\end{equation}
where the bath is in a thermal state at inverse temperature $\beta$ and the system is in the electronic ground state, which does not couple to the environment, since excited-state populations at equilibrium are negligible due to the very low intensity of the laser in such a setup~\citep{Mukamel, MayKuehn}.

\begin{figure}
	\centering
	\input{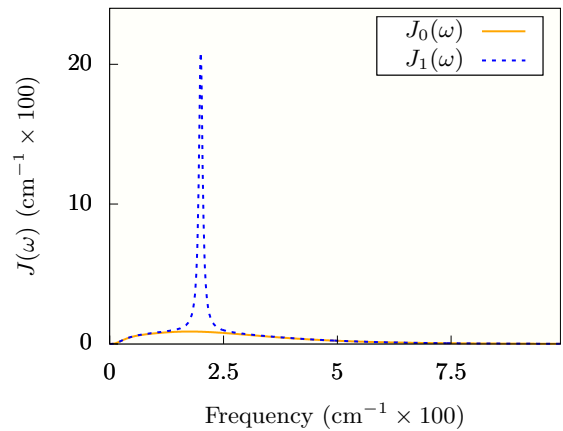}
	\caption{\label{Fig:DimerJ}The spectral densities $J_0(\omega)$ and $J_1(\omega)$ of the environments we used in our dimer model.}
\end{figure}

Specifically, the correlation function of interest is given by the scalar product of the dipole operator $\vec{\mu}$, applied at times $t_0=0$ and $t$: in terms of the overall unitary evolution, one has
\begin{equation}\label{eq:DipoleCorrelation}
	C_\mu(t)\coloneqq\mathrm{Tr}\left[U^\dagger(t)(\vec{\mu}^\dagger_1
	+\vec{\mu}^\dagger_2)U(t)\cdot(\vec{\mu}_1+\vec{\mu}_2)
	\rho_{0\mathrm{Abs}}\right].
\end{equation}
Note that this is formally a two-time object: we can compute it using an effective environment because the first operator acts on the system at equilibrium, so the hypotheses of Theorem~\ref{TSO_Theorem} are not violated. The unitary dynamics acts on $\left(\vec{d}_1\ket{E_1}+\vec{d}_2\ket{E_2}\right)\!\bra{g}\rho_{0\mathrm{Abs}}$, which is still a factorized object with the environment in a thermal state, so the equivalence with a suitable effective Lindblad dynamics remains well defined; however, note that this time the common-mode part of the total environment does not decouple from the problem and one needs to simulate the system along with both local baths, as pictured in Fig.~\ref{Fig:Dimer} (c).

We set the ansatz $\vec{d}_1=\vec{d}_2=\vec{d}$ for the geometry of the dimer in order to simplify the form of the dipole correlation function. Expressed in units of $|\vec{d}|^2$, $C_\mu(t)$ becomes
\begin{multline}\label{eq:SimplifiedDipoleCorrelation}
	C_\mu(t)=\mathrm{Tr}\left[
	U^\dagger(t)\ket{g}\!\left(\bra{E_1}+\bra{E_2}\right)U(t)\right.
	\\
	\left.\left(\ket{E_1}+\ket{E_2}\right)\!\bra{g}\rho_{0\mathrm{Abs}}
	\right].
\end{multline}
The absorption spectrum is then given by
\begin{equation}\label{eq:Spectrum}
	S_\mathrm{Abs}(\omega)\coloneqq\omega\,\Im\!
	\lim_{t_\mathrm{max}\rightarrow\infty}
	\int^{t_\mathrm{max}}_0\!\!\!\!\!\mathrm{d}t\,
	iC_\mu(t)e^{i\omega t}.
\end{equation}

In order to compare the effect on the absorption spectrum of a strongly coupled, underdamped vibrational mode in the environment, we will now consider two spectral densities: $J_0(\omega)\coloneqq J_\mathrm{AR}(\omega)$ and $J_1(\omega)\coloneqq J_\mathrm{AR}(\omega)+J_\mathrm{AL}(\Omega,\Gamma,S;\omega)$, with $\Omega=200\,\mathrm{cm}^{-1}$, $\Gamma=10\,\mathrm{cm}^{-1}$ and $S=0.25$. A plot of the spectral densities is given in Fig.~\ref{Fig:DimerJ}: as the figure shows, the contribution of the underdamped peak is much stronger in this new setup.

\begin{figure*}
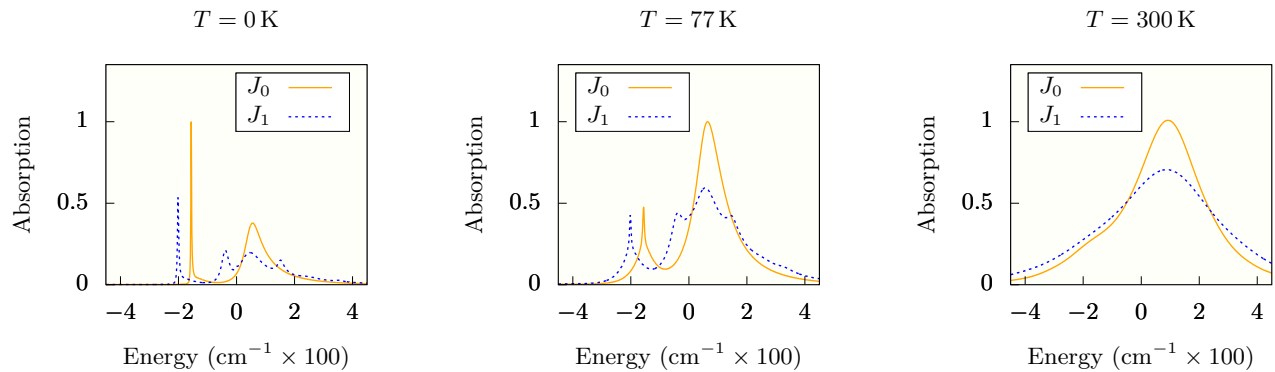

	\centering
	\begin{minipage}{.329\textwidth}
		\centering
		\input{DimerT0}
	\end{minipage}
	\hfill
	\begin{minipage}{.329\textwidth}
		\centering
		\input{DimerT77}
	\end{minipage}
	\hfill
	\begin{minipage}{.329\textwidth}
		\centering
		\input{DimerT300}
	\end{minipage}
	\caption{\label{Fig:AbsorptionSpectra}\textbf{Normalized absorption spectra for the model dimer.} The maxima appear at the eigenenergies of the system Hamiltonian minus the reorganization energy of the bath considered (solid orange lines correspond to the spectral density $J_0(\omega)$, dashed blue lines to $J_1(\omega)$). The upper eigenstate gives a broader peak, since it can decay to the lower one or lose energy to the bath. At higher temperatures, this peak prevails and eventually broadens to the point of erasing most of the spectral structure. The sharp mode in the environmental spectral density $J_1(\omega)$ causes additional lines and sidebands to appear in the absorption spectra at frequencies consistent with combined excitations of the system and the mode.}
\end{figure*}

Integrating the effective Lindblad equation with the initial pseudo-state $\tilde{\rho}_0\coloneqq\left(\ket{E_1}+\ket{E_2}\right)\!\bra{g}\rho_{0\mathrm{Abs}}$, one obtains the dipole correlation from Eq.~\eqref{eq:SimplifiedDipoleCorrelation} as
\begin{equation}\label{eq:LindbladDipoleCorrelation}
	C_\mu(t)=\mathrm{Tr}\left[
	\ket{g}\!\left(\bra{E_1}+\bra{E_2}\right)e^{\mathcal{L}t}
	\left[\tilde{\rho}_0\right]\right],
\end{equation}
where $\mathcal{L}$ is the Lindblad superoperator given by the TSO with both local environments included.

\begin{table}
	\centering
	\begin{tabular*}{.48\textwidth}{@{\extracolsep{\fill}}cccccc}
	\toprule
	$T$ $(\mathrm{K})$
	& \multicolumn{2}{c}{$t_\mathrm{max}$ $(\mathrm{ps})$}
	& \multicolumn{2}{c}{$|C_\mu(t_\mathrm{max})|$}
	& $\chi$
	\\
	\midrule
	& $J_0(\omega)$	& $J_1(\omega)$	& $J_0(\omega)$	& $J_1(\omega)$	& \\ 
	\cmidrule{2-3}
	\cmidrule{4-5}
	$0$		& $20.0$	& $11.0$	& $<10^{-3}$	& $<10^{-3}$	& $12$ \\
	$77$	& $6.75$	& $5.5$	& $<10^{-3}$	& $\sim10^{-3}$	& $12$ \\
	$300$	& $0.27$	& $0.17$	& $\sim10^{-3}$	& $\sim10^{-3}$	& \textit{variable}  \\
	\bottomrule
	\end{tabular*}
	\caption{\label{tb:DimerCFDecayTimes}Total simulation times, absolute values of the dipole correlation function $C_\mu(t)$ of the dimer at the final time and DAMPF bond dimensions (if applicable) for $J_0(\omega)$ and $J_1(\omega)$, respectively. The simulations at $T=300\,\mathrm{K}$ were performed with a time-adaptive bond dimension.}
\end{table}

Computing the dipole correlation function is much more challenging than simulating the reduced dynamics in the single-excitation subspace, because this time both sets of surrogate modes need to be explicitly accounted for and the local dimensions are quite high, as shown in the relevant parameter tables in Appendix~\ref{app:Parameters}. In order to keep the total Hilbert space dimension manageable, we employed a variation of a recently published tensor network--based technique called Dissipation-Assisted Matrix Product Factorization (DAMPF)~\citep{Somoza_DAMPF} to simulate the dimer. DAMPF, which was originally developed using non-interacting pseudomodes, is extremely efficient for vibronic aggregates in the single-excitation manifold, making it an ideal candidate for scaled-up simulations of systems involving many sites with local surrogate-oscillator baths, as will be shown in the next subsection.

For each temperature, we simulated the dimer until $C_\mu(t)$ had decayed to values small enough for the limit in Eq.~\eqref{eq:Spectrum} to be approximately satisfied (the initial value in our units is $C_\mu(0)=2$, as can be seen from Eq.~\eqref{eq:LindbladDipoleCorrelation}): the final times and corresponding absolute values of $C_\mu(t)$ reached in our simulations are reported in Table~\ref{tb:DimerCFDecayTimes}, and the resulting absorption spectra---obtained via a discrete Fourier transform and centered around the midpoint frequency $12\,400\,\mathrm{cm}^{-1}$ of the single-excitation subspace---are shown in Fig.~\ref{Fig:AbsorptionSpectra}.

The spectra show the expected features: the result for $J_0(\omega)$ displays absorption lines corresponding to the single-excitation eigenstates $\ket{\varepsilon_{1,2}}$ of $H_\mathrm{1ex}$ and appearing at the corresponding energy values redshifted by the bath reorganization energy; the line corresponding to the higher eigenstate is broadened due to the decay channels of that state, which couples to the environment and the lower excited state, whereas the latter gives a very narrow zero-temperature peak since it is not coupled to any lower-lying state it could decay to. At higher temperatures, the contribution from the upper level becomes larger than the lower one, but the energies associated with the---now markedly broadened---spectral lines no longer represent energy eigenstates of the system, since the dressed system-environment energy eigenbasis is very different from a tensor product basis in this regime, as hinted at by the fact that the environmental reorganization energy corresponding to the thermalized spectral density $C^E_\beta(\omega)/2$ is comparable to $J$. At room temperature, hardly any structure is discernible but for the fact that the spectrum rises slowly and somewhat irregularly to the left of the maximum.

Adding the strong peak to the spectral density, the spectra are shifted to the left by the added reorganization energy, and the expected new spectral lines associated with excitations of both the dimer and the coupled vibrational mode appear. At lower temperatures, higher sidebands are also visible as small bumps to the right of the main spectral curves; they are washed out by the strong broadening at room temperature.

It should be noted that at temperatures up to $T=77\,\mathrm{K}$ the timescale at which the reduced dynamics of the system reaches a steady state is of the order of about one picosecond for both $J_0(\omega)$ and $J_1(\omega)$ (most of the dissipation is due to the broad Adolphs--Renger background, since the Lorentzian mode has a long lifetime); the decay times of the dipole correlation functions, on the other hand, were found to be significantly longer. In order for $C_\mu(t)$ to reach values small enough to avoid visible spurious effects from an incomplete decay in the Fourier transform, some of the simulations had to run up to times of order $t=10\,\mathrm{ps}$ (see again Table~\ref{tb:DimerCFDecayTimes}). Such long-time simulations are only possible with methods whose cost scales slowly, e.g.\ linearly, in the simulation time, such as DAMPF (or, for smaller Hilbert spaces, MCWF or even direct integration routines). Convergence in DAMPF is achieved for sufficiently high local dimensions as well as bond dimensions $\chi$ (we refer the reader to Ref.~\citep{Schollwock_tDMRG} for more
details on parameters in the tensor-network setup); we found that both needed to be quite high at nonzero temperatures, and used a time-adaptive bond dimension for the $T=300\,\mathrm{K}$ case to optimize the effective Hilbert space dimension throughout those simulations in order to save time. The local dimensions of each mode are given in the relevant parameter tables in Appendix~\ref{app:Parameters}, and the bond dimensions are shown in Table~\ref{tb:DimerCFDecayTimes} for those simulations in which they were kept fixed.

\subsection{Excitation transport in organic polymers}

In this subsection, we will give a more concrete demonstration of the full potential of surrogate environments for physically sound and numerically efficient simulation of systems. To this end, we will apply our method to a problem involving an organic polymer modeled as a chain consisting of many sites with realistic local environments strongly coupled to each of them.

Organic polymers have been gaining growing attention from the condensed- matter, OQS and many-body-physics communities due to their considerable technological potential, e.g.\ in devising novel photovoltaic and other electronic components~\citep{Clarke_OPV, Proctor_OPV}. Such systems are often modeled in the same tight-binding approximation used for photosynthetic complexes in biological physics, and simulating charge transfer or separation processes in chains of organic monomers interacting with local non-Markovian environments is a notoriously challenging task even with state-of-the-art techniques such as HEOM, as mentioned previously~\citep{Yamagata_QuantumWires, Tempelaar_Coherence, Spano_Aggregates, Hestand_Aggregates}.

Typical treatments of such organic systems often employ strong coarse-graining of the environmental spectral features~\citep{Blau_dimer, Chenel_OPV}, in order to save computational resources for the simulation of a system which may consist of a large number of sites. We will now show how our surrogate environments can be used to calculate the reduced dynamics of extended vibronic systems consisting of multiple sites, with each site coupled to a realistic thermal bath comprising both a sharp mode and an Ohmic background. Interactions, both among sites and between each site and its local environment, are strong, with the baths characterized by high reorganization energies, and we will consider the system at both zero and room temperature.

We considered a homogeneous polymer Hamiltonian of the form
\begin{equation}\label{eq:ChainH}
	H_S=\sum^{K-1}_{n=1}J\left(\ket{E_n}\!\bra{E_{n+1}}
	+\ket{E_{n+1}}\!\bra{E_n}\right),
\end{equation}
where $K$ is the number of sites, $J$ is the site-site coupling and the on-site energies are assumed equal and set to zero. The ground state is disregarded since it does not couple to the single-excitation subspace we are working in, and we set $K=10$ and $J=200\,\mathrm{cm}^{-1}$.

\begin{figure*}
	\centering
	\begin{minipage}{.48\textwidth}
		\centering
		\input{Chain_T0}
		\caption{\label{Fig:Polymer0K}\textbf{Model polymer reduced dynamics at $T=0\,\mathrm{K}$.} Populations of three sites (top) and coherences between three pairs of sites (bottom) of the 10-site chain with structured local environments, with $\Im[\rho_{ij}]$ represented by a dashed line of the same color as the corresponding $\Re[\rho_{ij}]$. Note the propagation of the initial population along the chain: after a brief transient in which the excitation remains localized, traveling until it is reflected back by the opposite end of the chain, eventually it spreads out, settling for a delocalized steady state with more population in the middle.}
	\end{minipage}
	\hfill
	\begin{minipage}{.48\textwidth}
		\centering
		\input{Chain_T200}
		\caption{\label{Fig:Polymer300K}\textbf{Model polymer reduced dynamics at $T=288\,\mathrm{K}$.} Populations of three sites (top) and coherences between three pairs of sites (bottom) of the 10-site chain with structured local environments, with $\Im[\rho_{ij}]$ represented by a dashed line of the same color as the corresponding $\Re[\rho_{ij}]$. The system quickly relaxes to equilibrium, but displays fast oscillations from its interaction with the local high-frequency vibrational modes during the transient evolution as it approaches the steady state. The final populations are more uniformly distributed than at zero temperature.}
	\end{minipage}
\end{figure*}

Each site couples to a local thermal bath in the same way as in Eq.~\eqref{eq:DimerH}, and the spectral density of the baths is a sum of an Ohmic background of the form Eq.~\eqref{eq:OhmicJ}, with cutoff frequency $\Omega_c=200\,\mathrm{cm}^{-1}$ and rescaled by an overall factor $\kappa=0.25$, and an underdamped peak at $\Omega=1\,000\,\mathrm{cm}^{-1}$ with $\Gamma=20\,\mathrm{cm}^{-1}$ and Huang--Rhys factor $S=0.25$. The total reorganization energy is
\[
	\lambda=\kappa\Omega_c+S\Omega=50\,\mathrm{cm}^{-1}
	+250\,\mathrm{cm}^{-1}=300\,\mathrm{cm}^{-1},
\]
a very high value.

We simulated the evolution of this system up to $t=1.25\,\mathrm{ps}$ from an initial state $\rho_{0S}=\ket{E_1}\!\bra{E_1}$ at $T=0\,\mathrm{K}$ and at $T=\Omega_c=288\,\mathrm{K}$, again separating the background and the peak in the TSO. For the Ohmic spectral density, we used the surrogate environments already introduced in Sections~\ref{sec:TSO} and~\ref{sec:SpinBoson}. The results are shown in Figs.~\ref{Fig:Polymer0K} and~\ref{Fig:Polymer300K}, respectively. The local dimensions for the Ohmic background needed to be higher for this simulation than for the spin-boson case discussed in Section~\ref{sec:SpinBoson} (the population dynamics in the surrogate oscillators depends both on the coupling strength and on the internal dynamics of the system they are interacting with in any given problem), and we saw no relevant changes in the results for $\chi>9$ at zero temperature and $\chi>12$ at room temperature. The zero-temperature simulation took a few hours and the room-temperature one was completed over the course of several days on a 16-core node of the JUSTUS cluster at Ulm University.

Simulations such as these on a conventional architecture (i.e.\ one not boosted by the use of graphical processing units, which would further enhance the numerical efficiency of both other schemes and our own) are beyond the reach of any simulation technique we know of. The strength of the coupling, especially to the high-frequency mode, would be critical for a HEOM treatment with as few as two sites~\citep{Somoza_DAMPF}, and the system size rules out TEDOPA, QUAPI or any other method for non-Markovian open systems regardless of that property of the environment. Regarding DAMPF, which overcomes the problem of ever-growing bond dimensions in tensor network--based methods, it can also be used with independent auxiliary oscillators. However, this comes at the price of using a much greater number of modes for the same accuracy, again driving up the simulation cost. With the interacting modes given by the TSO, we can be certain that our results are closer to the reduced dynamics of the unitary model than anything that can be done with the same number of independent pseudomodes, due to the far smaller correlation function error our coupled modes entail. Therefore, this combination of accuracy and numerical efficiency would not be possible otherwise.

\section{Discussion}
\label{sec:Discussion}

After introducing and demonstrating our new simulation method, let us now recapitulate its main theoretical and technical points, discussing its strengths, limitations and error sources in order to give a clear and concise summary of its current state and possible future improvements.

\subsection{Theoretical basis and general remarks}

Our method is part of a category of hybrid approaches based on rephrasing microscopic OQS models as effective Markovian problems, in which the memory of the environment is accounted for by an \emph{ad hoc} auxiliary system. Although this divide-and-conquer strategy between Markovian and non-Markovian effects is a shared feature of several existing methods, the flexibility and quantitative control allowed for by the rigorous theoretical groundwork underlying our construction~\citep{TSO_Theorem, SpinBosonBounds} are, as far as we know, unprecedented for an approach of this type.

The transformation procedure we described in Section~\ref{sec:TSO} exploits the generality of a broad, physically well-defined class of effective environments to tailor them in a systematic way to fit the microscopic ones given: isolating the correlation function as the single property of the environment which needs to be replicated as accurately as possible, we take advantage of the added versatility from using interacting effective modes to make this fitting procedure more accurate while keeping the number of effective degrees of freedom lower than would be possible in chain- or star-configuration schemes. It should be noted that we took one particular ansatz for the effective environment because we found it the most convenient for our needs, but many other choices (non-hopping linear couplings, interactions beyond nearest neighbors, different damping and initial stationary state, etc.) are possible.

Another relevant feature is that the system on which the environment acts does not enter in this part of the procedure at all. Therefore, once the effective parameters corresponding to a given unitary bath are determined, they can be used in all problems featuring that particular bath, as we showed in Section~\ref{sec:Applications}. This makes determining the surrogate environment a one-off task, which can be very convenient in any field in which standard spectral densities appear in many different situations.

The second part of the method is the simulation of the system coupled to the effective environment. Here any of the analytical or numerical techniques for Lindblad master equations already developed in the literature can be used; the system part of the problem is completely unrestricted, so different strategies can be adopted depending on the problem at hand and available computational resources. We demonstrated the method on two-level systems interacting with small sets of up to six effective modes, for which simple and clear solution methods like direct integration of the master equation or MCWF are still suitable, and on more complex systems such as chains of monomers in the tight-binding approximation, with each local site interacting with its own surrogate bath, for which an integration scheme compressing the total Hilbert space dimension is necessary.

In summary, we have shown that although the mathematical question of how to generalize the use of independent damped oscillators as effective environments is highly nontrivial, finding ways to do so can be extremely beneficial to the modeling of non-Markovian OQS. Our recipe for the construction of few-body Gaussian environments with interacting surrogate modes proved a valuable technique to encode complex environmental effects in surprisingly compact effective models, with remarkable computational advantages.

\subsection{Numerical complexity and costs}

As we have shown in the examples given in the preceding sections, our method is very versatile and applies in principle to any non-Markovian, non-perturbative OQS problem involving a Gaussian bosonic bath, at any temperature and coupled arbitrarily strongly to a system. Clearly, some problem classes and physical regimes are more suited than others to this type of treatment: we will now summarize the key elements determining the computational effort required for any given application.

Concerning the simulation of systems, the most important variable to look at is the Hilbert space dimension, which includes both the system size and the number and local dimensions of the surrogate modes; temperature and coupling strength affect the overall complexity indirectly, mainly by determining the minimum local dimensions needed to accurately simulate the modes. The number of effective oscillators and their parameters come from the TSO and depend on the spectral density and temperature of the unitary environment. The spectral density has a prominent role in determining the number of oscillators required; higher temperatures can contribute too but typically result in an effectively stronger coupling to the system instead: this does cause the oscillators to become more populated, making higher local dimensions necessary for the reduced dynamics to converge, but the added computational cost is usually less than that entailed by adding a new mode.

When simulating highly structured systems or environments, the Hilbert space dimensions involved are such that memory, rather than time, typically becomes the main computational concern. Hilbert space dimensions of order $10^5$ require memory of order $1\,\mathrm{GB}$ per thread with the MCWF implementation we used in the aforementioned simulation, which can be managed on desktop-level hardware or individual nodes of a cluster. Larger Hilbert spaces, such as those of the dimer and polymer problems we considered, must be compressed by suitable optimization techniques such as matrix product operator--based methods~\citep{Mascarenhas_MPO, CuiCiracBanuls_MPO, Somoza_DAMPF} or reduction to Krylov subspaces, a topic of current relevance in the study of large systems of numerical differential equations~\citep{Minchev_MatExp, VoSidje_Krylov, Tokman_KIOPS}. We used the newly developed DAMPF technique because it could be easily adapted from its original form in order to accommodate coupled surrogate modes, and showed that the scaling of the simulations with the size and complexity of the system studied is quite favorable, in some cases outperforming any known method and thus attaining results hitherto out of reach.

In general, the cost of simulating a Lindblad dynamics scales linearly in the total evolution time for methods such as direct integration or MCWF, which allow for the full Hilbert space dimension to be fixed upfront. This makes them well suited for the study of long-time dynamics and relaxation to equilibrium. When the total effective Hilbert space of a problem is too large for any such technique, one needs to resort to time-adaptive truncation schemes, which can scale quite unfavorably in time. However, novel methods such as DAMPF exploit the damping in the simulated dynamics to bound the maximum effective dimensions they use, thus reducing these nonlinear additional costs to the point of recovering an approximately linear scaling which allowed us to simulate even a polymer up to arbitrarily long times. As ambiguous as performance assessments can become, depending on physical regimes and scales in the models studied, it seems quite clear nonetheless that there are situations in which using surrogate modes to reduce the number of effective degrees of freedom needed for accurate results is of paramount computational advantage.

As to determining the parameters of the surrogate modes in the first place, the inversion problem from the target correlation function is, in general, a mathematically difficult task. Our TSO algorithm uses a randomized parametrization as a variational method to reduce the number of variables in the problem and unlock a part of the solution, which is then fed back into the inversion problem to determine the values still missing; the solution found is the best possible for the random initial values given, and a minimization on the sample according to a suitable figure of merit is carried out \emph{a posteriori}.

This rather involved procedure gives satisfactory results but scales poorly with the number of modes; for more than five interacting oscillators, it is already very expensive. This, however, is not a major setback for several reasons. First of all, complex environmental correlation functions typically originate from spectral densities comprising several simple terms, which can be addressed---and recycled for other problems if needed---individually, as demonstrated in our example applications; secondly, keeping the number of effective oscillators as low as possible is also a priority for simulation purposes and does not put significant constraints on accessible coupling or temperature regimes; finally, the cost of the TSO is not fixed but depends on the form chosen for the effective environment, so the complexity of our particular algorithm is not universal.

Possible future improvements to the variational algorithm could come from employing different methods such as simulated annealing or importance sampling in the parameter search or machine-learning techniques to minimize the distance between original and effective correlation functions with respect to the parameters; finding a way to work with the map from effective environment to correlation function in the direct rather than the reverse direction, if possible, would be a major simplification.

\subsection{Accuracy and error sources}

To complete the discussion of our method, we must now turn to the sources of error affecting the reduced dynamics, and the control we have over them.

The most important error to be addressed is of physical origin and comes from the TSO. This is the error in the correlation function, and its impact on the reduced dynamics and operator expectation values at any time is rigorously bounded~\citep{SpinBosonBounds} (the paper focuses on the spin-boson model in particular, but similar bounds can be derived for other finite systems following the same steps).

Though under control, this error is worth a more careful analysis because it is actually a sum of two errors, one from a fundamental feature of our method, the other from a technical constraint.

The former source is the very form of any correlation function defined as in Eq.~\eqref{eq:2tAverage_L}: it has been shown that no infinite, unitary thermal environment can have a correlation function of the form obtained via the quantum regression formula for a finite, Lindblad-damped auxiliary bath, because the fluctuation-dissipation theorem~\citep{CallenWeltonFDT, BreuerPetruccione}, which holds for continuous, unitary thermal environments, is incompatible, strictly speaking, with the regression hypothesis~\citep{Talkner_NoQRT, FordOConnell_NoQRT}. This is reflected quantitatively in the fact that no zero-temperature correlation function obtained through the latter is exactly zero on the whole negative frequency domain; however, the violation of the fluctuation-dissipation theorem can always be reduced by adding effective modes, until the unavoidable residual error is comparable to other errors in the model at hand (except in pathological cases, such as the weakly coupled spin-boson model with pure dephasing: this system is only sensitive to the limit of the spectral density at zero frequency, where the analytical differences between unitary and effective environments emerge most clearly, as discussed in Section~\ref{sec:TSO}).

The second source contributing to the correlation function error are the constraints on the parameters in the master equation: for a fixed number of modes, not every linear combination of complex exponentials can be derived from a valid set of effective bath parameters via Eqs.~\eqref{eq:CR} and~\eqref{eq:ws} (for example, expressions obtained by setting at least one of the master equation rates $\Gamma_n$ to a negative value are out of physical scope). Therefore, the closest physically possible correlation function to the one given is generally not the best unconstrained fit with complex exponentials.

It should also be mentioned at this point that spectral densities of microscopic models are ultimately derived from experimental results in many applications of current interest~\citep{AdolphsRenger, Bennett_Dimer, Pelzer_Transport, JelezkoPlenio_NV}, so any error in our TSO resulting in a correlation function still compatible with the data is immaterial in practice.

Finally, the last error source in our method is strictly numerical and comes from the integration schemes used to solve the Lindblad equation, which necessarily involve some truncation of the Hilbert space. This error is not under rigorous control, but requires method-dependent convergence checks like any other numerical solution technique.

\subsection{Impact}

Finally, let us sum up the salient features of our simulation method and highlight its distinguishing qualities among existing schemes for general OQS.

First of all, we wish to emphasize that our aim in proposing this approach is to offer the level of accuracy and reliability of a fully microscopic simulation while retaining the benefits of working with two-tiered effective environments, particularly their simple mathematical structure and efficient numerical simulation.

Our scheme fills the gap between exact and simplified effective methods by providing auxiliary environments with a quantitatively certified link to the microscopic ones they stand in for, and enables very efficient simulation of nontrivial environments by keeping the number of modes much lower---thanks to the interactions among them---than any similar techniques we are aware of. For example, noninteracting pseudomodes are a special case of our surrogate oscillators, but we found that in order to reproduce an environment such as the one we discussed in our dimer example application in Section~\ref{sec:Applications}, we would have needed at least 20 pseudomodes to attain the same accuracy given by our TSO with 6 oscillators: while each independent pseudomode contributes a term
\[
	C^R_n(t)=w_ne^{\lambda_nt}
\]
with a real and positive $w_n$---a Lorentzian, in the frequency domain---to the correlation function, our interacting effective modes contribute terms of the form
\[
	C^R_n(\omega)=-2\frac{\Re[w_n]\Re[\lambda_n]+\Im[w_n]
	(\omega+\Im[\lambda_n])}{\Re[\lambda_n]^2+(\omega+\Im[\lambda_n])^2}
\]
thanks to the fact that the $w_n$ are complex, and these functions turn out to be far more flexible for fitting purposes. This difference is even more dramatic with Ohmic environments such as the one we considered in the polymer simulations, because the frequency-domain correlation function for an Ohmic bath at high temperature has a maximum at the origin: fitting such a shape with Lorentzians would result in a large number of underdamped low-frequency pseudomodes, which would become highly populated during the dynamics and require impractically high local dimensions. An accurate simulation of both our example systems would thus have been much more expensive using noninteracting pseudomodes.

We have also compared our simulations with calculations performed using microscopic methods and found that our accuracy is on par with numerically exact results, e.g. from TEDOPA, at least for the relatively simple application we used as a benchmark: long-time dynamics are much easier to compute by solving our effective Lindblad equation in all coupling and temperature regimes due to the nonlinear scaling of TEDOPA in the evolution time; on the other hand, spectral densities with complicated shapes requiring a large number of effective oscillators are hard with our method (though the solution of our effective Lindblad equation can be optimized by techniques such as DAMPF) while TEDOPA is much less sensitive to the shape of the spectral density. In our example application, a simple integration method running on a laptop performed better than TEDOPA at all temperatures for medium to long times (at zero temperature even for very short times) and comparably well for short times at nonzero temperature, for a moderately structured spectral density; the scaling in system size and complexity is very similar for the two schemes.

The HEOM method~\citep{Tanimura_HEOM, TanimuraKubo_HEOM} is more akin to our approach in spirit, since it is also based on exponential fitting of $C^E(t)$. Much like our number of surrogate modes, the number of exponentials needed for an adequate fit is one of the main factors determining complexity of HEOM simulations, the other being the tier at which the hierarchy needs to be truncated. This number should be the same for the two methods if one requires the same accuracy and uses efficient expansion techniques for the correlation function~\citep{Duan_EfficientHEOM, Duan_EfficientT0HEOM} (these overcome the well-known problem of the more traditional Matsubara-frequency expansion~\citep{MeierTannor_DecompositionHEOM, HuLuo_DecompositionHEOM}, which at low temperatures needs a very large number of exponential terms in order to converge). HEOM scales with the complexity of the spectral density in a similar manner as our method and can likewise account implicitly for temperature through the approximate correlation function. Long evolution times are also not problematic for most regimes; however, they can be in the presence of certain environmental features, e.g.\ narrow peaks with high Huang--Rhys factors corresponding to strongly coupled environmental modes, which make the hierarchy of equations converge very slowly, significantly increasing the simulation cost. We have shown in the last of our example simulations that including the effect of such terms in our effective baths does not affect our simulation costs as dramatically as it does HEOM's; in fact, the spectral density we considered poses a serious challenge for HEOM even with just two sites. Combined with the scaling of HEOM in the size of systems such as our dimer and polymer, this singles out at least one class of problems where both high- and low-temperature simulations make surrogate modes the most efficient if not the only viable option.

\section{Conclusions and outlook}
\label{sec:Conclusions}

In this paper, we introduced a new non-perturbative approach for the description and simulation of arbitrary open quantum systems in Gaussian bosonic environments. The method is based on the use of networks of dissipative auxiliary oscillators as a means to account for nontrivial environmental effects, and puts no restrictions on temperature, non-Markovianity, system-environment coupling strength or system structure. It generalizes previously existing schemes employing independent fictitious modes, and we demonstrated that such a generalization is both sensible from a methodological point of view and extremely useful in terms of practical results.

We devised a systematic recipe to build effective environments of a very versatile class, tuning their parameters in order to capture the effects of any given unitary bath, using as few degrees of freedom as possible and with a clear measure of the error involved. This procedure is grounded in rigorous theoretical results, specifically the equivalence conditions between unitary and dissipative Gaussian baths proved in Ref.~\citep{TSO_Theorem} and the relation between changes in the bath correlation function and in the reduced dynamics and single-time averages of the system derived in Ref.~\citep{SpinBosonBounds}, which give our approach a unique standing as a modeling technique based on an effective-environment concept but retaining the benefits of fully microscopic methods in terms of accuracy and rigor.

Our scheme maps a general open-system problem onto a Lindblad master equation for the system coupled to one or more small networks of interacting effective modes; the reduced dynamics of the system is then simulated by integrating this equation using standard numerical methods for Markovian problems and tracing out the oscillators. The surrogate modes are always at zero temperature regardless of the temperature of the original environment, giving the Lindblad equation a simple structure, and the interactions among them make a smaller number of modes necessary to account for the specific effect of any given environment than would be the case if they were all independent, with clear computational advantages. Not all modes need to be interacting; environmental spectral densities consisting of several terms may be reproduced using separate clusters of oscillators for each term, simplifying the calculation of their parameters while still exploiting the versatility of interacting oscillators in the rendering of each individual contribution.

As a first example of realistic use, we applied the method to a non-perturbative problem of a kind relevant for current research on coherent dynamics in biological molecular aggregates, obtaining accurate predictions (as confirmed by cross-checks with simulations performed using the well-established TEDOPA method) across the temperature range from absolute zero to room temperature with desktop-level computational resources. In addition, it was shown that by mapping non-Markovian problems to Markovian ones obeying a Lindblad quantum master equation, our approach can deal with short as well as long evolution times at comparable costs, making it a suitable tool for the simulation of long-lived dynamical features and relaxation to equilibrium. Next, we tested the performance of the technique on a much more challenging model inspired by the organic photovoltaics literature, demonstrating how the use of interacting surrogate modes allows for efficient simulation of systems so far only studied under much coarser approximations due to computational constraints. We have thus proved that the method is accurate, powerful and reliable and that there are classes of problems which no other approach known to us can successfully treat.

Future work on this project will be aimed mainly at improving the recipe for determining effective environment parameters, enhancing simulation efficiency and carrying out more case studies in order to better assess accuracy and performance, as well as applications to more systems of theoretical or experimental interest. Regarding the conceptual part of this work, we plan to undertake further analyses of the mapping problem from unitary to dissipative environments, in order to make the transformation more straightforward and possibly develop new TSO algorithms for different surrogate environment geometries. A deeper, system-dependent understanding of the error propagation from the environmental correlation function to the reduced dynamics could also be helpful in determining optimal figures of merit for individual problems, which would be a useful development for situations prioritizing accuracy over portability of the effective environment parameters. As a long-term goal, a fermionic extension of the entire approach based on the recent development of the relevant theoretical basis~\citep{Chen_FermionicTSO} is possible. Concerning the numerical implementations, it is our intention to refine and improve the codes for both the transformation and the simulations using various methods and possibly make them publicly accessible, as well as to investigate the possibility to integrate the two stages of our approach, for instance by choosing the surrogate environment geometry depending on the simulation routines for maximum efficiency.

\acknowledgments

We wish to thank A. Mattioni, F. Caycedo-Soler, J. Lim, G. Gasbarri, R. Puebla and M. Paternostro for useful discussions, suggestions and feedback about the work presented in this paper. We further acknowledge support by the State of Baden-W\"urttemberg through bwHPC for the use of the BwUniCluster and the German Research Foundation (DFG) through grant No. INST 40/467-1 FUGG for the use of the JUSTUS cluster. This work was supported by the ERC Synergy Grant BioQ, the EU Projects HYPERDIAMOND and AsteriQs, the BMBF projects DiaPol and NanoSpin, the Center for Integrated Quantum Science and Technology (IQST) and the Foundation BLANCEFLOR Boncompagni Ludovisi, n\'ee Bildt.

\bibliography{References}

\begin{thebibliography}{100}

\bibitem{BreuerPetruccione}
H.-P. Breuer and F.~Petruccione.
\newblock {\em The theory of open quantum systems}.
\newblock Oxford University Press, 2002.

\bibitem{GardinerZoller}
C.~Gardiner and P.~Zoller.
\newblock {\em Quantum Noise: A Handbook of Markovian and Non-Markovian Quantum
  Stochastic Methods with Applications to Quantum Optics}.
\newblock Springer, 2004.

\bibitem{Weiss}
U.~Weiss.
\newblock {\em Quantum Dissipative Systems}.
\newblock World Scientific, third edition, 2008.

\bibitem{RivasHuelga}
{\'A}.~Rivas and S.~F. Huelga.
\newblock {\em Open Quantum Systems. An Introduction}.
\newblock Springer, 2012.

\bibitem{Leggett_SpinBoson}
A.~J. Leggett, S.~Chakravarty, A.~T. Dorsey, M.~P.~A. Fisher, A.~Garg, and
  W.~Zwerger.
\newblock Dynamics of the dissipative two-state system.
\newblock {\em Reviews of Modern Physics}, 59:1--85, 1987.

\bibitem{CaldeiraLeggettModel}
A.~O. Caldeira and A.~J. Leggett.
\newblock Path integral approach to quantum {B}rownian motion.
\newblock {\em Physica {A}: Statistical Mechanics and its Applications},
  121(3):587--616, 1983.

\bibitem{GoriniKossakowskiSudarshan}
V.~Gorini, A.~Kossakowski, and E.~C.~G. Sudarshan.
\newblock Completely positive dynamical semigroups of {N}-level systems.
\newblock {\em Journal of Mathematical Physics}, 17(5):821--825, 1976.

\bibitem{GoriniFrigerioVerriKossakowskiSudarshan}
V.~Gorini, A.~Frigerio, M.~Verri, A.~Kossakowski, and E.~C.~G. Sudarshan.
\newblock Properties of quantum {M}arkovian master equations.
\newblock {\em Reports on Mathematical Physics}, 13(2):149--173, 1978.

\bibitem{Lindblad}
G.~Lindblad.
\newblock On the generators of quantum dynamical semigroups.
\newblock {\em Communications in Mathematical Physics}, 48(2):119--130, 1976.

\bibitem{Nakajima}
S.~Nakajima.
\newblock On quantum theory of transport phenomena: steady diffusion.
\newblock {\em Progress of Theoretical Physics}, 20(6):948--959, 1958.

\bibitem{Zwanzig}
R.~Zwanzig.
\newblock Ensemble method in the theory of irreversibility.
\newblock {\em The Journal of Chemical Physics}, 33(5):1338--1341, 1960.

\bibitem{Prigogine}
I.~Prigogine.
\newblock {\em Non-Equilibrium Statistical Mechanics}.
\newblock Interscience Publishers, 1962.

\bibitem{Shibata_TCL}
F.~Shibata, Y.~Takahashi, and N.~Hashitsume.
\newblock A generalized stochastic {L}iouville equation. {N}on-{M}arkovian
  versus memoryless master equations.
\newblock {\em Journal of Statistical Physics}, 17(4):171--187, 1977.

\bibitem{NEGF_Review}
J.~Rammer and H.~Smith.
\newblock Quantum field-theoretical methods in transport theory of metals.
\newblock {\em Reviews of Modern Physics}, 58:323--359, 1986.

\bibitem{Tanimura_HEOM}
Y.~Tanimura.
\newblock Nonperturbative expansion method for a quantum system coupled to a
  harmonic-oscillator bath.
\newblock {\em Physical Review A}, 41:6676--6687, 1990.

\bibitem{Makri_QUAPIletter}
N.~Makri.
\newblock Improved {F}eynman propagators on a grid and non-adiabatic
  corrections within the path integral framework.
\newblock {\em Chemical Physics Letters}, 193(5):435--445, 1992.

\bibitem{ChinPlenio_TEDOPA}
A.~W. Chin, {\'A}.~Rivas, S.~F. Huelga, and M.~B. Plenio.
\newblock Exact mapping between system-reservoir quantum models and
  semi-infinite discrete chains using orthogonal polynomials.
\newblock {\em Journal of Mathematical Physics}, 51(9), 2010.

\bibitem{PriorPlenio_TEDOPA}
J.~Prior, A.~W. Chin, S.~F. Huelga, and M.~B. Plenio.
\newblock Efficient simulation of strong system-environment interactions.
\newblock {\em Physical Review Letters}, 105:050404, 2010.

\bibitem{DiosiStrunz_QuantumStateDiffusion}
L.~Di{\'o}si and W.~T. Strunz.
\newblock The non-{M}arkovian stochastic {S}chr{\"o}dinger equation for open
  systems.
\newblock {\em Physics Letters A}, 235(6):569--573, 1997.

\bibitem{Piilo_NonMarkovianQuantumJumpsPRL}
J.~Piilo, S.~Maniscalco, K.~H\"ark\"onen, and K.-A. Suominen.
\newblock Non-{M}arkovian quantum jumps.
\newblock {\em Physical Review Letters}, 100:180402, 2008.

\bibitem{Davies_WeakCoupling}
E.~B. Davies.
\newblock Markovian master equations.
\newblock {\em Communications in Mathematical Physics}, 39(2):91--110, 1974.

\bibitem{DumckeSpohn_WeakCoupling}
R.~D{\"u}mcke and H.~Spohn.
\newblock The proper form of the generator in the weak coupling limit.
\newblock {\em Zeitschrift f{\"u}r Physik B Condensed Matter}, 34(4):419--422,
  1979.

\bibitem{SmirneVacchini_NakajimaVsTCL}
A.~Smirne and B.~Vacchini.
\newblock Nakajima-{Z}wanzig versus time-convolutionless master equation for
  the non-{M}arkovian dynamics of a two-level system.
\newblock {\em Physical Review A}, 82:022110, 2010.

\bibitem{BreuerKapplerPetruccione_TCL}
H.-P. {Breuer}, B.~{Kappler}, and F.~{Petruccione}.
\newblock The time-convolutionless projection operator technique in the quantum
  theory of dissipation and decoherence.
\newblock {\em Annals of Physics}, 291:36--70, 2001.

\bibitem{Minchev_MatExp}
B.~V. Minchev and W.~M. Wright.
\newblock A review of exponential integrators for first order semi-linear
  problems, 2005.

\bibitem{PlenioKnight_MCWF}
M.~B. Plenio and P.~L. Knight.
\newblock The quantum-jump approach to dissipative dynamics in quantum optics.
\newblock {\em Reviews of Modern Physics}, 70:101--144, 1998.

\bibitem{GisinPercival_QuantumStateDiffusion}
N.~Gisin and I.~C. Percival.
\newblock The quantum-state diffusion model applied to open systems.
\newblock {\em Journal of Physics A: Mathematical and General},
  25(21):5677--5691, 1992.

\bibitem{RivasHuelgaPlenio_NonMarkovianity}
{\'A}.~Rivas, S.~F. Huelga, and M.~B. Plenio.
\newblock Quantum non-{M}arkovianity: characterization, quantification and
  detection.
\newblock {\em Reports on Progress in Physics}, 77(9):094001, 2014.

\bibitem{BreuerVacchini_NonMarkovianity}
H.-P. Breuer, E.-M. Laine, J.~Piilo, and B.~Vacchini.
\newblock Colloquium: {N}on-{M}arkovian dynamics in open quantum systems.
\newblock {\em Reviews of Modern Physics}, 88:021002, 2016.

\bibitem{DeVegaAlonso_NonMarkovianity}
I.~de~Vega and D.~Alonso.
\newblock Dynamics of non-{M}arkovian open quantum systems.
\newblock {\em Reviews of Modern Physics}, 89:015001, 2017.

\bibitem{LiHallWiseman_NonMarkovianity}
L.~Li, M.~J.~W. Hall, and H.~M. Wiseman.
\newblock Concepts of quantum non-{M}arkovianity: A hierarchy.
\newblock {\em Physics Reports}, 759:1--51, 2018.

\bibitem{TanimuraKubo_HEOM}
Y.~Tanimura and R.~Kubo.
\newblock Time evolution of a quantum system in contact with a nearly
  {G}aussian-{M}arkoffian noise bath.
\newblock {\em Journal of the Physical Society of Japan}, 58(1):101--114, 1989.

\bibitem{TopalerMakri_QUAPI}
M.~Topaler and N.~Makri.
\newblock Quasi-adiabatic propagator path integral methods. {E}xact quantum
  rate constants for condensed phase reactions.
\newblock {\em Chemical Physics Letters}, 210(1):285--293, 1993.

\bibitem{MakriMakarov_QUAPI1}
N.~Makri and D.~E. Makarov.
\newblock Tensor propagator for iterative quantum time evolution of reduced
  density matrices. {I. Theory}.
\newblock {\em The Journal of Chemical Physics}, 102(11):4600--4610, 1995.

\bibitem{MakriMakarov_QUAPI2}
N.~Makri and D.~E. Makarov.
\newblock Tensor propagator for iterative quantum time evolution of reduced
  density matrices. {II. Numerical methodology}.
\newblock {\em The Journal of Chemical Physics}, 102(11):4611--4618, 1995.

\bibitem{Danielewicz_NEGF}
P.~Danielewicz.
\newblock Quantum theory of nonequilibrium processes, {I}.
\newblock {\em Annals of Physics}, 152(2):239--304, 1984.

\bibitem{Diosi_QuantumStateDiffusion}
L.~Di{\'{o}}si.
\newblock Exact semiclassical wave equation for stochastic quantum optics.
\newblock {\em Quantum and Semiclassical Optics: Journal of the European
  Optical Society Part B}, 8(1):309--314, 1996.

\bibitem{Strunz_QuantumStateDiffusion}
W.~T. Strunz.
\newblock Linear quantum state diffusion for non-{M}arkovian open quantum
  systems.
\newblock {\em Physics Letters A}, 224(1):25--30, 1996.

\bibitem{Piilo_NonMarkovianQuantumJumpsPRA}
J.~Piilo, K.~H\"ark\"onen, S.~Maniscalco, and K.-A. Suominen.
\newblock Open system dynamics with non-{M}arkovian quantum jumps.
\newblock {\em Physical Review A}, 79:062112, 2009.

\bibitem{Lovett_TEMPO}
A.~Strathearn, P.~Kirton, D.~Kilda, J.~Keeling, and B.~W. Lovett.
\newblock Efficient non-{M}arkovian quantum dynamics using time-evolving matrix
  product operators.
\newblock {\em Nature Communications}, 9:3322, 2018.

\bibitem{Daley_tDMRG}
A.~J. Daley, C.~Kollath, U.~Schollwöck, and G.~Vidal.
\newblock Time-dependent density-matrix renormalization-group using adaptive
  effective {H}ilbert spaces.
\newblock {\em Journal of Statistical Mechanics: Theory and Experiment},
  2004(04):P04005, 2004.

\bibitem{Guifre_tDMRG}
G.~Vidal.
\newblock Efficient simulation of one-dimensional quantum many-body systems.
\newblock {\em Physical Review Letters}, 93:040502, 2004.

\bibitem{Schollwock_tDMRG}
U.~Schollw\"ock.
\newblock The density-matrix renormalization group.
\newblock {\em Reviews of Modern Physics}, 77:259--315, 2005.

\bibitem{TamascelliSmirne_ThermalizedTEDOPA}
D.~Tamascelli, A.~Smirne, J.~Lim, S.~F. Huelga, and M.~B. Plenio.
\newblock Efficient simulation of finite-temperature open quantum systems.
\newblock {\em Physical Review Letters}, 123:090402, 2019.

\bibitem{WoodsCramerPleino_TEDOPAerrorbars}
M.~P. Woods, M.~Cramer, and M.~B. Plenio.
\newblock Simulating bosonic baths with error bars.
\newblock {\em Physical Review Letters}, 115:130401, 2015.

\bibitem{WoodsPlenio_LiebRobinsonBounds}
M.~P. Woods and M.~B. Plenio.
\newblock Dynamical error bounds for continuum discretisation via gauss
  quadrature rules--{A} {L}ieb--{R}obinson bound approach.
\newblock {\em Journal of Mathematical Physics}, 57(2):022105, 2016.

\bibitem{SpinBosonBounds}
F.~Mascherpa, A.~Smirne, S.~F. Huelga, and M.~B. Plenio.
\newblock Open systems with error bounds: Spin-boson model with spectral
  density variations.
\newblock {\em Physical Review Letters}, 118:100401, 2017.

\bibitem{PiiloManiscalco_HarmonicNetworks}
J.~Nokkala, F.~Galve, R.~Zambrini, S.~Maniscalco, and J.~Piilo.
\newblock Complex quantum networks as structured environments: engineering and
  probing.
\newblock {\em Scientific Reports}, 6:26861, 2016.

\bibitem{Imamoglu_pseudomodes}
A.~Imamo\u{g}lu.
\newblock Stochastic wave-function approach to non-{M}arkovian systems.
\newblock {\em Physical Review A}, 50:3650--3653, 1994.

\bibitem{Garraway_pseudomodes}
B.~M. Garraway.
\newblock Nonperturbative decay of an atomic system in a cavity.
\newblock {\em Physical Review A}, 55(3):2290--2303, 1997.

\bibitem{Dalton_pseudomodes}
B.~J. Dalton, Stephen~M. Barnett, and B.~M. Garraway.
\newblock Theory of pseudomodes in quantum optical processes.
\newblock {\em Physical Review A}, 64:053813, 2001.

\bibitem{Lemmer_SpinBoson}
A.~Lemmer, C.~Cormick, D.~Tamascelli, T.~Schaetz, S.~F. Huelga, and M.~B.
  Plenio.
\newblock A trapped-ion simulator for spin-boson models with structured
  environments.
\newblock {\em New Journal of Physics}, 20(7):073002, 2018.

\bibitem{IlesSmith_ReactionCoordinate}
J.~Iles-Smith, N.~Lambert, and A.~Nazir.
\newblock Environmental dynamics, correlations, and the emergence of
  noncanonical equilibrium states in open quantum systems.
\newblock {\em Physical Review A}, 90:032114, 2014.

\bibitem{IlesSmith_StructuredEnvironments}
J.~Iles-Smith, A.~G. Dijkstra, N.~Lambert, and A.~Nazir.
\newblock Energy transfer in structured and unstructured environments: Master
  equations beyond the {B}orn--{M}arkov approximations.
\newblock {\em The Journal of Chemical Physics}, 144(4):044110, 2016.

\bibitem{Lambert_SpinBosonStudy}
N.~Lambert, S.~Ahmed, M.~Cirio, and F.~Nori.
\newblock Modelling the ultra-strongly coupled spin-boson model with unphysical
  modes.
\newblock {\em Nature Communications}, 10:3721, 2019.

\bibitem{Fruchtman_Perturbative}
A.~Fruchtman, N.~Lambert, and E.~M. Gauger.
\newblock When do perturbative approaches accurately capture the dynamics of
  complex quantum systems?
\newblock {\em Scientific Reports}, 6, 2016.

\bibitem{Falci_1overf}
E.~Paladino, Y.~M. Galperin, G.~Falci, and B.~L. Altshuler.
\newblock 1/f noise: {I}mplications for solid-state quantum information.
\newblock {\em Reviews of Modern Physics}, 86:361--418, 2014.

\bibitem{Schwarz_LindbladDrivenDiscretizedLeads}
F.~Schwarz, M.~Goldstein, A.~Dorda, E.~Arrigoni, A.~Weichselbaum, and J.~von
  Delft.
\newblock Lindblad-driven discretized leads for nonequilibrium steady-state
  transport in quantum impurity models: Recovering the continuum limit.
\newblock {\em Physical Review B}, 94:155142, 2016.

\bibitem{Luchnikov_TensorNetworkLindblad}
I.~A. Luchnikov, S.~V. Vintskevich, H.~Ouerdane, and S.~N. Filippov.
\newblock Simulation complexity of open quantum dynamics: Connection with
  tensor networks.
\newblock {\em Physical Review Letters}, 122:160401, 2019.

\bibitem{Faccioli_QTFT}
E.~Schneider, S.~a~Beccara, F.~Mascherpa, and P.~Faccioli.
\newblock Quantum propagation of electronic excitations in macromolecules: A
  computationally efficient multiscale approach.
\newblock {\em Physical Review B}, 94:014306, 2016.

\bibitem{TSO_Theorem}
D.~Tamascelli, A.~Smirne, S.~F. Huelga, and M.~B. Plenio.
\newblock Nonperturbative treatment of non-{M}arkovian dynamics of open quantum
  systems.
\newblock {\em Physical Review Letters}, 120:030402, 2018.

\bibitem{YurkeDenker_QuantumCircuits}
B.~Yurke and J.~S. Denker.
\newblock Quantum network theory.
\newblock {\em Physical Review A}, 29:1419--1437, 1984.

\bibitem{RibeiroVieira_Transport}
P.~Ribeiro and V.~R. Vieira.
\newblock Non-{M}arkovian effects in electronic and spin transport.
\newblock {\em Physical Review B}, 92:100302(R), 2015.

\bibitem{Scholes_QBioNature}
E.~Collini, C.~Y. Wong, K.~E. Wilk, P.~M.~G. Curmi, P.~Brumer, and G.~D.
  Scholes.
\newblock Coherently wired light-harvesting in photosynthetic marine algae at
  ambient temperature.
\newblock {\em Nature}, 463:644, 2010.

\bibitem{HuelgaPlenio_QBio}
S.~F. Huelga and M.~B. Plenio.
\newblock Vibrations, quanta and biology.
\newblock {\em Contemporary Physics}, 54(4):181--207, 2013.

\bibitem{Lax_QRT}
M.~Lax.
\newblock Formal theory of quantum fluctuations from a driven state.
\newblock {\em Physical Review}, 129:2342--2348, 1963.

\bibitem{Chen_FermionicTSO}
F.~Chen, E.~Arrigoni, and M.~Galperin.
\newblock Markovian treatment of non-{M}arkovian dynamics of open fermionic
  systems.
\newblock arXiv:1909.08658.

\bibitem{Talkner_NoQRT}
P.~Talkner.
\newblock The failure of the quantum regression hypothesis.
\newblock {\em Annals of Physics}, 167(2):390--436, 1986.

\bibitem{MayKuehn}
V.~May and O.~K{\"u}hn.
\newblock {\em Charge and energy transfer dynamics in molecular systems}.
\newblock John Wiley \& Sons, 2008.

\bibitem{DeVegaBanuls_Thermofield}
I.~de~Vega and M.~C. Ba\~nuls.
\newblock Thermofield-based chain-mapping approach for open quantum systems.
\newblock {\em Physical Review A}, 92:052116, 2015.

\bibitem{DiosiGisinStrunz_NonMarkovianity}
L.~Di\'osi, N.~Gisin, and W.~T. Strunz.
\newblock Non-{M}arkovian quantum state diffusion.
\newblock {\em Physical Review A}, 58:1699--1712, 1998.

\bibitem{Ritschel_AbsorptionSpectra}
G.~Ritschel, D.~Suess, S.~M{\"o}bius, W.~T. Strunz, and A.~Eisfeld.
\newblock Non-{M}arkovian quantum state diffusion for temperature-dependent
  linear spectra of light harvesting aggregates.
\newblock {\em The Journal of Chemical Physics}, 142(3):034115, 2015.

\bibitem{Carmichael}
H.~Carmichael.
\newblock {\em An Open Systems Approach to Quantum Optics}.
\newblock Springer, 1993.

\bibitem{Marple_Prony}
S.~L. Marple.
\newblock {\em Digital spectral analysis: with applications}, volume~5.
\newblock Prentice-Hall Englewood Cliffs, NJ, 1987.

\bibitem{FordLewisOConnell_DampedOscillator}
G.~W. Ford, J.~T. Lewis, and R.~F. O'~Connell.
\newblock Independent oscillator model of a heat bath: Exact diagonalization of
  the {H}amiltonian.
\newblock {\em Journal of Statistical Physics}, 53(1):439--455, 1988.

\bibitem{CaldeiraLeggett_PureDephasing}
A.~O. Caldeira and A.~J. Leggett.
\newblock Influence of damping on quantum interference: An exactly soluble
  model.
\newblock {\em Physical Review A}, 31(2):1059--1066, 1985.

\bibitem{Peropadre_UltrastrongCoupling}
B.~Peropadre, D.~Zueco, D.~Porras, and J.~J. Garc\'{\i}a-Ripoll.
\newblock Nonequilibrium and nonperturbative dynamics of ultrastrong coupling
  in open lines.
\newblock {\em Physical Review Letters}, 111:243602, 2013.

\bibitem{QuTiP1}
J.~R. Johansson, P.~D. Nation, and F.~Nori.
\newblock {QuTiP}: An open-source {P}ython framework for the dynamics of open
  quantum systems.
\newblock {\em Computer Physics Communications}, 183(8):1760--1772, 2012.

\bibitem{QuTiP2}
J.~R. Johansson, P.~D. Nation, and F.~Nori.
\newblock {QuTiP} 2: A {P}ython framework for the dynamics of open quantum
  systems.
\newblock {\em Computer Physics Communications}, 184(4):1234--1240, 2013.

\bibitem{Pelzer_Transport}
K.~M. Pelzer, A.~F. Fidler, G.~B. Griffin, S.~K. Gray, and G.~S. Engel.
\newblock The dependence of exciton transport efficiency on spatial patterns of
  correlation within the spectral bath.
\newblock {\em New Journal of Physics}, 15(9):095019, 2013.

\bibitem{DeSio_OPV}
A.~De~Sio and C.~Lienau.
\newblock Vibronic coupling in organic semiconductors for photovoltaics.
\newblock {\em Physical Chemistry Chemical Physics}, 19:18813--18830, 2017.

\bibitem{Haase_Metrology}
J.~F. Haase, P.~J. Vetter, T.~Unden, A.~Smirne, J.~Rosskopf, B.~Naydenov,
  A.~Stacey, F.~Jelezko, M.~B. Plenio, and S.~F. Huelga.
\newblock Controllable non-{M}arkovianity for a spin qubit in diamond.
\newblock {\em Physical Review Letters}, 121:060401, 2018.

\bibitem{UzdinLevyKosloff_QHE}
R.~Uzdin, A.~Levy, and R.~Kosloff.
\newblock Quantum heat machines equivalence, work extraction beyond
  {M}arkovianity, and strong coupling via heat exchangers.
\newblock {\em Entropy}, 18(4), 2016.

\bibitem{MitchisonPlenio_NonEquilibrium}
M.~T. Mitchison and M.~B. Plenio.
\newblock Non-additive dissipation in open quantum networks out of equilibrium.
\newblock {\em New Journal of Physics}, 20(3):033005, 2018.

\bibitem{JelezkoPlenio_NV}
Y.~Wu, F.~Jelezko, M.~B. Plenio, and T.~Weil.
\newblock Diamond quantum devices in biology.
\newblock {\em Angewandte Chemie International Edition}, 55(23):6586--6598,
  2016.

\bibitem{ChinHuelgaPlenio_Metrology}
A.~W. Chin, S.~F. Huelga, and M.~B. Plenio.
\newblock Quantum metrology in non-{M}arkovian environments.
\newblock {\em Physical Review Letters}, 109:233601, 2012.

\bibitem{SmirneKolodynski_Metrology}
A.~Smirne, J.~Ko\l{}ody{\'n}ski, S.~F. Huelga, and
  R.~Demkowicz{-}Dobrza{\'n}ski.
\newblock Ultimate precision limits for noisy frequency estimation.
\newblock {\em Physical Review Letters}, 116:120801, 2016.

\bibitem{HaaseSmirneKolodynski_Metrology}
J.~F. Haase, A.~Smirne, J.~Ko\l{}ody{\'n}ski, R.~Demkowicz{-}Dobrza{\'n}ski,
  and S.~F. Huelga.
\newblock Fundamental limits to frequency estimation: a comprehensive
  microscopic perspective.
\newblock {\em New Journal of Physics}, 20(5):053009, 2018.

\bibitem{Clark_OPV}
J.~Clark, C.~Silva, R.~H. Friend, and F.~C. Spano.
\newblock Role of intermolecular coupling in the photophysics of disordered
  organic semiconductors: Aggregate emission in regioregular polythiophene.
\newblock {\em Physical Review Letters}, 98:206406, 2007.

\bibitem{Tamura_OPV}
H.~Tamura, R.~Martinazzo, M.~Ruckenbauer, and I.~Burghardt.
\newblock Quantum dynamics of ultrafast charge transfer at an
  oligothiophene-fullerene heterojunction.
\newblock {\em The Journal of Chemical Physics}, 137(22):22A540, 2012.

\bibitem{Plenio_QBio}
M.~B. Plenio and S.~F. Huelga.
\newblock Dephasing-assisted transport: quantum networks and biomolecules.
\newblock {\em New Journal of Physics}, 10(11):113019, 2008.

\bibitem{PlenioAlmeidaHuelga_Dimer}
M.~B. Plenio, J.~Almeida, and S.~F. Huelga.
\newblock Origin of long-lived oscillations in {2D}-spectra of a quantum
  vibronic model: Electronic versus vibrational coherence.
\newblock {\em The Journal of Chemical Physics}, 139(23):235102, 2013.

\bibitem{AdolphsRenger}
J.~Adolphs and T.~Renger.
\newblock How proteins trigger excitation energy transfer in the {FMO} complex
  of green sulfur bacteria.
\newblock {\em Biophysical Journal}, 91(8):2778--2797, 2006.

\bibitem{Tamascelli_RSVD}
D.~Tamascelli, R.~Rosenbach, and M.~B. Plenio.
\newblock Improved scaling of time-evolving block-decimation algorithm through
  reduced-rank randomized singular value decomposition.
\newblock {\em Physical Review E}, 91:063306, 2015.

\bibitem{Kohn_RSVD}
L.~Kohn, F.~Tschirsich, M.~Keck, M.~B. Plenio, D.~Tamascelli, and
  S.~Montangero.
\newblock Probabilistic low-rank factorization accelerates tensor network
  simulations of critical quantum many-body ground states.
\newblock {\em Physical Review E}, 97:013301, 2018.

\bibitem{DumZollerRitsch_MCWF}
R.~Dum, P.~Zoller, and H.~Ritsch.
\newblock Monte {C}arlo simulation of the atomic master equation for
  spontaneous emission.
\newblock {\em Physical Review A}, 45:4879--4887, 1992.

\bibitem{DalibardCastinMolmer_MCWF}
J.~Dalibard, Y.~Castin, and K.~M\o{}lmer.
\newblock Wave-function approach to dissipative processes in quantum optics.
\newblock {\em Physical Review Letters}, 68:580--583, 1992.

\bibitem{Mukamel}
S.~Mukamel.
\newblock {\em Principles of Nonlinear Optical Spectroscopy}.
\newblock Oxford University Press, 1995.

\bibitem{Somoza_DAMPF}
A.~D. Somoza, O.~Marty, J.~Lim, S.~F. Huelga, and M.~B. Plenio.
\newblock Dissipation-assisted matrix product factorization.
\newblock {\em Physical Review Letters}, 123:100502, 2019.

\bibitem{Clarke_OPV}
T.~M. Clarke and J.~R. Durrant.
\newblock Charge photogeneration in organic solar cells.
\newblock {\em Chemical Reviews}, 110(11):6736--6767, 2010.
\newblock PMID: 20063869.

\bibitem{Proctor_OPV}
C.~M. Proctor, M.~Kuik, and T.-Q. Nguyen.
\newblock Charge carrier recombination in organic solar cells.
\newblock {\em Progress in Polymer Science}, 38(12):1941--1960, 2013.
\newblock Topical issue on Conductive Polymers.

\bibitem{Yamagata_QuantumWires}
H.~Yamagata and F.~C. Spano.
\newblock Vibronic coupling in quantum wires: Applications to polydiacetylene.
\newblock {\em The Journal of Chemical Physics}, 135(5):054906, 2011.

\bibitem{Tempelaar_Coherence}
R.~Tempelaar, F.~C. Spano, J.~Knoester, and T.~L.~C. Jansen.
\newblock Mapping the evolution of spatial exciton coherence through
  time-resolved fluorescence.
\newblock {\em The Journal of Physical Chemistry Letters}, 5:1505--1510, 2014.

\bibitem{Spano_Aggregates}
F.~C. Spano and C.~Silva.
\newblock H- and {J}-aggregate behavior in polymeric semiconductors.
\newblock {\em Annual Review of Physical Chemistry}, 65(1):477--500, 2014.
\newblock PMID: 24423378.

\bibitem{Hestand_Aggregates}
N.~Hestand and F.~C. Spano.
\newblock Expanded theory of {H}- and {J}-molecular aggregates: The effects of
  vibronic coupling and intermolecular charge transfer.
\newblock {\em Chemical Reviews}, 118:7069--7163, 2018.

\bibitem{Blau_dimer}
S.~M. Blau, D.~I.~G. Bennett, C.~Kreisbeck, G.~D. Scholes, and A.~Aspuru-Guzik.
\newblock Local protein solvation drives direct down-conversion in
  phycobiliprotein {PC645} via incoherent vibronic transport.
\newblock {\em Proceedings of the National Academy of Sciences},
  115(15):E3342--E3350, 2018.

\bibitem{Chenel_OPV}
A.~Chenel, E.~Mangaud, I.~Burghardt, C~Meier, and M.~Desouter-Lecomte.
\newblock Exciton dissociation at donor-acceptor heterojunctions: Dynamics
  using the collective effective mode representation of the spin-boson model.
\newblock {\em The Journal of Chemical Physics}, 140(4):044104, 2014.

\bibitem{Mascarenhas_MPO}
E.~Mascarenhas, H.~Flayac, and V.~Savona.
\newblock Matrix-product-operator approach to the nonequilibrium steady state
  of driven-dissipative quantum arrays.
\newblock {\em Physical Review A}, 92:022116, 2015.

\bibitem{CuiCiracBanuls_MPO}
J.~Cui, J.~I. Cirac, and M.~C. Ba\~nuls.
\newblock Variational matrix product operators for the steady state of
  dissipative quantum systems.
\newblock {\em Physical Review Letters}, 114:220601, 2015.

\bibitem{VoSidje_Krylov}
H.~D. Vo and Roger~B. Sidje.
\newblock Approximating the large sparse matrix exponential using incomplete
  orthogonalization and {K}rylov subspaces of variable dimension.
\newblock {\em Numerical Linear Algebra with Applications}, 24(3):e2090, 2017.

\bibitem{Tokman_KIOPS}
S.~Gaudreault, G.~Rainwater, and M.~Tokman.
\newblock {KIOPS}: A fast adaptive {K}rylov subspace solver for exponential
  integrators.
\newblock {\em Journal of Computational Physics}, 372:236--255, 2018.

\bibitem{CallenWeltonFDT}
H.~B. Callen and T.~A. Welton.
\newblock Irreversibility and generalized noise.
\newblock {\em Physical Review}, 83:34--40, 1951.

\bibitem{FordOConnell_NoQRT}
G.~W. Ford and R.~F. O'~Connell.
\newblock There is no quantum regression theorem.
\newblock {\em Physical Review Letters}, 77:798--801, 1996.

\bibitem{Bennett_Dimer}
D.~I.~G. Bennett, P.~Mal{\'y}, C.~Kreisbeck, R.~van Grondelle, and
  A.~Aspuru-Guzik.
\newblock Mechanistic regimes of vibronic transport in a heterodimer and the
  design principle of incoherent vibronic transport in phycobiliproteins.
\newblock {\em The Journal of Physical Chemistry Letters}, 9(10):2665--2670,
  2018.
\newblock PMID: 29683676.

\bibitem{Duan_EfficientHEOM}
C.~Duan, Q.~Wang, Z.~Tang, and J.~Wu.
\newblock The study of an extended hierarchy equation of motion in the
  spin-boson model: The cutoff function of the sub-{O}hmic spectral density.
\newblock {\em The Journal of Chemical Physics}, 147(16):164112, 2017.

\bibitem{Duan_EfficientT0HEOM}
C.~Duan, Z.~Tang, J.~Cao, and J.~Wu.
\newblock Zero-temperature localization in a sub-ohmic spin-boson model
  investigated by an extended hierarchy equation of motion.
\newblock {\em Physical Review B}, 95:214308, 2017.

\bibitem{MeierTannor_DecompositionHEOM}
C.~Meier and D.~J. Tannor.
\newblock {Non-Markovian} evolution of the density operator in the presence of
  strong laser fields.
\newblock {\em The Journal of Chemical Physics}, 111(8):3365--3376, 1999.

\bibitem{HuLuo_DecompositionHEOM}
J.~Hu, M.~Luo, F.~Jiang, R.-X. Xu, and Y.~Yan.
\newblock Pad\'e spectrum decompositions of quantum distribution functions and
  optimal hierarchical equations of motion construction for quantum open
  systems.
\newblock {\em The Journal of Chemical Physics}, 134(24):244106, 2011.

\end{thebibliography}
\bibliographystyle{unsrt}

\clearpage

\appendix

\section{Transformation to Surrogate Oscillators in detail}
\label{app:TSO_details}

In Section~\ref{sec:TSO}, we defined the general form of the effective environments used in our method, and sketched the transformation algorithm by which we obtain their parameters given a target correlation function. Here we will go through the procedure in detail, in order to give a clearer view of its more technical aspects.

\subsection{Effective correlation function}

The Hamiltonian for our effective oscillators in a chain configuration with hopping interactions is (with $\hbar=1$)
\begin{equation}\label{eq:H_R_2}
	H_R\coloneqq\sum^N_{n=1}\Omega_nb^\dagger_n b_n+\sum^{N-1}_{n=1}
	\left(g_nb_nb^\dagger_{n+1}+g^*_nb^\dagger_nb_{n+1}\right)
\end{equation}
and we consider a zero-temperature Lindblad dissipator
\begin{equation}\label{eq:D_R}
	\mathcal{D}_R[\rho_R]\coloneqq
	\sum^N_{n=1}\Gamma_n\left(b_n\rho_Rb^\dagger_n
	-\frac{1}{2}\left\{b^\dagger_nb_n,\rho_R\right\}\right)
\end{equation}
acting locally on each mode. The interaction term with the system has the form
\begin{equation}\label{eq:H_I}
	H'_I\coloneqq\sum^m_{k=1}A_{Sk}\otimes	F_{Rk},
\end{equation}
where the interaction operators $F_{Rk}$ of the environment are linear in the creation and annihilation operators:
\begin{equation}\label{eq:F_k}
	F_{Rk}\coloneqq\sum^N_{n=1}\left(c_{nk}b_n+c^*_{nk}b^\dagger_n\right).
\end{equation}

Assuming factorizing initial conditions $\rho_0=\rho_{0S}\otimes\rho_{0R}$ with $\rho_{0R}=\bigotimes^N_{n=1}\ket{0}\!\bra{0}_n$, which is Gaussian and stationary under this dynamics, meaning it satisfies
\[
	\mathcal{L}_R[\rho_{0R}]\coloneqq-i[H_R,\rho_{0R}]
	+\mathcal{D}_R[\rho_{0R}]=0,
\]
the correlation function
\begin{equation}\label{eq:CFkk}
	C^R_{kk'}(t+\tau,\tau)\coloneqq\langle F_{Rk}(t+\tau)F_{Rk'}(\tau)\rangle_R
\end{equation}
is independent of the first evolution time $\tau$. We will drop the $\tau$ time argument from now on and also restrict our analysis to a single interaction operator ($m=1$ in Eq.~\eqref{eq:H_I}), so in the following the correlation function~\eqref{eq:CFkk} will be denoted by $C^R(t)$. Writing it out explicitly in terms of the expression of $F_R$, we get
\begin{equation}\label{eq:C_R}
	C^R(t)=\sum^N_{m,n=1}c_mc^*_n\langle b_m(t)b^\dagger_n(0)\rangle_R,
\end{equation}
since terms with two creation or annihilation operators and contributions proportional to $\langle b^\dagger_m(t)b_n(0)\rangle_R$ are zero for our initial vacuum state.

It is easy to show that the hopping coupling constants $g_n$ can be assumed real and positive without loss of generality in $C^R(t)$: define the canonical transformation
\begin{equation}
	b_n\mapsto\;e^{i\delta_n}b_n
\end{equation}
for arbitrary real $\delta_n$. The creation operators $b^\dagger_n$ will transform with the opposite phase, preserving the canonical commutation relations. The free term in the Hamiltonian~\eqref{eq:H_R_2} and the dissipator~\eqref{eq:D_R} are invariant under this transformation; the hopping term in Eq.~\eqref{eq:H_R_2} and the interaction operator $F_R$ defined as in~\eqref{eq:F_k} are not:
\begin{align*}
	c_nb_n+c^*_nb^\dagger_n\mapsto\;&
	c_ne^{i\delta_n}b_n+c^*_ne^{-i\delta_n}b^\dagger_n
	\\
	g_nb_nb^\dagger_{n+1}+g^*_nb^\dagger_nb_{n+1}\mapsto\;&
	g_ne^{i(\delta_n-\delta_{n+1})}b_nb^\dagger_{n+1}
	\\
	&+g^*_ne^{-i(\delta_n-\delta_{n+1})}b^\dagger_nb_{n+1}.
\end{align*}
Taking $\delta_n$ such that $g_ne^{i(\delta_n-\delta_{n+1})}=|g_n|$, we may absorb the phase of the couplings in the still undetermined $c_n$, without restricting the physical picture in any way. Note that this leaves one of the $\delta_n$ still free as an overall phase in all the operator coefficients, which may be set e.g.\ so that $c_1$ or $c_N$ is real.

The free dynamics of the oscillators with no coupling to the system is given by the Lindblad equation
\begin{equation}\label{eq:Lindblad}
	\frac{\mathrm{d}}{\mathrm{d}t}\rho_R(t)
	=-i[H_R,\rho_R(t)]+\mathcal{D}_R[\rho_R(t)].
\end{equation}
Acting with the operator $b_n$ from the left on both sides and taking the trace, we get
\begin{equation}\label{eq:<b(t)>}
	\frac{\mathrm{d}}{\mathrm{d}t}\langle b_n(t)\rangle_R
	=\sum^N_{m=1}M_{nm}\langle b_m(t)\rangle_R,
\end{equation}
with
\begin{equation}
	\begin{split}
		M_{nm}&:=\alpha_n\delta_{nm}
		-i(g_m\delta_{n\,m+1}+g_{m-1}\delta_{n\,m-1})
		\\
		&=\begin{pmatrix}
			\alpha_1	&	-ig_1	&	0	&	\dots	& 0
			\\
			-ig_1	&	\alpha_2	& \ddots	& & \vdots
			\\
			0	&	\ddots	& \ddots	&	& 0
			\\
			\vdots	& & & \alpha_{N-1} & -ig_{N-1}
			\\
			0	& 	\dots & 0	& -ig_{N-1}	& \alpha_N
		\end{pmatrix},
	\end{split}
\end{equation}
where we have introduced the shorthand $\alpha_n\coloneqq-\frac{\Gamma_n}{2}-i\Omega_n$. Eq.~\eqref{eq:<b(t)>} can be solved formally by diagonalizing the tridiagonal matrix $M$. Since $M$ is not Hermitian, one has
\[
	M=S\Lambda S^{-1},
\]
with $\Lambda\coloneqq\mathrm{diag}(\lambda_1,\dots,\lambda_N)$ the diagonal matrix containing the eigenvalues, $S\coloneqq(\mathbf{u}^1,\dots,\mathbf{u}^N)$ a matrix made of arbitrarily normalized right eigenvectors $\mathbf{u}^n$ and $S^{-1}\coloneqq(\mathbf{v}^1,\dots,\mathbf{v}^N)^T$ its inverse, whose rows $(\mathbf{v}^n)^T$ are left eigenvectors. Since $M$ is a symmetric matrix, left and right eigenvectors are the same, so $S^{-1}$ is just the transpose of $S$ up to normalization of the rows in such a way that
\[
	\sum^N_{l=1}v^m_lu^n_l=\delta_{mn}.
\]
Assuming that none of the eigenvalues are degenerate, which is always the case in numerical applications since the $\Lambda$ matrices with equal diagonal elements are a zero-measure set, the evolution of the expectation value $\langle b_n(t)\rangle_R$ is thus
\begin{equation}
	\begin{split}
		\langle b_n(t)\rangle_R
		&=\sum^N_{m=1}(Se^{\Lambda t}S^{-1})_{nm}\langle b_m(0)\rangle_R
		\\
		&=\sum^N_{k,m=1}e^{\lambda_kt}
		u^k_nv^k_m\langle b_m(0)\rangle_R,
	\end{split}
\end{equation}
and extends to the correlation functions $\langle b_n(t)b^\dagger_m(0)\rangle_R$ by the quantum regression hypothesis, which is true by construction in the context of Theorem~\ref{TSO_Theorem}~\citep{TSO_Theorem}; since $\langle b_n(0)b^\dagger_m(0)\rangle_R=\delta_{nm}$ on our initial state, one has
\begin{equation}
	\langle b_n(t)b^\dagger_m(0)\rangle_R
	=\sum^N_{k=1}e^{\lambda_kt}u^k_nv^k_m,
\end{equation}
which can now be substituted into~\eqref{eq:C_R} to give the expression found in the main text:
\begin{equation}
	C^R(t)=\sum^N_{n=1}
	\left(\sum^N_{k,l=1}c_kc^*_l
	u^n_kv^n_l\right)e^{\lambda_nt},
\end{equation}
or
\begin{equation}\label{eq:CLambdaW}
	C^R(t)=\sum^N_{n=1}w_ne^{\lambda_nt}
\end{equation}
in terms of the coefficients
\begin{equation}\label{eq:Ws}
	w_n\coloneqq\sum^N_{k,l=1}c_kc^*_lu^n_kv^n_l.
\end{equation}

If degenerate eigenvalues $\lambda^\mathrm{d}_k$ are present, the time evolution in the corresponding subspace will be driven by $e^{\lambda^\mathrm{d}_kt}$ times growing powers of $t$; we did not consider this case for the sake of simplicity, but it may be useful to keep in mind that a mixed algebraic and exponential time dependence of correlation functions is not entirely ruled out by considering a Lindblad dynamics. If one wishes to explore this possibility in the TSO method, equality of two or more eigenvalues should be enforced at the level of the initial fit of the original correlation function $C^E(t)$ (see the next paragraph), since its spontaneous occurrence in the numerical procedure is virtually impossible.

\subsection{Inversion problem from a given correlation function}

To construct an effective environment whose $C^R(t)$ is as similar as possible to the $C^E(t)$ of a given unitary environment, we first fit $C^E(t)$ with a linear combination of $N$ complex exponentials $e^{\tilde{\lambda}_nt}$ weighted by complex coefficients $\tilde{w}_n$, with $N$ large enough to give an accurate fit, and then work backwards from Eq.~\eqref{eq:CLambdaW} to find the parameters that give the best approximation of the target function.

Since the real parameters in $C^R(t)$ are $4N-1$ (taking into account the fact that $\sum^N_{n=1}w_n=\sum^N_{n=1}|c_n|^2$ is real and positive by construction) and it takes $5N-2$ real parameters ($N$ frequencies, $N$ damping rates, $N-1$ couplings and $N$ complex coefficients $c_n$ minus one overall redundant phase) to identify an effective environment, this is a highly nontrivial inversion problem, because the map from effective environments to correlation functions is both nonlinear and many-to-one. This means that existence or uniqueness of a solution to our problem are not guaranteed in general; furthermore, we must require $\Gamma_n>0$ for all $n$ in order for our effective master equation to be meaningful, which sets another important constraint.

It is useful to break down the problem into two parts: first an inverse eigenvalue problem leading from the $\tilde{\lambda}_n$ to the dynamical matrix $M$, and then a system of equations relating the coefficients $\tilde{w}_n$ to the interaction operator parameters $c_n$. This allows us to deal with the sign constraints on the rates once and for all in the first half of the solution procedure, and to exploit the fact that the $c_n$ only appear in the second.

To determine the relation between the eigenvalues and elements of the matrix $M$, it is not convenient to look for symbolic expressions for each eigenvalue in terms of the parameters, since these would necessarily involve high-degree roots of complex polynomials. A simpler approach is to consider the characteristic polynomial of $M$
\[
	p_M(\lambda)\coloneqq\mathrm{det}(\lambda\mathbb{I}-M)
	=\prod^N_{n=1}(\lambda-\lambda_n),
\]
substitute the target eigenvalues $\tilde{\lambda}_n$ on the right-hand side and equate the coefficients of like powers of $\lambda$, which are geometrical invariants of any operator. The result is a system of equations of degrees 1 through $N$
\begin{equation}\label{eq:LambdatoAlpha}
	\left\{
		\begin{array}{cl}
			\displaystyle{\sum^N_{n=1}\alpha_n}
			&=\displaystyle{\sum^N_{n=1}\tilde{\lambda}_n}
			\\
			\displaystyle{\sum^N_{m\neq n}\alpha_m\alpha_n
			+\sum^{N-1}_{n=1}g^2_n}
			&=\displaystyle{\sum^N_{m\neq n}
			\tilde{\lambda}_m\tilde{\lambda}_n}
			\\
			&\;\vdots
			\\
			\displaystyle{\mathrm{det}(M)}
			&=\displaystyle{\prod^N_{n=1}\tilde{\lambda}_n}
		\end{array}
	\right.
\end{equation}
stating the invariance of the sums of principal minors order by order (the trace and the determinant appearing in the first and last equation being the simplest such invariants).

Now, Eq.~\eqref{eq:LambdatoAlpha} can be regarded as a parametric system of equations in the couplings $g_n$. With the $g_n$ fixed, it becomes an algebraic nonlinear system of $N$ equations in $N$ unknowns which can be solved numerically to give multiple sets of $\alpha_n$---i.e.\ frequencies $\Omega_n$ and rates $\Gamma_n$ whose sign can be checked directly---and therefore the entire dynamical matrix $M$.

Given a dynamical matrix $M$ obtained by choosing some set of $g_n$ and solving Eq.~\eqref{eq:LambdatoAlpha}, its eigenvectors $\mathbf{u}^n$ and $\mathbf{v}^n$ can be substituted into the $w_n$ as defined in Eq.~\eqref{eq:Ws}, which then become functions of the $c_n$ only and can be equated with the target values $\tilde{w}_n$
\begin{equation}\label{eq:WtoC}
	\left\{
		\begin{array}{cl}
			\displaystyle{\sum^N_{m,n=1}
			c^*_mv^1_mu^1_nc_n}
			&=\displaystyle{\tilde{w}_1}
			\\
			&\;\vdots
			\\
			\displaystyle{\sum^N_{m,n=1}
			c^*_mv^N_mu^N_nc_n}
			&=\displaystyle{\tilde{w}_N}
		\end{array}
	\right.
\end{equation}
to solve the second half of the problem. These $N$ complex equations are equivalent to $2N-1$ equations in $2N-1$ real unknowns, since the overall phase of all $c_n$ drops out of the left-hand side while on the right-hand side $\sum^N_{n=1}\tilde{w}_n=C^E(0)$ has no imaginary part.

\begin{figure*}
	\centering
	\includegraphics[scale=1]{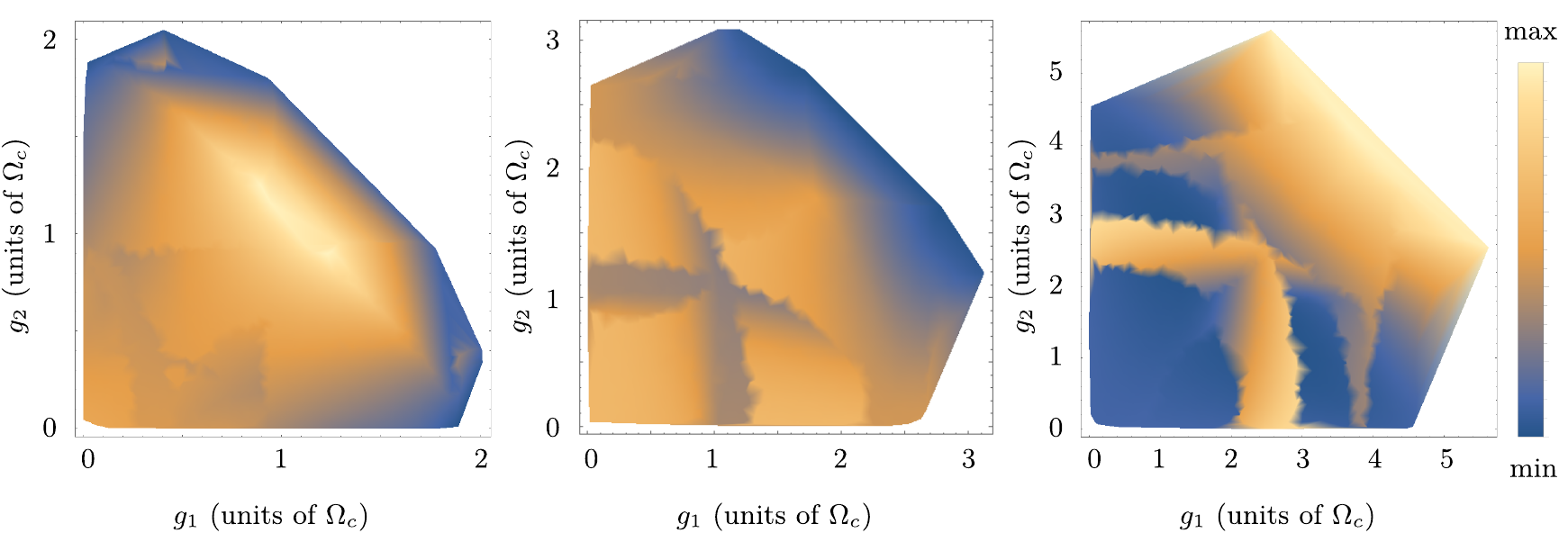}
	\caption{\label{Fig:gspace}Shape of the space of $g_n$ such that all $\Gamma_n$ are positive, with $N=3$ and three different thermal correlation functions $C^E_\beta(t)$ corresponding to $J(\omega)=\omega e^{-\omega/\Omega_c}$, $J(\omega)=(\omega^2/\Omega_c)e^{-\omega/\Omega_c}$ and $J(\omega)=(\omega^5/\Omega^4_c)e^{-\omega/\Omega_c}$, all at $\beta\Omega_c=0.85$. The axes show the values of $g_1$ and $g_2$ and the color denotes accuracy of the trial correlation function as estimated by $I_1(t_\mathrm{max})$ with $\Omega_ct_\mathrm{max}=25$ and normalized to the maximum accuracy obtained for each case, with blue areas representing smaller errors and yellow and orange ones indicating very vague resemblance.}
\end{figure*}

A set of $c_n$ solving Eq.~\eqref{eq:WtoC} does not always exist, so here we numerically minimize the Manhattan distance between the $w_n$ on the left-hand side and the $\tilde{w}_n$ instead. At this point, we have converted an arbitrary $(N-1)$-tuple of coupling constants into a trial effective correlation function
\[
	C^R_\mathrm{trial}(t)=\sum^N_{n=1}w_n(g_m,\alpha_m,c_m)e^{\lambda_n(g_m,\alpha_m)t}
\]
which can be compared to the target $C^E(t)$ according to some figure of merit. We used the integral
\begin{equation}\label{eq:fmerit1}
	I_1(t_\mathrm{max})\coloneqq\int^{t_\mathrm{max}}_0\!\!\!\!\!\!\!\!
	\mathrm{d}t'\!\!\int^{t'}_0\!\!\!\!\!\mathrm{d}t''\,
	|C^R_\mathrm{trial}(t'-t'')-C^E(t'-t'')|
\end{equation}
up to some final time $t_\mathrm{max}$ such that $C^E(t_\mathrm{max})\ll1$ in all cases where we had a closed expression for it, and
\begin{equation}\label{eq:fmerit2}
	I_2(t_\mathrm{max})\coloneqq\Delta t\sum^{N_\mathrm{max}}_{n=1}
	|C^R_\mathrm{trial}(n\Delta t)-C^E(n\Delta t)|
\end{equation}
for some number of points $N_\mathrm{max}$ and timestep $\Delta t=t_\mathrm{max}/N_\mathrm{max}$ when $C^E(t)$ was only known in integral form and needed to be evaluated for each value of the time argument. This whole procedure can be carried out for many values of the couplings in the physical parameter region, ranking the corresponding trial correlation functions by their values of the figure of merit in search of an optimum, in the spirit of the error bounds in Ref.~\citep{SpinBosonBounds} which relate the absolute difference between correlation functions to the changes in the reduced dynamics.

To summarize the steps described above, in order to find an effective environment corresponding to some correlation function $C^E(t)$, we first fit it with complex exponentials, and then overcome the mismatch between the number of variables from this fit and the number of parameters in the effective environment by setting up a variational problem in the $g_n$ couplings between neighboring surrogate modes. We sample multiple $(N-1)$-tuples $(g_1,\dots,g_{N-1})$ in a suitably sized open set $(0,g_\mathrm{max})^{N-1}$, solve Eq.~\eqref{eq:LambdatoAlpha} for each of them and then plug the eigenvectors of all physically acceptable matrices $M$ found into Eq.~\eqref{eq:WtoC} to determine the $c_n$. The trial correlation functions $C^R_\mathrm{trial}(t)$ constructed from each set of parameters are ranked according to the estimators~\eqref{eq:fmerit1} or~\eqref{eq:fmerit2}, depending on the original $C^E(t)$, and we search for the minimum of the figure of merit in the space of the $g_n$.

This variational problem is not convex in general: both the shape of the region in $g_n$-space leading to physically admissible solutions and the dependence of the cost functions defined in Eqs.~\eqref{eq:fmerit1} and~\eqref{eq:fmerit2} on the couplings can be highly nontrivial, with trenches, pointed features, local minima and gaps without any solutions at all appearing at unpredictable locations. We have also found no obvious patterns giving any hints as to the existence of a region of physically acceptable values of $g_n$ (for any $N$) such that there are exact solutions, or the form that such a region may have, based on the target parameters. All these mathematical features are very model-dependent; Fig.~\ref{Fig:gspace} shows some examples of parameter space shapes and cost function behavior for different correlation functions approximated using $N=3$ surrogate oscillators. Though none of the examples shown gave a $C^R(t)$ of sufficient accuracy for practical use due to the small number of effective modes, they nonetheless give a clear qualitative idea of the variety of possible outcomes. In general, sampling the parameter space efficiently is difficult, and we are looking for ways to improve this part of the algorithm.

\section{Special cases with exact solutions}
\label{app:ExactSols}

We will now show some examples, both general and related to the specific systems treated in the main text, of analytical solutions of the inversion problem in specific cases.

\subsection{One and two oscillators}

The simplest possible effective environment is a single damped oscillator ($N=1$ in Eq.~\eqref{eq:H_R_2}, with interaction operator $F_R=c(b+b^\dagger)$ since the phase of $c$ can be set to zero). This yields a correlation function
\[
	C^R(t)=c^2e^{-\frac{\Gamma}{2}|t|-i\Omega t}
\]
where the time dependence at $t<0$ is defined by $C^R(-t)\coloneqq C^{R*}(t)$ because the Lindblad equation only gives $C^R(t)$ for positive times, as discussed in the main text. The Fourier transform of this function is a Lorentzian of width $\Gamma$ centered in $\Omega$:
\[
	C^R(\omega)=c^2\frac{\Gamma}{\left(\Gamma/2\right)^2+\left(\omega-\Omega\right)^2}.
\]
A single sharp peak at zero temperature in the target spectral density can be mapped to a mode like this by simple nonlinear fitting of $C^E(t)$ with a complex exponential, as we did for the dimer simulations in the main text: in that case, the peaks were antisymmetrized Lorentzians so the frequency and damping rate of the effective mode matched those from the original spectral density almost exactly.

A less trivial, still exactly solvable case is given by two interacting oscillators and was already introduced by Garraway in Ref.~\citep{Garraway_pseudomodes} to show that not only sums but also differences of Lorentzians can be modeled by pseudomodes. In that paper, only one of the modes is coupled to the system (i.e.\ $c_2=0$); here we lift this assumption to show a more general result.

\begin{figure*}
	\centering
	\includegraphics[scale=1]{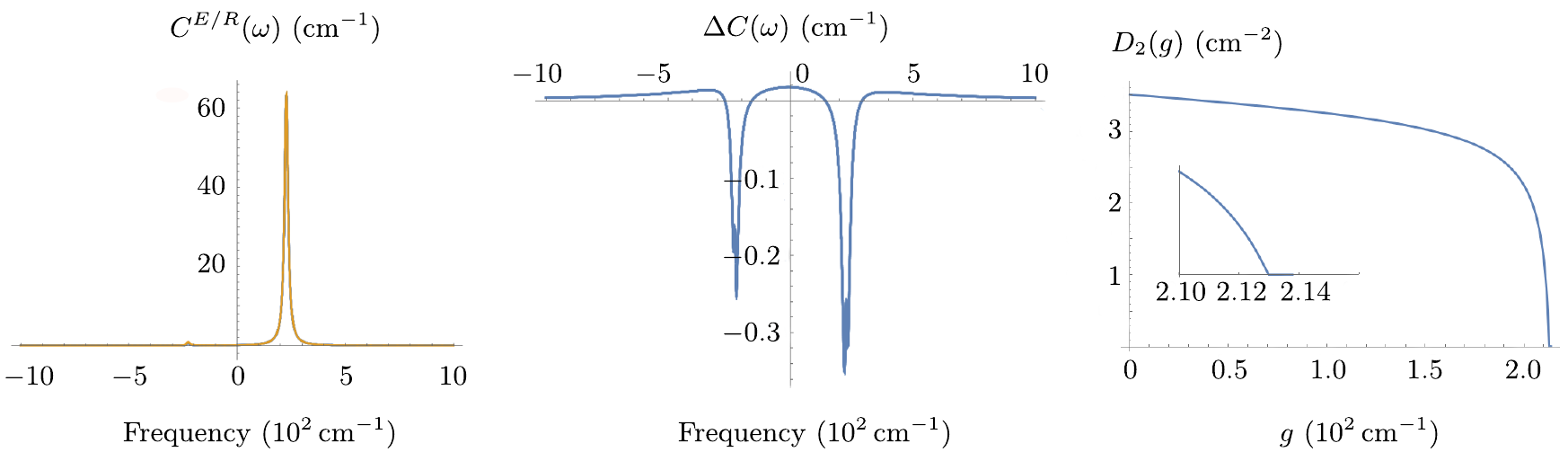}
	\caption{\label{Fig:2oscs}Left to right: overlapping plots of $C^E_\beta(\omega)$ and the effective $C^R_\beta(\omega)$ with $N=2$ for an antisymmetrized Lorentzian with peak frequency $\Omega_\mathrm{AL}=215\,\mathrm{cm}^{-1}$ and width $\Gamma_\mathrm{AL}=10\,\mathrm{cm}^{-1}$, at temperature $T=77\,\mathrm{K}$; a plot of the difference $\Delta C(\omega)\coloneqq C^E_\beta(\omega)-C^R_\beta(\omega)$; a plot of the minimum Manhattan distance $D_2(g)\coloneqq\sum^2_{n=1}|\tilde{w}_n-w_n(g)|$, with a very small region around $g=214\,\mathrm{cm}^{-1}$ (shown in the inset) in which the $\tilde{w}_n$ can only just be matched exactly, before the $\Gamma_n$ in the solutions change sign and higher values of $g$ no longer give acceptable solutions. Notice both the very steep descent of the error and the abrupt end of the physically admissible region on either side of this spot, as an example of how minima in our figures of merit quickly become hard to find in a more coarse-grained sampling in higher dimensions.}
\end{figure*}

The general correlation function for $N>1$ has the form
\begin{equation}
	C^R(t)=\sum^N_{n=1}w_ne^{\lambda_nt}
\end{equation}
with $\Re[\lambda_n]<0$, where the fact that the $w_n$ are complex changes the form in the frequency domain from a simple linear combination of Lorentzians to
\begin{equation}
	C^R(\omega)=-2\sum^N_{n=1}\frac{\Re[w_n]\Re[\lambda_n]
	+\Im[w_n](\omega+\Im[\lambda_n])}{\Re[\lambda_n]^2+(\omega+\Im[\lambda_n])^2},
\end{equation}
thus adding a linear frequency dependence in the numerator of the term associated with each mode. In phenomenological approaches where there is no intention of accurately simulating a specific correlation function, one may use \emph{ad hoc} combinations of weights and exponents to cancel terms in the numerator of the full $C^R(\omega)$ written as a single fractional polynomial and achieve a steeper fall-off in frequency than is possible with individual Lorentzians with positive coefficients (again, one example is given in Ref.~\citep{Garraway_pseudomodes}). Such strategies hardly generalize beyond specific applications but can be helpful to mitigate the error associated with the behavior of $C^R(\omega)$ near the origin, which does not comply with the fluctuation-dissipation theorem in general, as discussed in the main text.

In the case $N=2$, the eigenvalues $\lambda_{1,2}$ and weights $w_{1,2}$ depend on the effective environment parameters through the relations
\begin{equation}\label{eq:ParamsToCF}
	\begin{split}
		\lambda_{1,2}(\alpha_{1,2},g)&=\frac{\alpha_1+\alpha_2}{2}\pm
		\sqrt{\left(\frac{\alpha_1-\alpha_2}{2}\right)^2-g^2}
		\\
		w_{1,2}(\alpha_{1,2},g,c_{1,2})&=\frac{|c_1|^2+c^2_2}{2}
		\\
		&\quad\pm\frac{(|c_1|^2-c^2_2)(\alpha_1-\alpha_2)
		-4ig\Re[c_1]c_2}{2\sqrt{\left(\alpha_1-\alpha_2\right)^2-4g^2}},
	\end{split}
\end{equation}
where $c_2$ is taken to be real by fixing the overall phase mentioned in the preceding section. The first equation is readily inverted parametrically in $g$:
\begin{equation}\label{eq:AlphaFromLambda}
	\alpha_{1,2}=\frac{\tilde{\lambda}_1+\tilde{\lambda}_2}{2}\pm
	\sqrt{\left(\frac{\tilde{\lambda}_1-\tilde{\lambda}_2}{2}\right)^2+g^2}.
\end{equation}
The domain of physically admissible solutions is the set of all $g$ such that $\Gamma_{1,2}=-2\Re[\alpha_{1,2}]>0$ and can be found by using the formula for the square root of a complex number $z=z_\mathrm{R}+iz_\mathrm{I}$
\[
	\sqrt{z}=\sqrt{\frac{|z|+z_\mathrm{R}}{2}}+i\,\mathrm{sgn}(z_\mathrm{I})
	\sqrt{\frac{|z|-z_\mathrm{R}}{2}}.
\]
Using Eq.~\eqref{eq:AlphaFromLambda} in the equation for the weights, this becomes parametric in $g$ as well:
\begin{equation}\label{eq:WfromLambdaC}
	\begin{split}
		\tilde{w}_{1,2}&=\frac{|c_1|^2+c^2_2}{2}
		\\
		&\quad\mp\frac{(|c_1|^2-c^2_2)\sqrt{\left(\tilde{\lambda}_1
		-\tilde{\lambda}_2\right)^2+4g^2}
		-4ig\Re[c_1]c_2}{2(\tilde{\lambda}_1-\tilde{\lambda}_2)}.
	\end{split}
\end{equation}
This equation, which relates the four real quantities $g$, $\Re[c_1]$, $\Im[c_1]$ and $c_2$ to the three real numbers determining $\tilde{w}_{1,2}$, may or may not have a solution depending on the $g$ chosen, as discussed earlier: if the solution exists only for $g$ outside the physical region in which $\Gamma_{1,2}=-2\Re[\alpha_{1,2}]>0$, then it is necessary to operate variationally and minimize the distance $D_2(g)\coloneqq\sum^2_{n=1}|\tilde{w}_n-w_n(g)|$.

If one assumes $c_2=0$ in~\eqref{eq:ParamsToCF}, as is done in Ref.~\citep{Garraway_pseudomodes}, then the whole system can be inverted explicitly:
\begin{equation}\label{eq:OverExact2}
	\left\{
		\begin{array}{cl}
			|c_1|^2&=\tilde{w}_1+\tilde{w}_2
			\\[1.0em]
			\alpha_1&=\displaystyle{\frac{\tilde{w}_1\tilde{\lambda}_1
			+\tilde{w}_2\tilde{\lambda}_2}{\tilde{w}_1+\tilde{w}_2}}
			\\[1.0em]
			\alpha_2&=\displaystyle{\frac{\tilde{w}_1\tilde{\lambda}_2
			+\tilde{w}_2\tilde{\lambda}_1}{\tilde{w}_1+\tilde{w}_2}}
			\\[1.0em]
			g^2&=\displaystyle{\left(\frac{\tilde{\lambda}_1-\tilde{\lambda}_2}{2}\right)^2
			\left(\left(\frac{\tilde{w}_1-\tilde{w}_2}{\tilde{w}_1+\tilde{w}_2}\right)^2
			-1\right)}
		\end{array}
	\right.
\end{equation}
but the solution only exists if the $\tilde{\lambda}_{1,2}$, $\tilde{w}_{1,2}$ given are such that the expression on the right-hand side of the last equation has a vanishing imaginary part. This is because now the phase of $c_1$ has also decoupled from the problem, removing a second real degree of freedom and making the system overdetermined: the balance between equations and unknowns is thus restored by this real constraint appearing on the $\tilde{\lambda}_{1,2}$, $\tilde{w}_{1,2}$.

In our applications, we used pairs of effective modes to reproduce narrow antisymmetrized Lorentzians at nonzero temperature: we found no exact solution and had to minimize $D_2(g)$ in most cases, but e.g.\ for an antisymmetrized Lorentzian with peak frequency $\Omega_{AL}=215\,\mathrm{cm}^{-1}$ and width $\Gamma_{AL}=10\,\mathrm{cm}^{-1}$ at temperature $77\,\mathrm{K}$ the system can be solved exactly for $213.0\,\mathrm{cm}^{-1}<g<213.8\,\mathrm{cm}^{-1}$ (Fig.~\ref{Fig:2oscs}). Note that in all cases we considered, the best fit of such thermalized peaks with two modes was always obtained by mode frequencies close to zero and a strong coupling $g$ between the two; fitting the same function with two noninteracting modes at the positive and negative peak frequencies was consistently found to be a less accurate choice even for such seemingly obvious target functions.

One further possibility when dealing with very narrow high-frequency modes (so that the correlation function error around $\omega=0$ is very small) is to replace each such mode by a single oscillator, with $\Omega$ and $\Gamma$ equal to those of the antisymmetrized Lorentzian peak, and to initialize this mode in a Gibbs state at the bath temperature. Since independent oscillators or sets of oscillators have no correlations with each other in the initial state, this does not affect any other parts of the surrogate environment at hand in any way.

Such a mode would obey a full thermal Lindblad equation, with dissipator
\begin{multline}\label{eq:D_thermal}
	\mathcal{D}_{R\beta}[\rho_R]\coloneqq
	\Gamma^\uparrow_\beta\left(b^\dagger\rho_Rb
	-\frac{1}{2}\left\{bb^\dagger,\rho_R\right\}\right)
	\\
	+\Gamma^\downarrow_\beta\left(b\rho_Rb^\dagger
	-\frac{1}{2}\left\{b^\dagger b,\rho_R\right\}\right)
\end{multline}
comprising emission and absorption terms with rates obeying detailed balance:
\begin{align*}
	\Gamma^\uparrow_\beta&\coloneqq\Gamma\,n_\Omega(\beta)
	\\
	\Gamma^\downarrow_\beta&\coloneqq\Gamma\,(n_\Omega(\beta)+1),
\end{align*}
where $n_\Omega(\beta)\coloneqq(e^{\beta\Omega}-1)^{-1}$ is the Bose-Einstein distribution. Coupling a thermalized mode to a system via a coefficient $c$ results in a correlation function $C^R(t)$ combining two exponential contributions with weights proportional to the emission and absorption coefficients, which translate to two Lorentzians centered at $\pm\Omega$ in the frequency domain:
\begin{align}\label{eq:CF_ThermalMode}
	C^R(t)&=c^2\left((n_\Omega(\beta)+1)e^{-\frac{\Gamma}{2}|t|-i\Omega t}
	+n_\Omega(\beta)e^{-\frac{\Gamma}{2}|t|+i\Omega t}\right)
	\\
	C^R(\omega)
	&=c^2\left(\frac{\Gamma^\downarrow_\beta}{\left(\Gamma/2\right)^2
	+\left(\omega-\Omega\right)^2}
	+\frac{\Gamma^\uparrow_\beta}{\left(\Gamma/2\right)^2
	+\left(\omega+\Omega\right)^2}\right)\!.
\end{align}
This transformation can be convenient when a single oscillator with a thermalized population requires a lower local dimension than a pair of coupled surrogate modes initialized in the vacuum would, since this would limit the memory requirements of the simulation. We used this method to account for the strongly coupled high-frequency mode of the polymer simulations in Section~\ref{sec:Applications} of the paper.

\subsection{Three oscillators}

Adding a third oscillator, we found exact solutions for $c_2=0$, which we did not use in any of the simulations discussed in the main paper but can be useful in general.

For $N=3$, the system of eigenvalue equations is
\begin{equation}\label{eq:LambdatoAlpha3}
	\!\left\{\!\!
		\begin{array}{cl}
			\alpha_1+\alpha_2+\alpha_3
			&=\tilde{\lambda}_1+\tilde{\lambda}_2+\tilde{\lambda}_3
			\\[1.0em]
			\alpha_1\alpha_2+\alpha_2\alpha_3+\alpha_3\alpha_1+g^2_1+g^2_2\!\!
			&=\tilde{\lambda}_1\tilde{\lambda}_2+\tilde{\lambda}_2\tilde{\lambda}_3
			+\tilde{\lambda}_3\tilde{\lambda}_1
			\\[1.0em]
			\alpha_1\alpha_2\alpha_3+g^2_1\alpha_1+g^2_2\alpha_3
			&=\tilde{\lambda}_1\tilde{\lambda}_2\tilde{\lambda}_3
		\end{array}
	\right.
\end{equation}
and one may remove $\alpha_2$ from the last two equations by using the first, so that $\alpha_1$ and $\alpha_3$ can be regarded as effective functions of the real parameters $g_1$ and $g_2$.

With $c_2$ set to zero, the whole inversion problem is determined, since the equations for the $\tilde{w}_n$ will determine the values of $g_1$ and $g_2$ instead. Setting the overall phase so that $c_3$ is real, the equations can be written as
\begin{equation}\label{eq:WtoC3}
	\left\{
		\begin{array}{cl}
			|c_1|^2+c^2_3&=\tilde{w}_1+\tilde{w}_2+\tilde{w}_3
			\\[1.0em]
			|c_1|^2\alpha_1+c^2_3\alpha_3
			&=\tilde{w}_1\tilde{\lambda}_1+\tilde{w}_2\tilde{\lambda}_2
			+\tilde{w}_3\tilde{\lambda}_3
			\\[1.0em]
			2\Re[c_1]c_3g_1g_2&=-\tilde{w}_3(\tilde{\lambda}_1-\tilde{\lambda}_3)
			(\tilde{\lambda}_2-\tilde{\lambda}_3)
			\\[1.0em]
			&\quad-\displaystyle{\frac{(\alpha_1-\tilde{\lambda}_1)(\alpha_1
			-\tilde{\lambda}_2)(\alpha_3-\tilde{\lambda}_3)}{(\alpha_3-\alpha_1)
			(\alpha_1-\tilde{\lambda}_3)}|c_1|^2}
			\\[1.0em]
			&\quad-\displaystyle{\frac{(\alpha_3-\tilde{\lambda}_1)(\alpha_3
			-\tilde{\lambda}_2)(\alpha_1-\tilde{\lambda}_3)}{(\alpha_3-\alpha_1)
			(\alpha_3-\tilde{\lambda}_3)}c^2_3}
		\end{array}
	\right.
\end{equation}
where the last line again features a real expression on the left-hand side and a complex one whose imaginary part must be zero on the right-hand side. Since the first equation is real by construction, there are five real equations in the five real variables $g_1,g_2,\Re[c_1],\Im[c_1],c_3$ in Eq.~\eqref{eq:WtoC3}, so the existence of solutions is only subject to the constraint $\Gamma_n=-2\Re[\alpha_n]>0$.

If $c_3$ is also set to zero, then the system~\eqref{eq:WtoC3} becomes
\begin{equation}
	\left\{
		\begin{array}{cl}
			c^2_1&=\tilde{w}_1+\tilde{w}_2+\tilde{w}_3
			\\[1.0em]
			c^2_1\alpha_1
			&=\tilde{w}_1\tilde{\lambda}_1+\tilde{w}_2\tilde{\lambda}_2
			+\tilde{w}_3\tilde{\lambda}_3
			\\[1.0em]
			0&=-\tilde{w}_3(\tilde{\lambda}_1-\tilde{\lambda}_3)
			(\tilde{\lambda}_2-\tilde{\lambda}_3)
			\\[1.0em]
			&\quad-\displaystyle{\frac{(\alpha_1-\tilde{\lambda}_1)(\alpha_1
			-\tilde{\lambda}_2)(\alpha_3-\tilde{\lambda}_3)}{(\alpha_3-\alpha_1)
			(\alpha_1-\tilde{\lambda}_3)}c^2_1}.
		\end{array}
	\right.
\end{equation}
and can be inverted explicitly, giving $c^2_1$, $\alpha_1$ and $\alpha_3$. But now the system~\eqref{eq:LambdatoAlpha3} is overdetermined: the trace gives $\alpha_2$, and the last two complex equations can give $g_{1,2}$ only if the $\tilde{\lambda}_n$ and $\tilde{w}_n$ happen to satisfy two real relations among themselves (one because $c_3$ was removed from the problem, another because the phase of $c_1$ is now irrelevant). In particular, the expressions whose imaginary part must vanish now appear on the right-hand side of the last two lines of the full solution
\begin{widetext}
\begin{equation}\label{eq:OverExact3}
	\left\{
		\begin{array}{cl}
			c^2_1&=\tilde{w}_1+\tilde{w}_2+\tilde{w}_3
			\\[1.0em]
			\alpha_1&=\displaystyle{\frac{\tilde{w}_1\tilde{\lambda}_1
			+\tilde{w}_2\tilde{\lambda}_2+\tilde{w}_3\tilde{\lambda}_3}{
			\tilde{w}_1+\tilde{w}_2+\tilde{w}_3}}
			\\[1.0em]
			\alpha_2&=\displaystyle{\frac{(\tilde{w}_2+\tilde{w}_3)\tilde{\lambda}_1
			+(\tilde{w}_3+\tilde{w}_1)\tilde{\lambda}_2+(\tilde{w}_1+\tilde{w}_2)
			\tilde{\lambda}_3}{\tilde{w}_1+\tilde{w}_2+\tilde{w}_3}}
			\\[1.0em]
			&\quad-\displaystyle{\frac{\tilde{w}_2\tilde{w}_3(\tilde{\lambda}_2
			-\tilde{\lambda}_3)^2\tilde{\lambda}_1+\tilde{w}_3\tilde{w}_1(\tilde{\lambda}_3
			-\tilde{\lambda}_1)^2\tilde{\lambda}_2+\tilde{w}_1\tilde{w}_2(\tilde{\lambda}_1
			-\tilde{\lambda}_2)^2\tilde{\lambda}_3}{
			\tilde{w}_2\tilde{w}_3(\tilde{\lambda}_2-\tilde{\lambda}_3)^2
			+\tilde{w}_3\tilde{w}_1(\tilde{\lambda}_3-\tilde{\lambda}_1)^2
			+\tilde{w}_1\tilde{w}_2(\tilde{\lambda}_1-\tilde{\lambda}_2)^2}}
			\\[1.0em]
			\alpha_3&=\displaystyle{\frac{\tilde{w}_2\tilde{w}_3(\tilde{\lambda}_2
			-\tilde{\lambda}_3)^2\tilde{\lambda}_1+\tilde{w}_3\tilde{w}_1(\tilde{\lambda}_3
			-\tilde{\lambda}_1)^2\tilde{\lambda}_2+\tilde{w}_1\tilde{w}_2(\tilde{\lambda}_1
			-\tilde{\lambda}_2)^2\tilde{\lambda}_3}{
			\tilde{w}_2\tilde{w}_3(\tilde{\lambda}_2-\tilde{\lambda}_3)^2
			+\tilde{w}_3\tilde{w}_1(\tilde{\lambda}_3-\tilde{\lambda}_1)^2
			+\tilde{w}_1\tilde{w}_2(\tilde{\lambda}_1-\tilde{\lambda}_2)^2}}
			\\[1.0em]
			g^2_1&=\displaystyle{-\frac{\tilde{w}_2\tilde{w}_3(\tilde{\lambda}_2
			-\tilde{\lambda}_3)^2+\tilde{w}_3\tilde{w}_1(\tilde{\lambda}_3
			-\tilde{\lambda}_1)^2+\tilde{w}_1\tilde{w}_2(\tilde{\lambda}_1
			-\tilde{\lambda}_2)^2}{(\tilde{w}_1+\tilde{w}_2+\tilde{w}_3)^2}}
			\\[1.0em]
			g^2_2&=\displaystyle{-\frac{\tilde{w}_1\tilde{w}_2\tilde{w}_3
			(\tilde{\lambda}_2-\tilde{\lambda}_3)^2(\tilde{\lambda}_3-\tilde{\lambda}_1)^2
			(\tilde{\lambda}_1-\tilde{\lambda}_2)^2(\tilde{w}_1+\tilde{w}_2+\tilde{w}_3)
			}{\left(\tilde{w}_2\tilde{w}_3(\tilde{\lambda}_2-\tilde{\lambda}_3)^2
			+\tilde{w}_3\tilde{w}_1(\tilde{\lambda}_3-\tilde{\lambda}_1)^2
			+\tilde{w}_1\tilde{w}_2(\tilde{\lambda}_1-\tilde{\lambda}_2)^2\right)^2}}.
		\end{array}
	\right.
\end{equation}

\section{Effective parameters}
\label{app:Parameters}

We list here several sets of effective parameters used in the simulations discussed in the main text, along with the local dimensions of each mode in each set at convergence. The corresponding spectral densities are defined in Eq.~\eqref{eq:OhmicJ}, Eq.~\eqref{eq:J_AdRe} and Eq.~\eqref{eq:J_AntiL}, respectively.

\FloatBarrier

\begin{table}
	\centering
	\textbf{Ohmic spectral density} \\[2ex]
	\begin{tabular*}{.85\textwidth}{@{\extracolsep{\fill}}cccccccc}
	\toprule
	& Mode 1	& &	Mode 2	& &	Mode 3	& &	Mode 4 \\
	\midrule
	$\Omega_n$	&	$2.70796$	& &	$2.13014$	& &	$1.15884$	& &	$0.310906$ \\[0.5ex]
	$g_n$ & &	$3.38195$	& &	$1.43514$	& &	$0.491546$ & \\[0.5ex]
	$\Gamma_n$ & $11.9298$	& &	$0.573494$	& &	$0.0317143$	& &	$0.000795693$ \\[0.5ex]
	\multirow{2}{*}{$c_n$} & $-0.0333215$	& &	$0.319$	& &	$0.760716$	& &
	\multirow{2}{*}{$0.579218$} \\
	& $-0.0121362i$	& &	$+0.0811955i$	& &	$+0.0175762i$	& & \\
	\midrule
	$d_\mathrm{loc}$ (spin)	& $3$	& & $4$	& & $5$	& & $7$ \\
	$d_\mathrm{loc}$ (chain)	& $4$	& & $4$	& & $5$	& & $7$ \\
	\bottomrule
	\end{tabular*}
	\caption{\label{tb:OhmicT0}Ohmic spectral density with cutoff frequency $\Omega_c$, temperature $T=0$: parameters in units $\Omega_c$ and local dimensions.}
\end{table}

\begin{table}
	\centering
	\begin{tabular*}{.85\textwidth}{@{\extracolsep{\fill}}cccccccc}
	\toprule
	& Mode 1	& &	Mode 2	& &	Mode 3	& &	Mode 4 \\
	\midrule
	$\Omega_n$	&	$0.512683$	& &	$2.53779$	& &	$4.53293$	& &	$0.151433$ \\[0.5ex]
	$g_n$ & &	$1.82454$	& &	$3.20774$	& &	$1.60194$ & \\[0.5ex]
	$\Gamma_n$ & $0.056336$	& &	$4.42709$	& &	$15.7371$	& &	$0.110104$ \\[0.5ex]
	\multirow{2}{*}{$c_n$} & $-0.962917$	& &	$-0.227707$	& &	$0.231179$	& &
	\multirow{2}{*}{$0.818093$} \\
	& $+0.819128i$	& &	$+0.0701249i$	& &	$-0.137866i$	& & \\
	\midrule
	$d_\mathrm{loc}$ (spin)	& $5$	& & $4$	& & $4$	& & $7$ \\
	$d_\mathrm{loc}$ (chain)	& $7$	& & $4$	& & $3$	& & $8$ \\
	\bottomrule
	\end{tabular*}
	\caption{\label{tb:OhmicT1}Ohmic spectral density with cutoff frequency $\Omega_c$, temperature $T=\Omega_c$: parameters in units $\Omega_c$ and local dimensions.}
\end{table}

\begin{table}
	\centering
	\begin{tabular*}{.85\textwidth}{@{\extracolsep{\fill}}cccccccccc}
	\toprule
	& Mode 1	& &	Mode 2	& &	Mode 3	& &	Mode 4 & & Mode 5\\
	\midrule
	$\Omega_n$	& $0.306859$	& & $0.361308$	& & $0.167597$	& & $0.0297981$	& &
	$0.00236395$ \\[0.5ex]
	$g_n$ & &	$4.17718$	& & $2.1243$	& & $0.673391$	& & $0.166947$ \\[0.5ex]
	$\Gamma_n$ & $16.0093$	& & $2.76375$	& & $0.00358704$	& & $0.0949691$	& &
	$0.0517414$ \\[0.5ex]
	\multirow{2}{*}{$c_n$} & $-0.166675$	& &	$0.21927$	& &	$1.61933$	& &	$0.187388$
	& & \multirow{2}{*}{$1.1553$} \\
	& $-0.0342019i$	& & $+0.103791i$	& & $-0.00703994i$	& & $-1.07416i$	& &
	\\
	\midrule
	$d_\mathrm{loc}$	& $3$	& & $3$	& & $4$	& & $4$	& & $6$ \\
	\bottomrule
	\end{tabular*}
	\caption{\label{tb:OhmicT2.5}Ohmic spectral density with cutoff frequency $\Omega_c$, temperature $T=\frac{5}{2}\Omega_c$: parameters in units $\Omega_c$ and local dimensions.}
\end{table}

\begin{table}
	\centering
	\textbf{Adolphs--Renger spectral density} \\[2ex]
	\begin{tabular*}{.7\textwidth}{@{\extracolsep{\fill}}cccccccc}
	\toprule
	& Mode 1	& &	Mode 2	& &	Mode 3	& &	Mode 4 \\
	\midrule
	$\Omega_n$	&	$0.718918$	& &	$3.06543$	& &	$2.96082$	& &	$0.667101$ \\[0.5ex]
	$g_n$ & &	$2.10958$	& &	$3.91248$	& &	$1.56527$ & \\[0.5ex]
	$\Gamma_n$ & $0.00554063$	& &	$15.4881$	& &	$0.00291091$	& &	$0.294244$ \\[0.5ex]
	\multirow{2}{*}{$c_n$} & $-0.57271$	& &	$-0.0147923$	& &	$0.725729$	& &
	\multirow{2}{*}{$0.409762$} \\
	& $+0.06491i$	& &	$+0.0820348i$	& &	$+0.0119678i$	& & \\
	\midrule
	$d_\mathrm{loc}$	& $6$	& & $4$	& & $4$	& & $4$ \\
	\bottomrule
	\end{tabular*}
	\caption{\label{tb:AR0K}Adolphs--Renger spectral density, temperature $T=0$: parameters in units $u=100\,\mathrm{cm}^{-1}$ and local dimensions.}
\end{table}

\begin{table}
	\centering
	\begin{tabular*}{.7\textwidth}{@{\extracolsep{\fill}}cccccccc}
	\toprule
	& Mode 1	& &	Mode 2	& &	Mode 3	& &	Mode 4 \\
	\midrule
	$\Omega_n$	&	$3.05106$	& &	$2.74196$	& &	$0.00670418$	& &	$0.00780109$\\[0.5ex]
	$g_n$ & &	$2.74161$	& &	$2.01796$	& &	$0.33975$ & \\[0.5ex]
	$\Gamma_n$ & $0.0284151$	& &	$11.6481$	& &	$0.00549033$	& &	$0.0184315$ \\[0.5ex]
	\multirow{2}{*}{$c_n$} & $-0.910465$	& &	$-0.135049$	& &	$0.524001$	& &
	\multirow{2}{*}{$0.114767$} \\
	& $-0.0164266i$	& &	$-0.0104797i$	& &	$+0.317767i$	& & \\
	\midrule
	$d_\mathrm{loc}$	& $5$	& & $4$	& & $6$	& & $8$ \\
	\bottomrule
	\end{tabular*}
	\caption{\label{tb:AR77K}Adolphs--Renger spectral density, temperature $T=77\,\mathrm{K}$: parameters in units $u=100\,\mathrm{cm}^{-1}$ and local dimensions.}
\end{table}

\begin{table}
	\centering
	\begin{tabular*}{.7\textwidth}{@{\extracolsep{\fill}}cccccccc}
	\toprule
	& Mode 1	& &	Mode 2	& &	Mode 3	& &	Mode 4 \\
	\midrule
	$\Omega_n$	&	$0.788783$	& &	$0.414407$	& &	$-0.0300357$	& &	$-0.034035$ \\[0.5ex]
	$g_n$ & &	$3.10576$	& &	$0.978945$	& &	$0.294823$ & \\[0.5ex]
	$\Gamma_n$ & $10.4575$	& &	$0.0934767$	& &	$0.00983292$	& &	$0.0167273$ \\[0.5ex]
	\multirow{2}{*}{$c_n$} & $0.189405$	& &	$1.23326$	& &	$0.0221509$	& &
	\multirow{2}{*}{$0.365249$} \\
	& $+0.0639657i$	& &	$+0.451035i$	& &	$+0.962709i$	& & \\
	\midrule
	$d_\mathrm{loc}$	& $3$	& & $4$	& & $7$	& & $7$ \\
	\bottomrule
	\end{tabular*}
	\caption{\label{tb:AR300K}Adolphs--Renger spectral density, temperature $T=300\,\mathrm{K}$: parameters in units $u=100\,\mathrm{cm}^{-1}$ and local dimensions.}
\end{table}

\begin{table}
	\centering
	\textbf{Antisymmetrized Lorentzian spectral densities} \\[2ex]
	\begin{tabular*}{.7\textwidth}{@{\extracolsep{\fill}}cc|ccc|ccc}
	\toprule
	& $T=0\,\mathrm{K}$	& \multicolumn{3}{c}{$T=77\,\mathrm{K}$}
	& \multicolumn{3}{c}{$T=300\,\mathrm{K}$} \\
	& &	Mode 1	& & Mode 2	& Mode 1	& & Mode 2 \\
	\midrule
	$\Omega_n$	& $2.275$ &	$0.662126$	& & $-0.667153$	&	$-0.00139464$	&
	& $0.0013106$ \\[0.5ex]
	$g_n$ & $-$	& &	$2.1788$	& &	& $2.2772$ \\[0.5ex]
	$\Gamma_n$ & $0.197195$ & $0.264596$	& & $0.0788813$	&	$0.00326568$	&
	& $0.396252$ \\[0.5ex]
	\multirow{2}{*}{$c_n$} & \multirow{2}{*}{$0.440408$} & $0.333222$	& &
	\multirow{2}{*}{$0.296358$}	&	$0.578109$	& &	\multirow{2}{*}{$0.169995$} \\
	& & $-0.000005i$	& &	&	$-0.176482i$	& &	\\
	\midrule
	$d_\mathrm{loc}$	& $5$ &	$4$	& & $4$	&	$4$	&
	& $4$ \\
	\bottomrule
	\end{tabular*}
	\caption{\label{tb:J1_227.5}Antisymmetrized Lorentzian spectral density with $\Omega=227.5\,\mathrm{cm}^{-1}$, $\Gamma=20\,\mathrm{cm}^{-1}$, $S=0.0379$: parameters in units $u=100\,\mathrm{cm}^{-1}$ and local dimensions.}
\end{table}

\begin{table}
	\centering
	\begin{tabular*}{.7\textwidth}{@{\extracolsep{\fill}}cc|ccc|ccc}
	\toprule
	& $T=0\,\mathrm{K}$	& \multicolumn{3}{c}{$T=77\,\mathrm{K}$}
	& \multicolumn{3}{c}{$T=300\,\mathrm{K}$} \\
	& &	Mode 1	& & Mode 2	& Mode 1	& & Mode 2 \\
	\midrule
	$\Omega_n$	& $2.00$ &	$-0.318699$	& &	$0.316331$	&	$-0.00048954$	& &	$0.000480821$
	\\[0.5ex]
	$g_n$ & $-$	& &	$1.976$	& &	& $2.00052$ \\[0.5ex]
	$\Gamma_n$ & $0.098296$ & $0.045988$	& &	$0.138442$	&	$0.00953908$	& &	$0.190362$ \\[0.5ex]
	\multirow{2}{*}{$c_n$} & \multirow{2}{*}{$0.992322$} & $0.764199$	& &
	\multirow{2}{*}{$0.676024$}	&	$1.45733$	& &	\multirow{2}{*}{$0.343374$} \\
	& & $+0.000002i$	& &	&	$+0.000003i$	& & \\
	\midrule
	$d_\mathrm{loc}$	& $6$ &	$5$	& & $6$	&	$8$	& & $8$ \\
	\bottomrule
	\end{tabular*}
	\caption{\label{tb:J1_200}Antisymmetrized Lorentzian spectral density with $\Omega=200\,\mathrm{cm}^{-1}$, $\Gamma=10\,\mathrm{cm}^{-1}$, $S=0.25$: parameters in units $u=100\,\mathrm{cm}^{-1}$ and local dimensions.}
\end{table}

\end{widetext}

\clearpage

\end{document}